%% file: main.tex
\documentclass[conference]{IEEEtran}
\usepackage{ifpdf}
\usepackage{cite}

\ifCLASSINFOpdf
  \usepackage[pdftex]{graphicx}
\else
\fi

\usepackage{amsmath}

\usepackage{array}
\ifCLASSOPTIONcompsoc
 \usepackage[caption=false,font=normalsize,labelfont=sf,textfont=sf]{subfig}
\else
  \usepackage[caption=false,font=footnotesize]{subfig}
\fi

\usepackage{stfloats}
\usepackage{url}
\usepackage{graphicx}
\usepackage{caption}
\usepackage{subcaption}
\usepackage{import}
\usepackage{comment}
\usepackage{xurl}
\usepackage{multirow}
\usepackage{tablefootnote}
\usepackage{listings}
\usepackage{xcolor}
\usepackage{tikz}
\usepackage{amsmath}
\usepackage{fontawesome5}

\usepackage[utf8]{inputenc}

\input{commands}
\newcommand{\seonghun}[1]{{\color{red}{SS: #1}}}

\usepackage{placeins}       % for \FloatBarrier
\usepackage{listings}
\usepackage{xcolor}

\definecolor{codegray}{gray}{0.5}
\definecolor{codegreen}{rgb}{0,0.6,0}
\definecolor{codered}{rgb}{0.6,0,0}

\lstdefinestyle{mystyle}{
    frame=tb,
    backgroundcolor=\color{white},       % white background
    commentstyle=\color{codegreen},
    keywordstyle=\color{blue},
    numberstyle=\tiny\color{codegray},
    stringstyle=\color{codered},
    basicstyle=\ttfamily\scriptsize,
    breakatwhitespace=false,         
    breaklines=true,                 
    captionpos=b,                    
    keepspaces=true,                 
    numbers=left,                    
    numbersep=5pt,                  
    showspaces=false,                
    showstringspaces=false,
    showtabs=false,                  
    tabsize=1,    
    belowskip=0pt, 
    xleftmargin=-2pt,  
    xrightmargin=-2pt,
    morekeywords={nohup,ovrgpuprofiler},
    comment=[l]{\#}
}

\definecolor{codegreen}{rgb}{0,0.6,0}
\definecolor{codegray}{rgb}{0.5,0.5,0.5}
\definecolor{codepurple}{rgb}{0.58,0,0.82}
\definecolor{backcolour}{rgb}{0.95,0.95,0.92}
\definecolor{codered}{rgb}{0.6,0,0}
\definecolor{codeblue}{rgb}{0,0,0.8}

\begin{document}


\title{Side-channel Inference of User Activities in AR/VR Using GPU Profiling}
%Inferring User Activities in XR using GPU Profiling
%Inferring User Activities in XR Systems Using GPU Profiling
%Please confirm the title with Prof. Reham.
%Inferring AR/VR Application Usage via GPU Performance Profiling}

%\title{\Large \bf OVRWatcher:\\ 
%\update{Inferring AR/VR Application Usage via GPU Performance Profiling}
%Inferring User Activities through GPU Profiling on AR/VR Devices
%}


\begin{comment}

\author{
{\rm Anonymous}
}
\end{comment}

\author{%
Seonghun Son\textsuperscript{*} \quad
Chandrika Mukherjee\textsuperscript{$\diamond$} \quad
Reham Mohamed Aburas\textsuperscript{$\dagger$} \quad
Berk Gulmezoglu\textsuperscript{*} \quad
Z.\ Berkay Celik\textsuperscript{$\diamond$}
\\
%[0.75em]
\textsuperscript{*}Iowa State University \quad
\textsuperscript{$\diamond$}Purdue University \quad
\textsuperscript{$\dagger$}American University of Sharjah 
\\
%[0.5em]
Emails:\{seonghun, bgulmez\}@iastate.edu,\ \{cmukherj, zcelik\}@purdue.edu,\ raburas@aus.edu
}

\begin{comment}

\author{\IEEEauthorblockN{Michael Shell}
	\IEEEauthorblockA{Georgia Institute of Technology\\
		someemail@somedomain.com}
	\and
	\IEEEauthorblockN{Homer Simpson}
	\IEEEauthorblockA{Twentieth Century Fox\\
		homer@thesimpsons.com}
	\and
	\IEEEauthorblockN{James Kirk\\ and Montgomery Scott}
	\IEEEauthorblockA{Starfleet Academy\\
		someemail@somedomain.com}}
\end{comment}


%Telephone: (800) 555--1212, Fax: (888) 555--1212}


\IEEEoverridecommandlockouts
\makeatletter\def\@IEEEpubidpullup{6.5\baselineskip}\makeatother
\IEEEpubid{\parbox{\columnwidth}{
		Network and Distributed System Security (NDSS) Symposium 2026\\
		23 - 27 February 2026 , San Diego, CA, USA\\
		ISBN 979-8-9919276-8-0\\  
		https://dx.doi.org/10.14722/ndss.2026.231302\\
		www.ndss-symposium.org
}
\hspace{\columnsep}\makebox[\columnwidth]{}}

\maketitle


\begin{comment}

\begin{abstract}
The abstract goes here.
\end{abstract}
\end{comment}

\import{Sections/}{0_Abstract.tex}


%\ifCLASSOPTIONpeerreview
% \fi
\IEEEpeerreviewmaketitle

\import{Sections/}{1_Introduction.tex}
\import{Sections/}{2_Background.tex}

\import{Sections/}{3_ProblemStatement.tex}
\import{Sections/}{4_System.tex}
\import{Sections/}{5_Evaluation.tex}

\import{Sections/}{5-1_UserStudy.tex}
%\import{Sections/}{6_Evaluation.tex}
% \import{Sections/}{6_Countermeasure.tex}
%\import{Sections/}{7_RelatedWork.tex}
\import{Sections/}{8_Discussion.tex}
\import{Sections/}{9_Conclusion.tex}
\import{Sections/}{10_Acknowledgment.tex}
%\newpage

\import{Sections/}{EthicalConsiderations.tex}
\bibliographystyle{plain}
%\bibliography{sample-base}
%\import{Sections/}{EthicalConsiderations.tex}
\bibliography{Ref.bib}
%\newpage
%\appendix
\appendices
\import{Sections/}{Appendix.tex}

\end{document}

%% file: commands.tex
\usepackage{amsthm}
\usepackage{comment}
\usepackage{threeparttable}
\usepackage{multirow}
\theoremstyle{definition}
\makeatletter
\def\thm@space@setup{\thm@preskip=1pt
\thm@postskip=1pt}
\makeatother

\usepackage{graphicx}
\usepackage{amsmath}
\usepackage{algpseudocode}% http://ctan.org/pkg/algorithmicx

\usepackage{tikz}
\usepackage{amsfonts}
\usepackage{ifthen}

\usepackage{enumitem}

\usepackage{booktabs}

\usepackage{mdframed}
\mdfsetup{skipabove=0.5pt,skipbelow=0.5pt}
\newmdenv{allfour}
\newmdenv[leftline=false,rightline=false, linecolor=darkgray, linewidth = 0.2mm, startinnercode={\baselineskip=0cm}]{topbot}
\newmdenv[topline=false,rightline=false]{leftbot}
\newmdenv[topline=false,bottomline=false]{leftright}

\usepackage{pifont}
\usepackage{makecell}

\usepackage{booktabs}
\usepackage{moresize}
\usepackage{tabularx}
\usepackage{float}
\usepackage{caption}
\usepackage{listings}

\usepackage{tabularray}
\definecolor{Silver}{rgb}{0.752,0.752,0.752}

\newboolean{commentsOn}
\setboolean{commentsOn}{true}
\ifthenelse{\boolean{commentsOn}}{
\newcommand{\berkay}[1]{{\color{purple}{\bf Berkay:} #1}}}
{\newcommand{\berkay}[1]{}}

\ifthenelse{\boolean{commentsOn}}{\newcommand{\new}[1]{{\color{orange}#1}}}
{\newcommand{\new}[1]{}}

\ifthenelse{\boolean{commentsOn}}{\newcommand{\reham}[1]{{\color{purple}{\bf RM:} #1}}}

\ifthenelse{\boolean{commentsOn}}{}

\ifthenelse{\boolean{commentsOn}}{\newcommand{\update}[1]{{\color{black}#1}}}

\usepackage{url}
\usepackage{hyperref}

\usepackage{xspace}
\newcommand{\system}{{\textsc{\small{OVRWatcher}}}\xspace}

\newcommand{\sysprof}{{\texttt{ovrgpuprofiler}}\xspace}

\DeclareRobustCommand*\circled[1]{\tikz[baseline=(char.base)]{ \node[shape=circle,draw,color=white,fill=black,inner sep=0.5pt] (char){#1};}}

\def\ie{{i.e.},~}
\def\eg{{e.g.},~}

\usepackage{microtype}
\usepackage{amsmath,stackengine}

\newcommand{\shortsectionBf}[1]{\vspace{2pt}
\noindent {\bf #1}}

\newcommand{\Cone}{${\textbf{C1}}$\xspace}
\newcommand{\Ctwo}{${\textbf{C2}}$\xspace}
\newcommand{\Cthree}{${\textbf{C3}}$\xspace}

%% file: Sections/0_Abstract.tex
\begin{abstract}
%Area
Over the past decade, AR/VR devices have drastically changed how we interact with the digital world. Users often share sensitive information, such as their location, browsing history, and even financial data, within third-party apps installed on these devices, assuming a secure environment protected from malicious actors. 
Recent research has revealed that malicious apps can exploit such capabilities and monitor benign apps to track user activities, leveraging fine-grained profiling tools, such as performance counter APIs. However, app-to-app monitoring is not feasible on all AR/VR devices (e.g., Meta Quest), as a concurrent standalone app execution is disabled. 
In this paper, we present \system, a novel side-channel primitive for AR/VR devices that infers user activities by monitoring low-resolution (1Hz) GPU usage via a background script, unlike prior work that relies on high-resolution profiling.
\system captures correlations between GPU metrics and 3D object interactions under varying speeds, distances, and rendering scenarios, without requiring concurrent app execution, access to application data, or additional SDK installations.
We demonstrate the efficacy of \system in fingerprinting both standalone AR/VR and WebXR applications.
\system also distinguishes virtual objects, such as products in immersive shopping apps \update{selected by real users} and the number of participants in virtual meetings, thereby revealing users’ product preferences and potentially exposing confidential information from those meetings.
\update{\system achieves over 99\% accuracy in app fingerprinting and over 98\% accuracy in object-level inference.}

\end{abstract}

%% file: Sections/1_Introduction.tex
%\berkay{Intro is locked.}
%\seonghun{From the Storyline}\\
\section{Introduction}\label{sec:introduction}
%Area!!
Augmented Reality (AR) and Virtual Reality (VR) platforms have reshaped industries and are redefining how we interact with the digital and physical worlds~\cite{pertus2025future, programace2025vrtrends}. These immersive systems integrate virtual elements into real-world environments (AR) or create entirely simulated environments (VR). This offers new ways of interaction, visualization, and engagement with other people and the environment. Several companies have designed their own AR/VR devices, \eg Meta Quest~\cite{Quest2,Quest3}, Microsoft HoloLens~\cite{hololens}, and Apple Vision Pro~\cite{apple_pro}. These devices comprise CPUs, GPUs, neural engines, various sensors (\eg visible light cameras and infrared cameras), and audio components for real-time processing of multimodal sensory information and rendering~\cite{news1, news2}. %\berkay{please add references}

%\chandrika{here existing side channel attacks. don't mention GPU first. }
%%AR/VR SIDE CHANNEL ATTCKS!!!!!

The myriad of sensors and powerful processors in AR/VR devices enable users to become deeply immersed in digital content, whether in fully virtual environments or blended physical–digital scenes. 
However, recent research has shown that these capabilities also introduce novel side-channel vulnerabilities in these devices.
These works include virtual keystroke inference using WiFi signals~\cite{al2021vr} or embedded 2D infrared sensors~\cite{ni2024non}, as well as avatar typing observation~\cite{yang2024can}, head movement analysis~\cite{slocum2023going}, or a combination of computer vision methods and motion sensors~\cite{ling2019know}.
Additionally, other studies have explored the inference of visual and audio activities during device charging~\cite{li2024dangers}, and the reconstruction of high-quality vital signals and speech content, employing embedded motion sensors in AR/VR headsets~\cite{zhang2023facereader,cayir2025speak}.
%
%Some works have demonstrated the feasibility of recovering speech content using motion sensors~\cite{cayir2025speak, shi2021face} in AR/VR devices.
%

%
A recent work~\cite{zhang2023s} proposed side-channel attacks that recover gestures, voice commands, keystrokes, and detect bystanders using a concurrent malicious app on AR/VR devices.
These attacks exploit memory allocation APIs and performance counters exposed through SDKs provided by game engines such as Unity and Unreal.
Although the aforementioned work~\cite{zhang2023s} uses CPU and GPU frame rates obtained via SDKs as attack vectors, it faces three key limitations.

First, it requires a concurrently running standalone background app to access these performance counters.
However, devices like the Meta Quest restrict concurrent app execution, apart from a few Meta-approved apps (\eg Messenger and Meta Chat), rendering the attack infeasible.
Second, their approach relies on high-resolution profiling (60 Hz), which makes it detectable and easily prevented by lowering the resolution of the profiling tool.
%
% AR/VR devices focus on rendering complex 3D objects and handling multiple concurrent tasks, but those capabilities expose novel side-channel vulnerabilities. 
% %
% A diverse set of side channels has been proposed on AR/VR headsets, including physical access to devices~\cite{ni2024non,al2021vr,zhang2023facereader,li2024dangers}, VR avatar-based inference~\cite{yang2024can}, motion sensors (Accelerometer, Gyroscope)~\cite{slocum2023going,cayir2025speak,ling2019know,shi2021face}, and performance counters~\cite{zhang2023s}.
% %
% Those side channels can capture personal data, such as personal data (\eg passwords) and app usage patterns. 
% %RECENT WORK AND LIMITATIONS.
% Recent work~\cite{zhang2023s} has proposed a performance counter based side-channel attack that exploits user activity (gestures, voice, and keystrokes) by fingerprinting usage of CPU and GPU. 
% %
% While that approach achieves impressive accuracy, it suffers from three key limitations. 
%
% First, it needs SDKs to create performance counters in user space, which can easily be prevented by restricting access to user space performance counters. 
% %
% Second, their work requires concurrent app execution. This requirement relies on running a malicious profiler alongside a target app, which is impossible on devices that enforce concurrent standalone app execution (\eg, Meta Quest) 
% %
% Third, their approach relies on high-resolution profiling (60 Hz), making it detectable and easily prevented by lowering the resolution of the profiling tool as proposed.
%
Lastly, their work focuses on a small set of attack attributes and performance counters without modeling the detailed relationship between 3D object rendering and GPU usage. Hence, limited sensitive data, such as built-in voice commands and simple gestures, can be captured.

\update{
%\textbf{[R1]} 
In this paper, we revisit the threat landscape of GPU-based side-channel attacks in immersive environments by focusing on Meta devices, which expose GPU metrics with unusually high and fine-grained correlation to application behavior.
We demonstrate that even low-resolution (1Hz) profiling of these metrics can reveal fine-grained information and leak sensitive application activity.
To this end, we introduce \system which leverages the built-in GPU profiling tool~\cite{ovrgpuprofiler} to fingerprint user activity with over 98\% accuracy, without relying on concurrent app execution or additional SDKs.
We show that even profiling a single GPU metric is sufficient to leak fine-grained information, which underscores the practicality and stealthiness of this attack vector.
These findings thus reveal critical gaps in existing mitigation strategies~\cite{naghibijouybari2018rendered}, and highlight the urgent need for more robust protections on current and future XR platforms. 
}

%WHAT WE DO IN THIS PAPER 
% Therefore, in this paper, we address these key gaps in prior work~\cite{zhang2023s} by proposing \system.
% %
% Our method leverages a built-in GPU profiler and achieves over 98\% accuracy in fingerprinting user activity in immersive environments.
% %
% Our proposed attack does not rely on the concurrent app execution and does not require additional SDKs. 
% %
% Also, we demonstrate the feasibility of an attack with low-resolution (1Hz) GPU metrics profiling, thereby presenting a more practical attack vector and circumventing existing mitigation techniques~\cite{zhang2023s,naghibijouybari2018rendered}.
% %
% We further demonstrate that high accuracy can be achieved by profiling only a single GPU metric, thereby reducing overhead.
%
% %
% We further demonstrate that high accuracy can be achieved using a single GPU metric among the provided 72 GPU-specific metrics with less overhead.

%OUR SYSTEM Evaluation with Numbers.

%\berkay{the connection between the paragraph below and above is lost. You concluded in the previous paragraph by saying that "These findings thus reveal ..." Now, below, you continue giving some more numbers. I'm lost.}\seonghun{fixed}
%
We therefore quantify these gaps by evaluating \system on XR platforms.
\system achieves over 99\% accuracy in fingerprinting standalone AR/VR applications, which are primarily rendered in 3D or occupy the full immersive space, thus exerting a stronger impact on GPU metrics.
%
%We also evaluate \system on the challenging task of fingerprinting WebXR applications rendered within a 2D browser window.
%
Despite the constrained setting of 2D, it achieves 99\% accuracy in fingerprinting WebXR apps, and with two GPU metrics yielding more than 94\% accuracy.

As suggested by prior works in the mobile domain~\cite{istelan, mohamed2024attention}, identifying user interests can be leveraged for targeted advertising.
Motivated by this, we demonstrate the feasibility of identifying virtual objects within immersive environments.
Through case studies, we show that \system can fingerprint realistic products in immersive shopping apps and detect participant counts in virtual meeting apps with over 98\% accuracy.
%
%\chandrika{one sentence on user study: }
%\seonghun{sounds good!}
%
We further validate the effectiveness of \system through a user study, where participants interact with the real-world application Meta Layout~\cite{meta_layout_experience}, achieving accuracy of up to 88\%.
%}
%
These attacks broaden our understanding of the threat models faced by emerging immersive AR/VR systems.

Our attack operates even on platforms such as the Meta Quest by leveraging built-in GPU profiling tools, specifically, Meta’s \sysprof tool~\cite{ovrgpuprofiler}.
This attack requires no elevated privileges, physical access, or additional SDK installations, and it overcomes the key limitation of prior work that depends on concurrent application execution.
In contrast to existing AR/VR side-channel attacks that primarily detect keystrokes or application usage, \system is the first to reveal low-level GPU metrics that can expose fine-grained information within an app's immersive environment.

In summary, we make the following contributions:
\begin{comment}
\berkay { I'm not sure it is a good idea to use the ``attack'' as an subject -- I have never seen this. This might be OK in side channel community.\\

I think we can do a better job in writing contributions with using the subject "we" as follows\\

(1) We demonstrate that an attacker can...
(2) We introduce  \system, 
(3) We evaluate \system  ... (you need to give numbers here.)

please check IotCupid or Reham's ATT paper to see how they write the contributions.\\

Lastly, I do not understand why this is a contribution by itself: "identifies VR furniture objects in shopping apps
across both AR and VR environments"
}
    
\end{comment}

\begin{itemize}
    \item We present \system, a method that uses a built-in GPU profiler with low (1Hz) resolution to perform side-channel attacks on AR/VR devices, without requiring additional SDKs or concurrent app execution.
    % \item We present \system, a method that leverages a built-in GPU profiler with a low one-second resolution without requiring additional SDKs or concurrent app execution to perform side-channel attacks on AR/VR devices.
    %\item We demonstrate that an attacker can leverage a built-in GPU profiler at one-second resolution for the first time.
    \item We systematically analyze how virtual object rendering in AR/VR affects GPU metrics, revealing indicators correlated with scene complexity and user interactions.
    % \item We systematically analyze how the virtual object rendering process in AR/VR environments affects GPU metrics, revealing key indicators that correlate with 3D scene complexity and user interactions.
    %
    \item Through case studies, we show that \system achieves near 100\% accuracy in tracking foreground standalone AR/VR and WebXR apps, and inferring fine-grained user activities, such as identifying products in a shopping app and participant counts in a private meeting app, highlighting the significant privacy risks posed by low-resolution GPU side channels in AR/VR.
    \item \update{We demonstrate that virtual objects selected by real participants can be distinguished with up to 88\% accuracy despite the noise introduced by the surrounding environment.}
    \item We evaluate \system in a cross-device setting, showing its effectiveness even with limited GPU metrics, achieving over 93\% accuracy across all case studies.
\end{itemize}

%Given our study exposes side-channel vulnerabilities,
\shortsectionBf{Responsible Disclosure.} We have disclosed our findings with the Meta Quest development team through Meta Bug Bounty~\cite{meta_bug_bounty}.
\update{
%\textbf{[R1]} 
The Meta Security Team acknowledged these findings in the Meta Quest series and recognized our contribution with a Meta Bounty Award (Detailed in Section~\ref{sec:ethical}).}
%The Meta Security Team has acknowledged the presence of these vulnerabilities in the Meta Quest series and recognized our contribution with a Meta Bounty Award.

%% file: Sections/2_Background.tex
\section{Background}\label{sec:background}
\subsection{Extended Reality (XR)}
%\shortsectionBf{Extended Reality (XR).}
%
% Extended Reality (XR) is an umbrella term that encompasses technologies (VR, AR, MR) that alter reality by incorporating digital elements into the real-world environment to varying degrees.
% %
% Virtual Reality (VR) completely separates users from their physical environment, completely immersing them in an alternate reality, as exemplified by gaming apps such as Beat Saber~\cite{beat_saber_meta}.
% Extended Reality (XR) encompasses technologies such as Virtual Reality (VR), Augmented Reality (AR), and Mixed Reality (MR) that integrate digital elements into the real-world environment to varying extents.
% %
% XR is rapidly evolving, with applications in gaming, healthcare, education, training, remote work, and data visualization.
%
Extended Reality (XR) encompasses technologies such as Virtual Reality (VR), Augmented Reality (AR), and Mixed Reality (MR) that integrate digital elements into the real-world environment to varying extents.
%
% Extended Reality (XR), including Virtual Reality (VR), Augmented Reality (AR), and Mixed Reality (MR), integrates digital elements into the real world.
%
XR is rapidly advancing with applications in gaming, healthcare, education, training, and data visualization.
VR fully immerses users in an alternate reality, isolating them from their physical surroundings, as demonstrated by gaming apps such as Beat Saber~\cite{beat_saber_meta}.
%
% In contrast, Augmented Reality (AR) overlays digital elements on the user's physical environment, typically without enabling direct interaction between the digital and physical elements, as seen in apps such as the Pokémon Go game app\cite{pokemon_go} and the IKEA Place e-Commerce app\cite{ikeaPlace}.
% %
% Mixed Reality (MR), as a subset of XR, combines aspects of VR and AR, allowing users to view and interact with virtual objects within their own physical environment. 
In contrast, AR overlays digital elements onto the user's physical environment, usually without allowing direct interaction between the digital and physical elements, as seen in apps such as Pokémon Go\cite{pokemon_go} and IKEA Place\cite{ikeaPlace}.
MR, a subset of XR, merges VR and AR, enabling users to view and interact with virtual objects within their physical environment.
For example, Figmin XR~\cite{figmin_xr} allows users to create, collect, and play in an MR environment. In addition to these standalone apps, both Meta Quest and HoloLens 2 support WebXR~\cite{webxr_samples}, a web-based immersive environment.
\subsection{GPU Performance Counter}
%\shortsectionBf{GPU Performance Counter.}
To assist developers in tracking and optimizing the performance of their apps, XR frameworks provide GPU performance counters. 
Metrics provided by the GPU performance counter reflect the graphical complexity of objects rendered on the display.
Similarly, to analyze GPU usage, development frameworks offer tools to access real-time metrics and GPU profiling data provided for Meta Quest headsets.
For example, Meta's \sysprof tool exposes 72 GPU metrics (78 for Quest 3S) at 1 Hz sampling resolution.
%and can be driven from a background script without privilege elevation.
%
Apple Vision Pro's Xcode Instruments~\cite{apple2025gpuoptimization} offer over 150 GPU counters at up to 60Hz, and Microsoft HoloLens supplies memory and GPU utilization metrics via Windows Performance Recorder~\cite{microsoft2025wpr} and PIX~\cite{microsoft2025pix} at 10Hz, accessible in the background.

Because maintaining high frame rates and low latency is critical to user comfort and immersion in AR/VR, GPU profilers are essential to identify performance bottlenecks. This makes them impossible to remove without severely degrading the experience.
Each metric provides insight into different GPU subsystems. For instance, in \sysprof, GPU counter cover utilization, memory bandwidth, and geometry throughput.

%indicate overall GPU utilization and scheduling overhead. \texttt{Vertex Memory Read (Bytes/Second)} and \texttt{Global Memory Load Instructions} measures memory bandwidth usage. \texttt{Pre-clipped} \texttt{Polygons/Second} and \texttt{Prims Trivially Rejected} reflect geometry throughput and discard rates.
\update{
%\textbf{[R4]} 
%\berkay{The previous sentence starts with For instance, and the sentence below starts with For example. Is something missing between these two sentences?}\seonghun{fixed}

%
Users can also install the OVR Metrics Tool~\cite{ovr_metrics_tool} from the Meta Horizon Store, which runs entirely in user-space and exposes various performance counters. 
%
%\chandrika{Why Developer? If that app is available in store, it can be installed by a user.}\seonghun{yes you are right, fixed}
%
This app provides basic \sysprof metrics, such as GPU utilization and frame timing, to users either as an in-app HUD overlay or as CSV reports, while developers can access advanced metrics~\cite{ovrmetricstool}
within their apps.}

%% file: Sections/3_ProblemStatement.tex
\section{Related Work}\label{sec:Related Work}

%\berkay{Please separate related work from motivation. Have a section called Related work, another Section for Motivation. Have the design challenges in the motivation section with a \shortsectionBf{Design Challenges}. Here you can start the design challenges with modifying the following sentence that you use at the end of the motivation section: " Achieving this performance at low sampling rates, however, requires overcoming several technical challenges, as described in the following sections" and then list the challenges, C1, C2, and C3.}\seonghun{fixed}

\begin{table*}[ht]
\caption{Comparison of related AR/VR side-channel attacks and \textbf{\system}. Each row presents (i) the type of side channel, (ii) the specific sensor or API used, (iii) the extracted attributes and the corresponding number of labels, (iv) the sampling resolution (Hz), (v) whether a standalone or concurrent app is required, and (vi) the tested AR/VR environment. Symbols are used to denote an AR (\faMask) and a VR (\faStreetView). The attack denoted by $^\dagger$ is not feasible on the Meta Quest series.
}
\scriptsize
\setlength{\tabcolsep}{0.5em}
\def\arraystretch{1.3}
\centering
\begin{tabular}{|l|l|l|c|c|c|}
\hline
\textbf{Side-Channel Type} & \textbf{Side-Channel Primitive} & \textbf{Extracted Attributes (\# Labels)} & \textbf{Resolution (Hz)} & \textbf{Standalone} & \textbf{AR/VR} \\
\hline \hline

\multirow{2}{*}{Physical} 
 & Facial Vibration Monitoring Belt \cite{zhang2023facereader} & Gender (2), User (27), Body fat (-) & 203 & $\times$  & \faStreetView\\ 
 & Power Monitoring device\cite{li2024dangers} & App (10), Website (10), Audio (5) & 100 & $\checkmark$ & \faStreetView \\ 
\hline

\multirow{4}{*}{Motion/Gesture Sensors} 
 & Controller \cite{wu2023privacy} & Keystrokes (38) & 60 & $\checkmark$ & \faStreetView \\ 
 & IMU \cite{slocum2023going} & Keystrokes (60) & 72 & $\times$  & \faStreetView \\ 
 & Camera \cite{wang2024gazeploit} & Keystrokes (4) & 30 & $\checkmark$ &  \faStreetView \\ 
 & IMU (Accelerometer/Gyro)  \cite{cayir2025speak} & Digits (10) & 1000 & $\times$ & \faStreetView \\ 
\hline

\multirow{1}{*}{System-level APIs} 
 & Performance Counters (Unity/Unreal) \cite{zhang2023s} & Gestures (5), Voice (5), Digits (10), App$^\dagger$(12) & 60 & $\times$ & \faMask, \faStreetView \\ 

\hline

\textbf{\system} 
 & \textbf{GPU Profiler (in-built)} & \textbf{VR Object (35), App (100), Website (100), Avatar (10)} & \textbf{1} & \textbf{$\checkmark$} & \textbf{\faMask, \faStreetView} \\ 
\hline
\end{tabular}
% \caption{Comparison of related AR/VR side-channel attacks and \textbf{\system}. Each row presents (i) the type of side channel, (ii) the specific sensor or API used, (iii) the extracted attributes and the corresponding number of labels, (iv) the sampling resolution (Hz), (v) whether a standalone or concurrent app is required, and (vi) the tested AR/VR environment. Symbols are used to denote an AR (\faMask) and a VR (\faStreetView). The attack denoted by $^\dagger$ is not feasible on the Meta Quest series.
% }
\label{table:RelatedWork}
\end{table*}

%\subsection{Related Work}\label{sec:Related Work}
\shortsectionBf{Side-Channel Attacks on GPU.}
%\shortsectionBf{Side-Channel Attacks on GPU.}
Researchers~\cite{naghibijouybari2018rendered, dutta2023spy, giner2024generic} have introduced new side-channel attacks exploiting the parallelism and resource sharing of modern GPUs in desktop and mobile environments.
Naghibijouybari et al.~\cite{naghibijouybari2018rendered} exploited GPU performance counters for website fingerprinting, keystroke detection, and neural network recovery.
These attacks, however, can be mitigated by lowering the sampling rate and reducing website fingerprinting accuracy to less than 40\% at 2 samples per second. % (2Hz).
Dutta et al.~\cite{dutta2023spy} extended side-channel attacks to multi-GPUs using a Prime+Probe method targeting L2 cache contention to infer sensitive data transfer patterns in multi-GPU interconnects.
These cross-GPU attacks can be mitigated by disabling peer memory access between untrusted GPUs and enforcing cache and memory partitioning across GPUs.
%\chandrika{What is the mitigation against the last point, if exists, one sentence can be added after the last sentence.}\seonghun{added}

\begin{comment}
Additional studies have explored recovering encrypted data via GPU cache and shared memory.
%
Karimi et al.~\cite{karimi2018timing} used Shader processor timing on mobile GPUs to extract AES-128 keys.
%
Jiang et al.\cite{jiang2017novel} used timing-based GPU techniques to leak cryptographic keys, targeting table-based AES encryption.
%
Giner et al.\cite{giner2024generic} created malicious JavaScript for cache-based attacks in WebGPU, leaking a full AES key in 6 minutes and emphasizing the need to restrict browser GPU access. 
%

In comparison, our work demonstrates the feasibility of low-resolution GPU-based side-channel attacks on XR.
\end{comment}

\shortsectionBf{Privacy Leakage Attacks in XR.}
Prior works
~\cite{farrukh2023locin, nair2023unique, zhang2023s, ling2019know, slocum2023going,  shi2021face} have demonstrated various privacy leakage attacks in XR environments.
Zhang et al.\cite{zhang2023s} leveraged leakage vectors, including memory allocation APIs and performance counters from Unity\cite{unity} and Unreal~\cite{unreal_engine}, to recover hand gestures, voice commands, virtual keyboard keystrokes, perform application fingerprinting, and estimate bystander distance.
Slocum et al.~\cite{slocum2023going} proposed a system to infer words or characters typed by a victim on an XR device using a concurrent application by extracting IMU motion signals.
Ling et al.\cite{ling2019know} used computer vision and motion sensors to infer keystrokes in virtual environments, while Shi et al.\cite{shi2021face} exploited AR/VR motion sensors to infer sensitive information from facial dynamics associated with speech (\eg identity, gender).

%
%\chandrika{If there is a difference between the above and bottom categories of privacy attacks, a sentence can be added in the starting of this paragraph representing the following types of privacy leakage.}
%

%\berkay{you start this paragraph with a paper, but not sure why you need a new paragraph. Is this work below different from the ones above? When you give an example, you need to provide a context first, then illustrate it with an example.}
%
Beyond keystroke recovery and app fingerprinting, XR also leaks higher-level attributes such as location and identity.
Farrukh et al.~\cite{farrukh2023locin} presented a location inference attack for MR devices, which employs geometric and semantic features from 3D spatial maps.
In an analysis of a publicly available dataset, Nair et al.~\cite{nair2023unique} found that hand and head motion data captured in a VR environment can uniquely identify a large number of users.
%Nair et al.~\cite{nair2023unique} demonstrated that a large number of users could be uniquely identified by using their hand and head motion data captured within a VR environment, utilizing a publicly available dataset. 
%
%They evaluated their proposed method using a publicly available dataset from a popular application.
%
Tricomi et al.~\cite{tricomi2023you} developed a generic profiling framework leveraging machine learning on behavioral data, including head, controller, and eye movements, to identify users and infer attributes such as age and gender. 
Unlike previous works, our approach uses GPU metrics without relying on motion sensors or camera data to analyze user activity in XR.

\update{
%\textbf{[R1]} 
Table~\ref{table:RelatedWork} compares our attack with existing side-channel attack vectors and their target environments.
While prior work relies on high-resolution profiling for accuracy, our results show that even 1Hz GPU sampling can effectively compromise user privacy.
We evaluate our attack in AR and VR environments, demonstrating its effectiveness in both these settings.
%
% \berkay{not sure what mode means here.}
%
% We evaluate our attack in both AR and VR environments %with real participants
% and demonstrate its effectiveness across both 
% these settings.
% modes \berkay{not sure what mode means here.}.
%%\berkay{why past tense? Please make it active and present tense. We evaluate...}
}
% Table~\ref{table:RelatedWork} presents a comparison between our attack and existing side-channel attack vectors along with their target environment.
% %
% All these methods rely on high-resolution methods to exploit side-channel leakage for the best accuracy. 
%However, Naghibijouybari et al.~\cite{naghibijouybari2018rendered} proposed a 2Hz resolution to successfully mitigate GPU-based side-channel attacks. 
% Our results show that even a lower resolution, 1Hz, of the GPU profiler can achieve high accuracy while compromising user privacy. 
% %
% Moreover, attacks employing concurrent apps are not allowed in the Meta Quest series~\cite{farrukh2023locin,zhang2023s} and do not offer the same feasibility for the exploitation of novel target vectors. Our attack was conducted through both AR and VR environments and proved across both AR and VR modes.

\section{Motivation}\label{sec:motivation}
%\reham{Edited, please revise and check the comments}\seonghun{fixed!}

Our goal is to demonstrate that sensitive user information in XR environments, such as active apps, virtual object properties, and the number of participants in private meetings, can be exploited by analyzing GPU metrics collected from the built-in GPU profiler.
%collected GPU metrics  \reham{we can clarify here that it is collected from the built-in GPU profiler by running a background script}. \seonghun{fixed}
The extracted sensitive data could enable an adversary to launch various attacks, including targeted advertising or the leakage of personal and organizational information~\cite{jin2016go,kim2019seeing,barbosa2021design}.

For example, in a virtual shopping app, an adversary could correlate the GPU usage metrics patterns with the rendering of complex virtual product assets, such as furniture items, generating fingerprints for each item. These fingerprints can be exploited to target users with recommendations or malicious advertisements based on their preferences. 
Similarly, a malicious background script could track periodic fluctuations in GPU metrics corresponding to the rendering of participant avatars in a virtual meeting environment.
% as each participant's avatar is rendered in a virtual meeting app. 
%
This allows inference of the number of participants and their spatial arrangement, which could be exploited by an adversary to craft tailored phishing messages based on the inferred context of the meeting.
% This allows inference of the number of attendees and even their arrangements, which an attacker could leverage to tailor phishing messages based on inferred meeting information.

%1. SDK counter + concurrent-app attacks
%
\update{
%\textbf{[R1]} 
Prior AR/VR side-channel attacks have used high-frequency performance counters (\eg from Unity or Unreal SDKs) to infer common gestures, voice commands, numeric inputs, and application usage with up to 95\% accuracy~\cite{zhang2023s}.
However, their effectiveness drops below $40\%$ at sampling rates under 10 Hz, and they rely on a malicious app running concurrently with the target app, a model infeasible on Meta Quest devices due to strict runtime isolation.
}
\update{
Beyond XR platforms, similar high-resolution side-channel attacks have been demonstrated on desktops and smartphones, such as website fingerprinting via GPU utilization~\cite{naghibijouybari2018rendered}, achieving around 90\% accuracy with high-resolution timers and GPU rasterization. 
However, their success rate drops to around 59\% when the performance counter resolution is reduced or rasterization is disabled.
}
\update{
%
%\chandrika{
Therefore, in this work, we aim to explore GPU profilers on AR/VR devices, operating at a low sampling rate of 1Hz without requiring additional SDKs, elevated privileges, or concurrent background app execution.
We further aim to investigate the relationship between GPU metrics and user activity in AR/VR environments.
To assess the extent to which such activity can be recognized, we aim to evaluate the efficacy of the attack through lab-controlled studies in both AR and VR settings.
Finally, we aim to examine the practicality of the attack through a user study in which participants interact with a real-world application.
%
%}
%\chandrika{Remove the following sentences until the start of design challenges.}
%
%\system overcomes prior limitations by using only the built-in GPU profiler on AR/VR devices, operating at a low sampling rate of 1Hz without requiring additional SDKs, elevated privileges, or concurrent background app execution.
%
%It accurately classifies active apps, identifies virtual objects, and estimates participant counts in virtual meetings, outperforming high-frequency approaches~\cite{zhang2023s, naghibijouybari2018rendered}.
%
%This makes it practical for real-world XR settings, as it remains effective despite low-resolution countermeasures and platform restrictions.
%
}

\shortsectionBf{Design Challenges.}\label{sec:DesignChallenges}
%\section{Design Challenges}\label{sec:DesignChallenges}
Achieving high accuracy in predicting user activity at low sampling rates, however, entails overcoming several technical challenges, as described below.
%
% Achieving high accuracy in predicting user activity 
% %Achieving high performance 
% at low sampling rates, however, requires overcoming a set of technical challenges, described below.

\shortsectionBf{(\Cone) Disabled Concurrent Applications.} 
% Prior work~\cite{zhang2023s} shows that malicious applications can operate concurrently with benign applications to facilitate profiler-based side-channel data collection. 
%
% However, unlike Microsoft HoloLens and Apple Vision Pro, Meta Quest devices restrict the concurrent application execution except for simple Messenger \reham{simple messaging apps?} apps such as Facebook Messenger or Meta Chat. 
Prior work~\cite{zhang2023s} shows that malicious apps can run concurrently with benign apps, which enables profiler-based side-channel data collection.
%
% However, unlike Microsoft HoloLens and Apple Vision Pro, Meta Quest devices strictly limit concurrent app execution, with limited exceptions only for lightweight Meta's messaging apps such as Facebook Messenger or Meta Chat rather than general third-party apps.
% However, unlike Microsoft HoloLens and Apple Vision Pro, Meta Quest devices strictly limit concurrent app execution, allowing exceptions only for lightweight Meta messaging apps like Facebook Messenger or Meta Chat, and not for general third-party applications.
%
\update{
%\textbf{[R1]}
However, unlike Microsoft HoloLens and Apple Vision Pro, Meta Quest devices restrict concurrent app execution, permitting only selected Meta apps like Messenger or Meta Chat. 
This limitation in Meta Quest devices makes prior attacks infeasible for third-party applications and presents a particularly challenging environment for side-channel attacks.
}
%\reham{Are the exceptions for messaging apps or apps from Meta?}\seonghun{fixed}
%
% Hence, Meta Quest devices present a particularly challenging environment for conducting profiler-based side-channel attacks since executing a malicious app alongside foreground apps becomes infeasible.
% Hence, Meta Quest devices create a challenging system to perform profiler-based side-channel attacks, as it is infeasible to execute a malicious application alongside foreground applications.
%
%\reham{We can remove the following sentence}\seonghun{fixed}
%In this study, we primarily focus on Meta Quest devices, as they are among the most secure and challenging targets within the AR/VR platforms.

% Therefore, we focus particularly on Meta Quest devices, which are among the most challenging targets across AR/VR platforms. %To address this, we employ a built-in GPU profiler to make it feasible to launch GPU profiler-based side-channel attacks.

\shortsectionBf{(\Ctwo) Efficient GPU Metric Selection.} Accessing the full spectrum of GPU performance metrics (72 available GPU metrics on Quest2 and 78 on Quest3) using the built-in GPU profiler introduces significant overhead due to high resource consumption. This often results in missing or inconsistent metric values. Consequently, Meta's official documentation~\cite{ovrgpuprofiler_detail} recommends not to request more than 30 real-time metrics simultaneously.
%
%This presents the challenge of ensuring that GPU metric values can be reliably profiled to perform side-channel attacks.
%
This limitation poses a challenge for reliably profiling the GPU metrics required for the side-channel attacks.
%
%Hence, an efficient metric selection strategy is crucial to identify a subset of the most informative metrics, thereby reducing profiling overhead and minimizing data loss.
% Therefore, an efficient GPU metric selection process, by identifying and focusing on a subset of informative metrics, is required to reduce the profiling overhead and minimize data loss while choosing the most informative metrics.
%

\shortsectionBf{(\Cthree) Low-Resolution Attack Channel. %\reham{Do we mean two have the word channel twice here? We can name "Low-Resolution Attack Channel"}\seonghun{changed}
}
\update{
%\textbf{[R1]} 
Prior GPU side-channel attacks rely on high-frequency sampling to achieve high accuracy~\cite{naghibijouybari2018rendered,zhang2023s}. 
In contrast, Meta Quest devices expose GPU metrics at a low sampling rate of 1Hz, making fine-grained activity inference more challenging.
%
%To address this, we use machine learning models to capture temporal patterns in 1Hz GPU readings and analyze the correlation between pixel values and GPU metrics, enabling accurate inference of user activities despite the low resolution.
}

\begin{figure}[t]
    \centering
    {
    
    \includegraphics[width=0.9\linewidth]{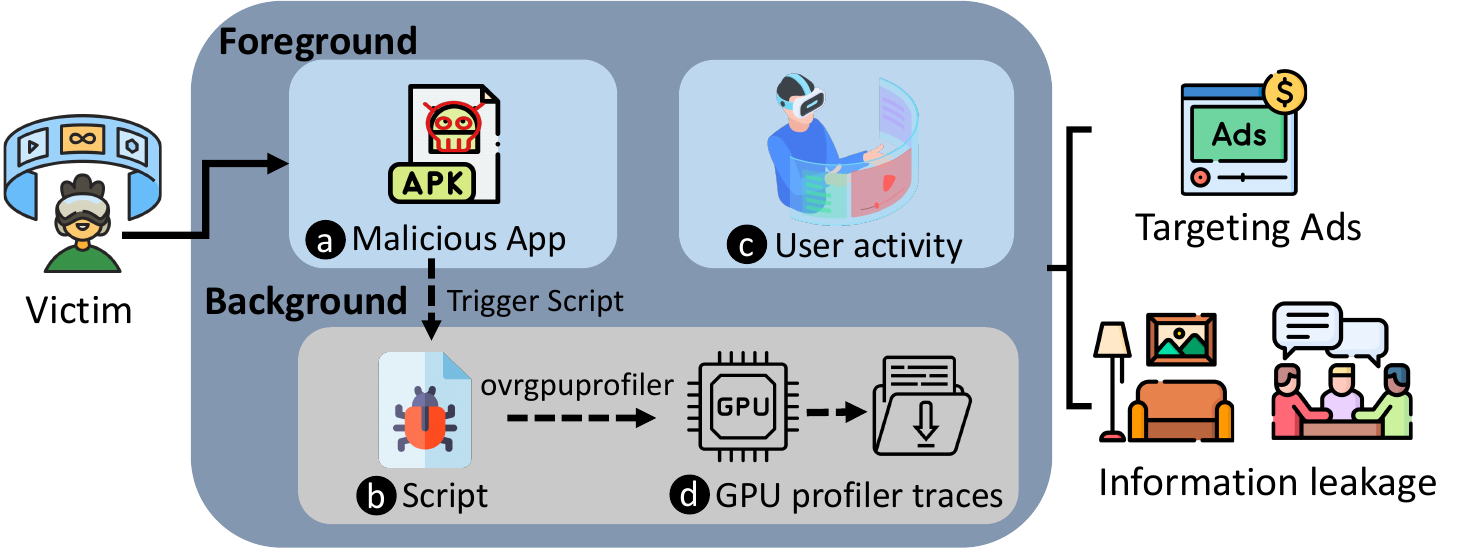}
    }
    \caption{Illustration of \system's threat model.  %\berkay{(1) you can increase the font size to match text font size. (2) You can shorten the caption by just saying ``The illustration of \sys's threat model''.}\seonghun{fixed}
    %\seonghun{changed the figure}
    %\reham{We need to modify this figure}
    %\seonghun{fixed by removing script attached in (a)}
    }
    \label{fig:threatmodel}
\end{figure}

\section{Threat Model}
\label{sec:ThreatModel}
We consider an attacker whose goal is to infer a user's privacy-sensitive activities by analyzing GPU performance metrics on an AR/VR device, as shown in Figure~\ref{fig:threatmodel}.
To achieve this, we assume that a user installs a standalone AR/VR app, such as games or productivity tools, provided by the adversary (\circled{a}).
While running in the foreground, this app initiates the device's GPU profiler tool to monitor and record GPU usage patterns (e.g., frame timing, shading statistics) as the user engages in user activity with other benign AR/VR apps (\circled{b}, \circled{c}).
\update{
%\textbf{[R4]} 
We load the adversary app in developer mode, which allows installation of APKs and invokes shell commands via Android's API to execute a script file.
%in shell commands. 
%
Similarly, cross-compiling and bundling the \sysprof \textit{Executable and Linkable Format} (ELF) binary with the Android NDK~\cite{android_ndk}, the user space app can launch the profiler directly at runtime~\cite{afonso2016going,ruggia2025dark}, which makes it accessible without root or special privileges. 
%Furthermore, basic GPU counters (\eg \texttt{gpu\_utilization\_percentage}) remain accessible to any user-space application via the OVR Metric Tool SDK~\cite{ovrmetricstool} without root or special privileges.
}
We note that AR/VR app stores (\eg Meta Store) allow apps with a GPU profiler for legitimate performance analysis~\cite{ovr_metrics_tool}
%\berkay{fix}\seonghun{fixed} 
and, once the adversary app is activated, the profiler operates in the background even after the app is terminated (See Section~\ref{sec:AttackWorkflow}).
%
%\berkay{hoping that we have experiments for the sentence below.}\seonghun{we need benchmarks to conduct this overhead}

The captured GPU metrics are then stored within the app or transmitted to a server. Both methods incur minimal energy and computational overhead, which allows the attack to remain undetected by the user and the OS (See Section~\ref{subsec:metric_selection}).
By analyzing the resulting GPU profiler traces, an adversary infers sensitive information about the user's activities (\circled{d}). This information includes the apps a user is using, e.g., gaming, chat, or shopping apps, and interactions with 3D objects, e.g., the products that the user is browsing in a shopping app, as well as in-app contextual information, e.g., the number of participants in a meeting (See Section~\ref{sec:casestudies}).

%% file: Sections/4_System.tex
\begin{figure}[t]
    \centering
    {
    \includegraphics[width=\linewidth,height=4.5cm]{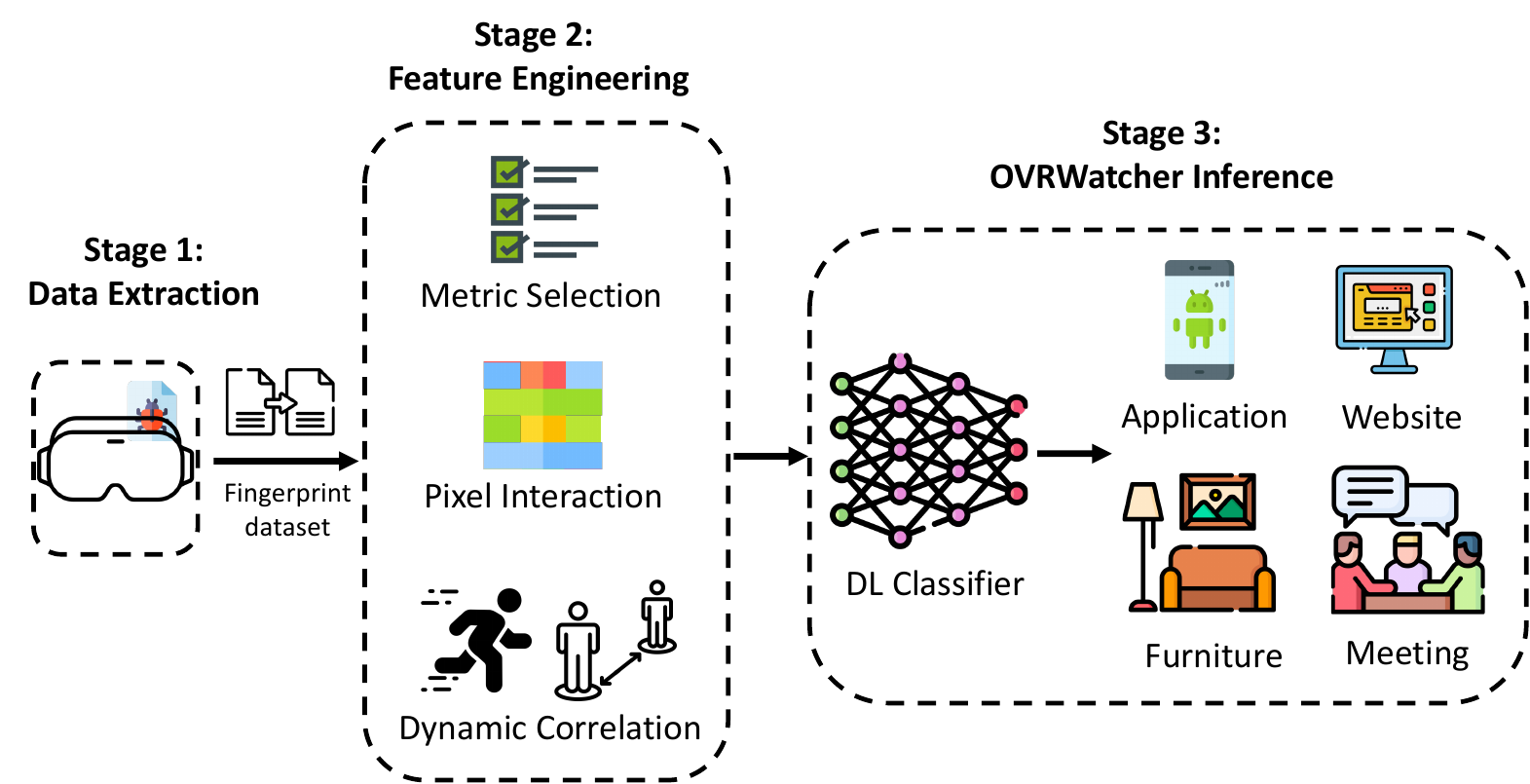}
    }
    \caption{Architecture of \system. 
    %\berkay{why do you have "period" here but not have it in Figure 1?. In addition, please make sure the location of Figures 1 and 2 makes sense in the paper.} \seonghun{fixed} %\reham{We can name the third stage OVRWatcher inference or something similar}
    %\seonghun{fixed}
    }
    \label{fig:systemfigure}
\end{figure}
\section{OVRWatcher Design}
\label{sec:OVRSeer}
We present \system, a side-channel attack that infers a user’s augmented and virtual reality environment by leveraging the built-in GPU performance monitoring tool.
%
% We present \system, a side-channel attack, which monitors the user's augmented and virtual reality environment by monitoring the built-in GPU performance monitoring tool. 
%
The tool not only allows \system to monitor rendered content and user interactions but also circumvents the need for elevated privileges or direct sensor access from XR devices.
% \berkay{kill the widow please.} \seonghun{fixed}
%to extract sensitive user information from AR/VR environments. %\berkay{this is your systems's headline, please make sure it looks good.}\seonghun{fixed}
%\shortsectionBf{System Overview.}
\subsection{System Overview}
Figure~\ref{fig:systemfigure} illustrates three main stages in the \system attack. 
First, \system creates a malicious script file that runs in the background even though the malicious app is terminated. The script file collects GPU metrics (i.e., frame timing, shading performance, GPU memory usage and texture details) %(\berkay{we can give some examples of such metrics}) \seonghun{fixed}
for a certain period of time to capture the user activity fingerprint dataset, which overcomes \Cone by bypassing Meta Quest's restriction on side-by-side VR app execution.%\reham{We can briefly explain why it overcomes C1}.\seonghun{fixed}

In the second stage, \system performs reverse engineering on GPU metrics using the collected fingerprints to analyze how they manifest in fully immersive (VR) and pass-through (AR) environments.
For example, an increase in GPU memory usage or texture usage obtained from GPU metrics can indicate that a new virtual object has been rendered in the immersive scene.
%\berkay{For instance, an example with one-two sentences would be good to demonstrate this}
% This accounts for the real-time effects of user interactions on GPU metrics,
% overcoming \Ctwo by reducing overhead. 
%
We address \Ctwo by selecting only the most informative GPU metrics that capture real-time user interactions. 
This reduces the profiling overhead of accessing all available metrics, which would otherwise stall the GPU through frequent counter reads, and also mitigates potential data loss.
%
% By selecting the most informative GPU metrics that capture real-time user interactions, this overcomes \Ctwo as it reduces the profiling overhead of serializing the profiler metric, which stalls the GPU by frequent counter reads
% %\reham{overhead of what}\seonghun{fixed}
% and mitigates potential data loss. %approach

The last stage illustrates how \system applies machine learning classifiers such as Convolutional Neural Network (CNN), Long Short-Term Memory (LSTM), Random Forest (RF), and Support Vector Machine (SVM) models to the 1-second sampled metric sequences. 
By learning patterns in low-resolution data, it overcomes \Cthree and accurately infers user activities such as app usage, virtual object interaction, and virtual meeting inference.

\begin{table*}[t!]
\caption{A total of $30$ selected GPU performance counters leveraged by \system.}
\centering
\scriptsize  
\setlength{\tabcolsep}{4pt} 
\renewcommand{\arraystretch}{1.1} 
\begin{tabular}{|l|l|}
\hline
\textbf{Category}      & \textbf{Metric}                         \\ \hline\hline
GPU Utilization        & GPU Frequency, GPU Bus Busy, Preemptions / second, Avg Preemption Delay\\ \hline
Stalls                 & Vertex Fetch Stall,  Texture Fetch Stall, Texture L2 Miss, Stalled on System Memory\\ \hline
Memory Access          & \begin{tabular}[c]{@{}l@{}}Vertex Memory Read (Bytes/Second), SP Memory Read (Bytes/Second), Global Memory Load Instructions, \\ Global Buffer Data Read Request BW (Bytes/sec), Global Buffer Data Read BW (Bytes/sec), Global Image Uncompressed Data Read BW (Bytes/sec), \\ Bytes Data Write Requested, Bytes Data Actually Written\end{tabular} \\ \hline
Shader/Instruction     & \begin{tabular}[c]{@{}l@{}}Vertex Instructions / Second, Local Memory Store Instructions, Avg Load-Store Instructions Per Cycle, Avg Bytes / Fragment, \\ L1 Texture Cache Miss Per Pixel\end{tabular}\\ \hline
Geometry/Rasterization & Pre-clipped Polygons/Second,  Prims Trivially Rejected, Prims Clipped, Average Vertices/Polygon, Average Polygon Area\\ \hline
Texture/Filtering      & Nearest Filtered, Anisotropic Filtered, Non-Base Level Textures \\ \hline
\end{tabular}
% \caption{A total of $30$ selected GPU performance counters leveraged by \system.}
%provided by \texttt{ovrgpuprofiler} tool.}
\label{table:MetricList}
\end{table*}

%\berkay{Instead of "Feature Engineering", this is too generic. Please see if you can have a title that is more unique to the paper. }\seonghun{fixed}

%\subsection{Feature Engineering}
\subsection{GPU Metrics for AR/VR Scene Analysis}
\label{sec:leakageVectors}
We explore the relationship between GPU metrics and user activity in AR/VR environments. 
First, we identify GPU metrics that can be exploited to infer such activity. 
Then, the selected metrics are analyzed to determine their behavior based on different pixel rendering in AR/VR scenes.
This analysis process lays the groundwork for understanding how low-level GPU behavior can reveal sensitive activity in AR/VR apps.

\shortsectionBf{GPU Metric Selection.}\label{subsec:metric_selection} Simultaneous monitoring of all 72 metrics using the built-in \sysprof tool generates substantial computational overhead, often straining the GPU profiling pipeline. As a result, metric data may be missing or inconsistent due to the excessive number of data buffer access requests.
%
% Monitoring all 72 metrics simultaneously from the built-in \sysprof tool introduces computational overhead and leads to missing or inconsistent metric values since GPU's profiling pipeline is overwhelmed with an excessive number of data buffer requests. 
%Specifically, iteratively attempting to capture all 72 metrics at once overwhelms the device's data buffers, resulting in missing samples.
%
%To minimize profiling overhead, we evaluate each metric individually, which allows us to identify and select the most informative metrics for \system attack purposes.
%To minimize the overhead, we evaluate each metric individually to select the most informative metrics for attack purposes.
%
We use a controlled baseline of rendering three basic 3D objects (cube, cylinder, and sphere) in Unity~\cite{unity} and monitor each metric during the rendering process. 
This approach enables us to systematically evaluate each metric and its effectiveness in accurately classifying rendered objects. 
 
% By starting with simple 3D objects, we can systematically test each metric to see how accurately classifies which object is rendered and helps us identify the most meaningful metrics to use in \system with less overhead. 
%
%
%we employed a selective approach by rendering three basic 3D objects (cube, cylinder, and sphere) provided by Unity~\cite{unity} and collecting a total of 72 \texttt{ovrgpuprofiler} metric values.
%

\begin{comment}

Rendering these simple objects serves as a controlled baseline to simulate a basic immersive environment %\chandrika{is it user interaction or a basic immersive environment} 
within an application. 
%
This enables us to observe how the GPU metrics respond to known elementary changes in the immersive scene created by the application. 
\end{comment}

%\seonghun{you have three different objects to distinguish. However, what are their sizes? for how long do they stay on screen? how many fingerprints do you collect? How many samples do you have in your fingerprints?}
Specifically, each 3D object is set to Unity's default size of 1 $unit$\footnote{Unity's $unit$~\cite{unit} corresponds to one meter in real-world space as a basic measurement in the scene}, placed on the screen for 5 secs. 
We repeat the process 20 times per object,  creating a comprehensive dataset covering all 72 metrics across 20 measurements for each of the 3D objects.
We then train a Convolutional Neural Network (CNN) model to distinguish which virtual object is being rendered. %In parallel, we monitored metric value changes using MQDH to verify that the metrics meaningfully capture changes in the scene.

Based on our preliminary analysis, we identified 30 individual metrics that,
%
%\chandrika{
%when tested individually, }\seonghun{added "individual"}
%
consistently achieved over 60\% classification accuracy and exhibited variation across different rendering objects within the AR/VR scene.
These 30 selected metrics, shown in Table~\ref{table:MetricList}, reduce computational overhead and ensure capturing fingerprints relevant to user interactions with higher accuracy. 
Specifically, limiting profiling to this subset prevents missing or inconsistent values as Meta recommends fewer than 30 simultaneous real-time counters~\cite{ovrgpuprofiler_detail}.
%\reham{rephrase this sentence to show why relying on 30 metrics doesn't result in missing or inconsistent values, can we give any numbers to evaluate the overhead? e.g. time, buffer size}\seonghun{we dont calculate those overhead but encoutner inconssitent value and this is mentioned in \Ctwo}
%
%By performing metric selection, we reduce the overhead required for \system and ensure the effectiveness of fingerprint datasets in detecting rendered objects without any missing or inconsistent values. 
%
Consequently, we further refine these metrics for each case study %, as shown in 
(Section~\ref{sec:casestudies}).

%\seonghun{How did you run, which tool?, Same Pearson from Case study 3? What is the final list?}
%
\update{
%\textbf{[R3]} 
Beyond selecting metrics from the accuracy, we perform a pairwise correlation analysis on GPU metric traces from a basic 3D cube. 
We compute Pearson correlation coefficients~\cite{benesty2009pearson} over each metric and measure highly redundant pairs ($|r|>0.90$).
%
%We prune one metric from each highly correlated pair. 
This process reduces the metric set from 30 to 11 while preserving the most informative traces (Appendix Table~\ref{table:pruned_gpu_metrics}).
%
%Throughout the paper, we demonstrate that a smaller set of metrics to achieve higher accuracy with lower computational overhead. 
}

\begin{comment}
    
Final kept metrics:
 - GPU % Bus Busy
 - % Vertex Fetch Stall
 - % Anisotropic Filtered
 - % Non-Base Level Textures
 - SP Memory Read (Bytes/Second)
 - Preemptions / second
 - Avg Preemption Delay
 - Global Memory Load Instructions
 - Local Memory Store Instructions
 - Avg Load-Store Instructions Per Cycle
 - Bytes Data Write Requested

 both list and case study 
Global Memory Load Instructions
SP Memory Read (Bytes/Second)
% Vertex Fetch Stall
% Anisotropic Filtered
% Non‑Base Level Textures
\end{comment}

\shortsectionBf{ Non-Base Level Textures Metric.}\label{subsec:None-base_level}
%
%CHANGED TO BELOW-SH
\begin{comment}
In AR/VR devices, virtual 3D objects are rendered as pixels within the immersive environment. 
%
Therefore, we posit that pixel-level analysis is a critical component for understanding user interactions with 3D content. 
Hence, we identify the GPU metrics that are sensitive to pixel variations.
\end{comment}
%Thus, we assume that pixel-level analysis is a vital step for understanding user interactions with 3D objects. To justify our assumption, we determine which GPU metrics are most sensitive to these pixel variations. 
%
%\shortsectionBf{Non-Base Level Textures.} 
%
We \update{systematically} compared all 72 metrics from \texttt{ovrgpuprofiler} under various rendering cases. Our experiment involves rendering basic 3D objects such as a cube and observing how each metric responds to changes in object size and position from the user's point of view. From this evaluation, we found that \texttt{Non-Base Level Textures} metric achieves the highest correlation among 72 metrics provided by the \texttt{ovrgpuprofiler}. 

The \texttt{Non-Base Level Textures} metric represents the percentage of textures that are not at the base
mipmap~\cite{williams1983pyramidal} level (level 0), which is simply the original, full-resolution texture image that the application uploads into GPU memory. 
%Mipmaps are sequences of pre-calculated, progressively lower-resolution images that optimize rendering by minimizing aliasing and moiré patterns~\cite{MoirePattern,gonzalez2009digital,akenine2019real}.
%They enhance rendering performance by selecting the appropriate texture resolution based on the distance of the object from the camera origin. 
The base mipmap level corresponds to the highest resolution texture, delivering maximum detail when a 3D object is viewed up close.
%
%The \texttt{Non-Base Level Textures} metric represents the percentage of textures that are not at the base
%\reham{please change it to base not based}\seonghun{fixed} mipmap~\cite{williams1983pyramidal} level, which is used for rendering virtual objects. Specifically, mipmaps are sequences of pre-calculated, optimized images improving rendering performance by reducing aliasing and moiré patterns~\cite{MoirePattern}\reham{do we have a better reference? paper, book, etc}. 
%
% The base mipmap level refers to the highest texture resolution which provides maximum detail when the 3D object is viewed up close.
%Therefore, the base mipmap level refers to the highest screen resolution %\reham{we can highest screen resolution} \seonghun{fixed}of texture, providing maximum detail when the 3D object is viewed up close. 
%
% For example, when the user interacts with large 3D objects that occupy significant screen space in the AR/VR devices, the rendering process prioritizes visual clarity for detail. To achieve this, the GPU utilizes lower mipmap levels (close to the base level) to ensure that the texture of objects in the AR/VR scene appears sharp and detailed. 
%
For example, when a user interacts with large 3D objects that occupy a significant portion of the screen in AR/VR devices, the rendering process prioritizes visual clarity by using lower mipmap levels (closer to the base level) to ensure the textures of the objects appear sharp and detailed.

As a result, a higher percentage of textures near the base mipmap level leads to an increase in the \texttt{Non-Base Level Textures} metric value, which indicates that the GPU is processing more detailed textures to render the 3D object accurately.
Conversely, smaller objects occupying fewer pixels on the screen require less texture detail, utilizing higher mipmap levels (further from the base level). Hence, there is a positive correlation between object size and the \texttt{Non-Base Level Textures} metric for 3D objects.
%
% Conversely, smaller objects occupying less number of pixels in the screen do not require the smaller level of texture detail and lead to higher mipmap levels (far from the base level). Hence, there is a positive correlation between the object size and the \texttt{\% Non-Base Level Textures} metric for the 3D objects.

\begin{figure}[t]
  \centering
  
  \subfloat[Meta Quest 2\label{fig:PixelAnalysis_subfig1}]{
    \includegraphics[width=0.22\textwidth,height=3.3cm]{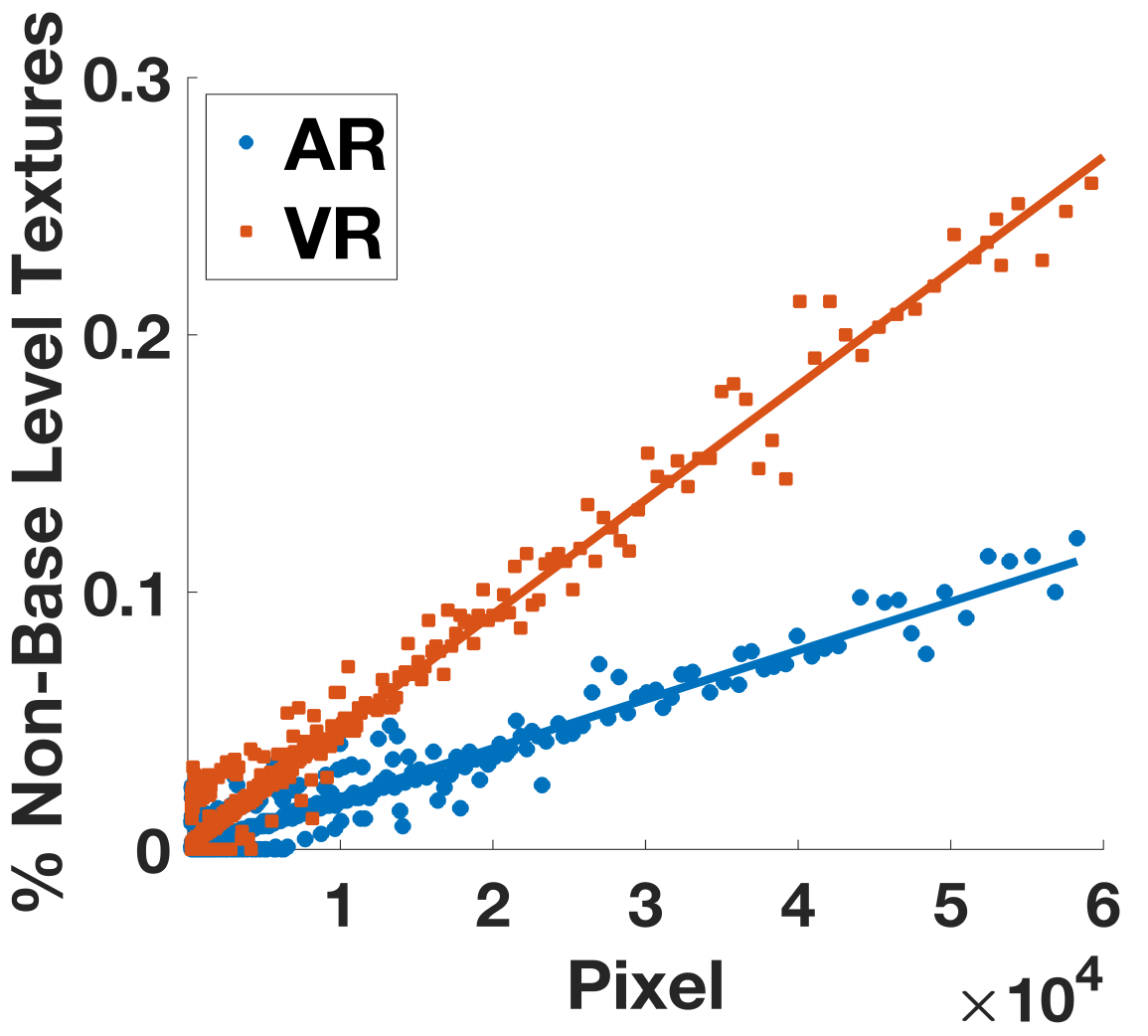}
  }
  \hfill
  \subfloat[Meta Quest 3\label{fig:PixelAnalysis_subfig2}]{
    \includegraphics[width=0.22\textwidth,height=3.3cm]{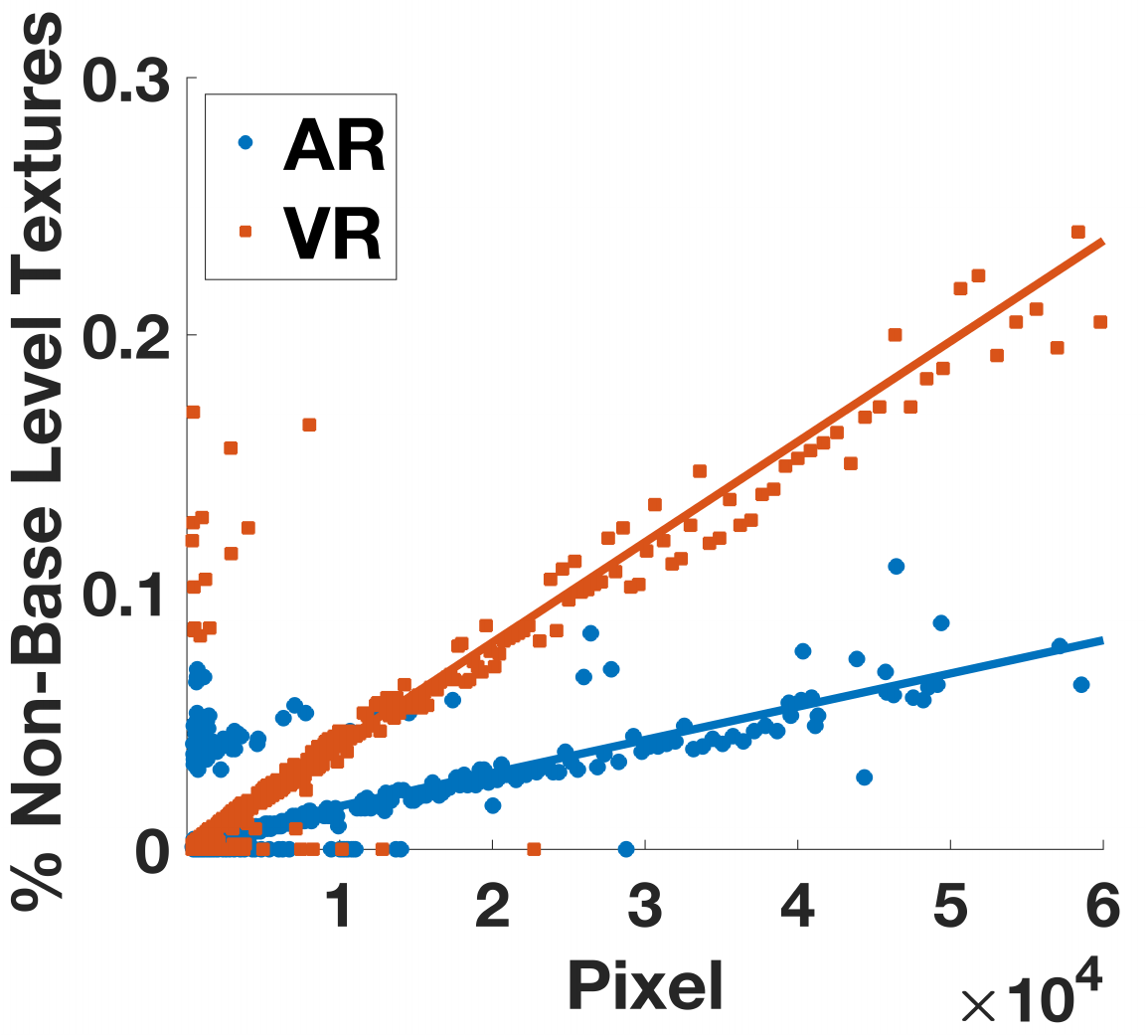}
  }

  \caption{Relation between pixel value and the \texttt{Non-Base Level Textures} metric in both (a) Meta Quest 2 and (b) Meta Quest 3. 
  %
  %Orange data points represent the experiments conducted in the VR (immersive) scene, while blue data points represent the AR (passthrough) scene. 
  %\reham{we need to have a better resolution for these figures.}\seonghun{fixed}
  }
  \label{fig:PixelAnalysis}
\end{figure}

\shortsectionBf{Correlating Pixel and GPU Metric.}\label{subsec:CorrelationPixel}
%\reham{can we use a better indicative title here}\seonghun{fixed}
In AR/VR devices, virtual 3D objects are rendered as pixels within the immersive environment. 
Therefore, we posit that pixel-level analysis is a critical component for understanding user interactions with 3D content. 
%Hence, we identify the GPU metrics that are sensitive to pixel variations.
%
We further characterize the exact number of pixels rendering on the screen with the selected \texttt{Non-Base Level Textures} metric. 
%
%To calculate precise pixel coverage on the screen, we create another temporary camera in Unity with a designated layer on the target VR object. 
%
This process assigns the target object to a specific layer to isolate it from the black background, $R, G, B=(0,0,0)$.
%as shown in Appendix Figure~\ref{fig:BlackBG}. 
%
Next, we apply the \texttt{RenderTexture} library to convert the 3D images into \texttt{Texture2D} objects since it allows us to obtain the number of pixels for each object.
%\reham{why?}.\seonghun{fixed}
We then apply a color threshold technique, which filters out the black background and retains only the pixels belonging to the target object, resulting in an accurate measure of the object's rendered pixels.
%Specifically, we iterate through each pixel in the \texttt{Texture2D} Unity class variable and determine whether the pixel is part of the object by checking if any RGB color component exceeds the predefined threshold, $\ R, G, B >0$. Therefore, we are able to count exact pixels as part of the target object. This approach produces a more accurate method to measure pixel coverage and allows us to establish a correlation between the number of pixels on the AR/VR scene and the \texttt{\% Non-Base Level Textures}. %, For example, if any $R, G, B >0$.

We collect 1,000 GPU metric readings by varying the size of the 3D object to observe a wide range of pixel coverage. 
To achieve the correlation between pixels and the \texttt{Non-Base Level Textures} metric, we employ \texttt{LinearRegression} from the Python \texttt{scikit-learn} library~\cite{pedregosa2011scikit} and \update{separately compute} Pearson correlation coefficient, $\rho_{X,Y}= {cov(X,Y)}/\sigma_X\sigma_Y,$ where $cov(X,Y)=\mathbb{E}[(X-\mu_X)(Y-\mu_Y)]$. %\reham{do you apply linear regression and then pearson correlation or separately?}. \seonghun{fixed}
We employ linear regression to quantify how well our chosen metric predicts pixel coverage and to validate that the selected GPU metric remains robust across both AR and VR scene configurations.  %\reham{Before referring to the library mention how much points we collect and that we apply linear regression and why? }

Our analysis yields a strong correlation in both AR and VR scenes. 
We evaluate our method on both Meta Quest 2 and Meta Quest 3 to demonstrate it applies to different hardware generations (Section~\ref{sec:ExperimentSetup}). 
Specifically, Meta Quest 3 incorporates an upgraded system-on-chip (SoC) and improved display hardware compared to Meta Quest 2, potentially influencing GPU performance metrics. %Additional details on hardware differences are provided in Section~\ref{sec:ExperimentSetup}.
%
%\reham{Before directly showing the result on Quest 2 and 3, we need to jusifty why we tested and both and what are the differences? If we mention it in the background, refer to it briefly}\seonghun{fixed}
In AR mode, Meta Quest 2 achieves an $R^2$ score of 0.90 with a correlation coefficient of 0.95, while Meta Quest 3 achieves an $R^2$ score of 0.71 with a correlation coefficient of 0.84.
%
% Specifically, for AR, in Meta Quest 2, we achieve a $R^2$ score of 0.90 with a correlation coefficient of 0.95. For the Meta Quest 3, with an $R^2$ score of 0.71 and a correlation coefficient of 0.84. 
Similarly, in VR mode, Meta Quest 2 achieves an $R^2$ score of 0.98 and a correlation coefficient of 0.99, whereas  Meta Quest 3 records an $R^2$ score of 0.89 with a correlation coefficient of 0.94.

The higher VR correlations came from the fully rendered virtual environment, where every pixel change reflects GPU work. However, AR passthrough mode mixes camera feed and overlays, introducing additional noise into the metric.
%
%and a $R^2$ score of 0.71 with a correlation coefficient of 0.84 for Meta Quest 3. Moreover, for VR scenes, the $R^2$ score is 0.98 with a 0.99 correlation coefficient for Meta Quest 2, and the $R^2$ score is 0.71 with a 0.84 correlation coefficient for Meta Quest 3. 
%
These values indicate a strong relationship between pixel coverage and the \texttt{Non-Base Level Textures} metric in both AR and VR scenes, as illustrated in Figure~\ref{fig:PixelAnalysis}.  %\reham{We can give a justification why VR is better than AR}\seonghun{fized}

%\seonghun{---------UNTIL HERE--------}

%%%%%%%%%%%%%%Quest 2%%%%%%%%%%%%%%
%Regression for 'AR': % Non-Base Level Textures = 0.0000 * Total + 0.0012, R² = 0.9016
%Regression for 'VR': % Non-Base Level Textures = 0.0000 * Total + 0.0025, R² = 0.9818
%Correlation between 'Total' and '% Non-Base Level Textures' for 'AR': 0.9495
%Correlation between 'Total' and '% Non-Base Level Textures' for 'VR': 0.9909
%%%%%%%%%%%%%%%%%%%%%%%%%%%%%%%%%%%
%%%%%%%%%%%Quest 3%%%%%%%%%%%%%%%%%%%%
%Regression for 'AR': % Non-Base Level Textures = 0.0000 * Total + 0.0039, R² = 0.7084
%Regression for 'VR': % Non-Base Level Textures = 0.0000 * Total + 0.0055, R² = 0.7059
%Correlation between 'Total' and '% Non-Base Level Textures' for 'AR': 0.8415
%Correlation between 'Total' and '% Non-Base Level Textures' for 'VR': 0.8402
%%%%%%%%%%%%%%%%%%%%%%%%%%%%%%%%%%%
\begin{figure}[t]
  \centering

  \subfloat[$v=1, z=2$\label{fig:SpeedDistacnesubfig1}]{
    \includegraphics[width=0.14\textwidth,height=2.3cm]{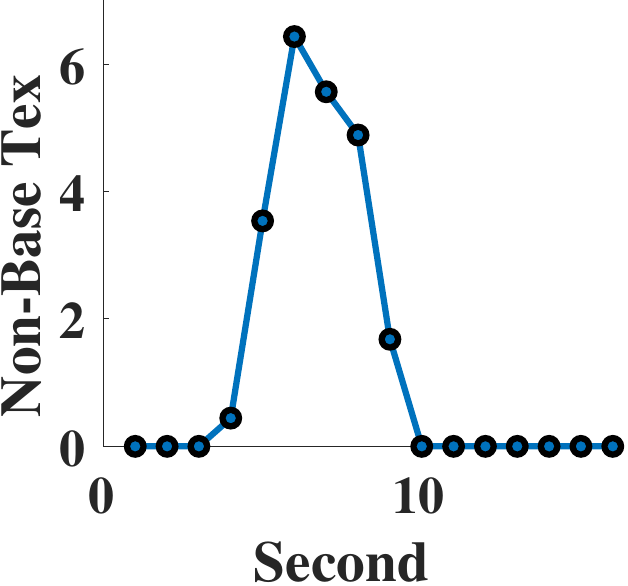}
  }
  \hfill
  \subfloat[$v=2, z=2$\label{fig:SpeedDistacnesubfig2}]{
    \includegraphics[width=0.14\textwidth,height=2.3cm]{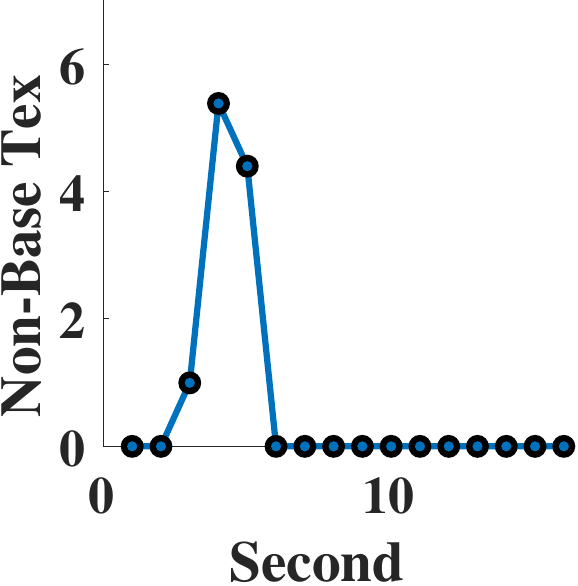}
  }
  \hfill
  \subfloat[$v=1, z=3$\label{fig:SpeedDistacnesubfig3}]{
    \includegraphics[width=0.14\textwidth,height=2.3cm]{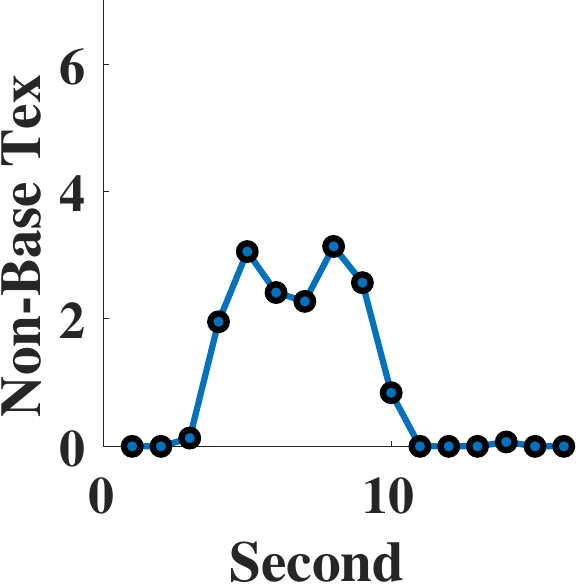}
  }

  \caption{%\reham{For this figure also, we need a better resolution ,you can make the font bigger and the lines more thick}\seonghun{changed}
  Correlation between \texttt{Non-Base Level Textures} metric and VR Cube object rendering with (a) the speed, $v$ = 1 $unit/second$ and distance coordinate $z=2\ unit$, (b) increasing speed to $v$ = 2 $unit/second$, and (c) move object further from the point of view, $z=3\ unit$.}
  \label{fig:SpeedDistacne}
\end{figure}

\shortsectionBf{GPU Metrics in Relation to Speed and Depth.} %\reham{We can find a more descriptive title for this subsection}
%
%We examine dynamic factors such as the speed of the 3D object and the depth between the camera and the 3D object. 
%
In this experiment, we automatically launch the basic Unity application, generating a single cube moving at two different speeds (1 and 2 $unit/second$). %\reham{We can add a footnote here to explain what is units/second}
The cube travels left to right over a total horizontal distance of 30 $units$ (x-coordinates from -15 to 15). 
We also vary the cube's depth %\reham{suggestion: we can use depth instead of distance} 
from the camera by placing it at three distinct z-coordinates (2, 2.5, and 3 $unit$). 
Each experiment runs for 30 seconds and is repeated 20 times to ensure sufficient data capture. 
This setup allows us to measure how moving speed and object depth affect the \texttt{Non-Base Level Textures} metric. More metrics relation is depicted in Appendix Figure~\ref{fig:SpeedDistacne_l2miss}. %\reham{Can we do similar experiments for other metrics and put it in Appendix?}\seonghun{yes, but I will leave this at the end}   

%We demonstrate the relationship between the GPU metric, \texttt{\% Non-Base Level Textures}, and the moving 3D cube object. 
When the cube moves slowly across the scene, the GPU's workload remains high for a longer duration, resulting in a broader width in the fingerprint created from \texttt{Non-Base Level Textures}.
% metric created fingerprint. 
Conversely, when any 3D object moves faster in the scene, the GPU metric value spikes appear narrower. 
%
%For example, at a speed of 1 $unit/second$, a wider width of the peak is observable, whereas at 2 $unit/second$ speed shows a slimmer width of the peak, as depicted in Figure~\ref{fig:SpeedDistacne}. 
For instance, at a speed of 1~$unit/second$, a broader peak width is observed, whereas at 2~$unit/second$, the peak appears narrower, as illustrated in Figure~\ref{fig:SpeedDistacnesubfig1} and \ref{fig:SpeedDistacnesubfig2}.
%\reham{I think it is shown in all Fig. 7 not just (a)}. 
%\reham{Edited the next sentence, please check and add the speed value}\seonghun{didn't measure the speed it depends on the screen size..which means if size is smaller and far, it takes more time to traverse the scene so there is no absolute number of speed we can report}\seonghun{we decide to do when the time allows.}\seonghun{fixed}
However, due to the low resolution of the GPU profiler (1Hz), our attack only detects objects that traverse the scene in more than one second. %This corresponds to a speed of \seonghun{xx}, which remains reasonable for many typical VR scene scenarios. \seonghun{We discussed}
% However, we can only distinguish 3D objects with relatively lower speeds that traverse the scene in more than 1 second due to the low resolution of the profiler.
%demonstrating these dynamic correlations has limitations due to the profiler's low resolution. 
%Specifically, since \texttt{ovrgpuprofiler} captures data at one-second intervals, it is difficult to distinguish fast-moving 3D objects that traverse the entire scene in less than one second. \reham{We can rephrase to say that due to the low resolution of the profiler, we can only distinguish 3D objects with relatively lower speed that traverse the seen in more than 1 second.}

%The depth between the user's viewpoint (camera) and the 3D object either amplifies or diminishes metric values. 
%
When the 3D object is rendered closer to the camera and occupies a substantial portion of the screen, it causes an increase in the \texttt{Non-Base Level Textures} metric value, along with other \texttt{ovrgpuprofiler} metrics that are affected by the depth between the viewpoint and 3D objects.
%\reham{need to mention them or refer to appendix}\seonghun{we only know based on the accuracy...I will add if i can do based on the earlier excel file }. %change in size of 3D object affect user activities \reham{that are affected by?}. 
%
This visibly higher peak within the sampled dataset indicates that the GPU actively renders more detailed textures in both AR/VR scenes. 
In contrast, if a 3D object is farther away and occupies fewer screen pixels due to the smaller sizes, it creates minor changes in the GPU metrics. 
For example, placing the cube object at coordinate $z=2$ significantly increases \texttt{Non-Base Level Textures} values, resulting in higher peaks, whereas $z=3$ produces smaller values, as shown in Figure~\ref{fig:SpeedDistacnesubfig1} and \ref{fig:SpeedDistacnesubfig3}. 
% 
%As demonstrated earlier in the pixel relation presented in Figure~\ref{fig:PixelAnalysis}, subtle changes in the \texttt{Non-Base Level Textures} metric values can be challenging to classify and exploit in an attack scenario\reham{Which part are you exactly referring to here? I think we can remove this sentence}. \seonghun{removed}
%
Consequently, these observations serve as a foundation for the detailed case studies discussed in Section~\ref{sec:casestudies}.

\begin{comment}

\subsubsection{Feasibility across different vendor AR/VR devices.}
\update{Although our primary evaluation centers on the Meta Quest series, we also explore the feasibility of deploying our methodology on additional AR/VR headsets to demonstrate broader applicability.
%
In particular, we tested the \system on the most recent version of Microsoft HoloLens even though Microsoft stopped releasing their devices online. 
%
By collecting \seonghun{A,B,C} available GPU metrics, we verified that \system can capture rendered objects in both AR and VR scenes.

Our analysis shows that different vendors AR/VR devices exhibit similar behavior to Meta Quest in terms of GPU profiler rifle usage patterns in both AR/VR scenes, as described in \seonghun{Figure~}. Even though HoloLens is no longer available to consumers, our method demonstrates how an attacker could exploit GPU profiler-based side channels across different vendor platforms to infer rendered content.
}

\end{comment}
%\update{feasibility on different AR/VR\\
%1. HoloLens (not available anymore in online store)
%2. Powershell }

\subsection{Experiment Setup}\label{sec:ExperimentSetup}
\shortsectionBf{Software Configuration.}\label{sec:deviceConfig}
We launch our experiments on the most recently updated software of Meta Quest devices, specifically, Meta Quest builds version 72.0, Android OS version 12~\cite{android12}, and Unity~\cite{haas2014history} version 2022.3.34f1. 

\shortsectionBf{Hardware Configuration.}\label{sec:HW_sec} Meta Quest 2 is equipped with a Qualcomm Snapdragon XR2 system on a chip (SoC). This chip is an advanced product of the Snapdragon 865 model, specifically designed for AR/VR devices. The SoC integrates the Adreno 650 GPU operating at 587 MHz. Moreover, Meta Quest 3 utilizes the Snapdragon XR2 Gen 2, an enhanced version with Adreno 740 running at 599 MHz. Meta Quest 3S also employs a Qualcomm Snapdragon XR2 Gen 2 processor with 8GB of RAM. 
\update{
%\textbf{[R6]} 
Our experiments are conducted on different dates and at multiple locations by different authors to validate cross-device and cross-environment.}
Furthermore, our experiments are conducted across all available Meta Quest series, which demonstrates the feasibility of our attack.

\shortsectionBf{Attack Workflow.}\label{sec:AttackWorkflow}
We explain the detailed step-by-step process of \system on the victim's device. 
%
\begin{comment}
\update{[R4] In our proof-of-concept, the user launches the attacker-controlled app on their AR/VR device in developer mode to invoke shell commands. Nevertheless, basic GPU metrics can be accessible to any user-space application~\cite{ovr_metrics_tool}.
%
This app starts a malicious bash shell script while being in the foreground, which triggers the GPU profiler. 
%
\system employs the \texttt{nohup}~\cite{nohup} command to ensure persistent GPU profiler execution using the bash script when the user terminates the malicious app, as shown in Listing~\ref{lst:gpu_profiler}.} 
%
\begin{lstlisting}[caption={Example GPU Profiling Script},label=lst:gpu_profiler]
#!/bin/bash
# Nohup command ensure persistent execution
nohup ovrgpuprofiler -r gpu_freq,texture_l2_miss > log.txt &
\end{lstlisting}
\end{comment}
%
%\seonghun{did you observe any differences between developer mode ovrgpuprofiler counter monitoring values and in-app monitoring values?}
%
\update{
%[R4] 
As an in-app attack workflow, the adversary downloads the OVRMetric Tool packages~\cite{meta_ovr_metrics_download}. Then uses the Android NDK to cross-compile a \sysprof by linking \texttt{libOVRMetricTool.so}~\cite{ruggia2025dark} with wrapper code, as shown in Listing~\ref{lst:CrossCompile}.
%
%This will create the \sysprof ELF binary.
}
\begin{figure}[t]
\centering
\begin{minipage}{\columnwidth}
\captionsetup{type=lstlisting,width=\columnwidth,margin=0pt}
  \captionof{lstlisting}{Cross-compile to create binary.}
  \label{lst:CrossCompile}

\lstset{
  xleftmargin=2em,          % shift listing right
  framexleftmargin=2em,     % keep frame aligned
  framexrightmargin=0pt,
 % belowskip=-12pt
}
\begin{lstlisting}[language=bash]
export NDK=<Path to your NDK>/Android_ndk
export TOOLCHAIN=$NDK/toolchains/llvm/prebuilt/darwin-x86_64  # MacOS

$TOOLCHAIN/bin/aarch64-linux-android21-clang++ \
    -std=c++11 -O2 -fPIE -pie \
    -I"$PWD/include" \
    -L"$PWD/libs/arm64-v8a" -lOVRMetricsTool -pthread \
    -o ovrgpuprofiler \
    my_profiler.cpp             # Wrapper code 
\end{lstlisting}
\end{minipage}
\end{figure}
% \begin{lstlisting}[caption={Cross-compile to create  binary},label=lst:CrossCompile]
% export NDK=<Path to your NDK>/Android_ndk
% export TOOLCHAIN=$NDK/toolchains/llvm/prebuilt/darwin-x86_64                          # MacOS

% $TOOLCHAIN/bin/aarch64-linux-android21-clang++ \
%     -std=c++11 -O2 -fPIE -pie \
%     -I"$PWD/include" \
%     -L"$PWD/libs/arm64-v8a" -lOVRMetricsTool -pthread \
%     -o ovrgpuprofiler \
%     my_profiler.cpp             # Wrapper code 
% \end{lstlisting}
%
\update{When the app launches, it copies the binary into the private data directory and then starts \sysprof by either invoking \texttt{ProcessBuilder}~\cite{java_processbuilder} in Java or calling a Unity JNI helper~\cite{unity_androidjnihelper} that performs a double-fork method~\cite{stackoverflow_detached_fork} as depicted in Listing~\ref{lst:inapp_profiler}.
This double-fork sequence (line 13-19) creates a detached background process. First, fork creates a child process and \texttt{setsid()} makes that child run its own session, separating it from the app. 
The second fork creates a grandchild while the first child exits immediately. Because the first child exists, the grandchild process is left without a parent and automatically adopted by Android's main init process (PID1).
Therefore, the profiler is no longer tied to the app's process and will continue writing GPU counter metrics to a file even within the sandbox environment. 
%This allows \sysprof to write GPU counter metrics to a file even with the sandbox environment.
}
\begin{figure}[t]
\centering
\begin{minipage}{\columnwidth}
  \captionsetup{type=lstlisting,width=\columnwidth,margin=0pt}
  \captionof{lstlisting}{In-app profiler example. 
  % \berkay{The space between Listing 1 and Listing 2's caption is too small. There is something wrong. Please fix.}
  }
  \label{lst:inapp_profiler}

\lstset{
  xleftmargin=2em,          % <--- shift listing right
  framexleftmargin=2em,     % match frame to margin
  framexrightmargin=0pt, 
  % belowskip=-12pt
}
\begin{lstlisting}[language=Python]
# Copy and prepare the profiler ELF
dst = getFilesDir() + "/ovrgpuprofiler"
copyFile(src, dst)
setExecutable(dst)

# Option A: via Java ProcessBuilder
args = [ dst, "-r", <counters> ]
ProcessBuilder(args).start()

# Option B: Unity JNI helper double-fork
nativeStartDetached(dst, ["-r", <counters>])

function nativeStartDetached(path, args):
    pid1 = fork()               # first fork
    if pid1 > 0: return         # parent returns
    setsid()                    # child detach
    pid2 = fork()               # second fork
    if pid2 > 0: exit(0)        # first child exits
    execv(path, [path] + args)  # Execute in grandchild
\end{lstlisting}
\end{minipage}
\end{figure}
% \begin{lstlisting}[caption={In‑App profiler example},label=lst:inapp_profiler]
% # Copy and prepare the profiler ELF
% dst = getFilesDir() + "/ovrgpuprofiler"
% copyFile(src, dst)
% setExecutable(dst)

% # Option A: via Java ProcessBuilder
% args = [ dst, "-r", <counters> ]
% ProcessBuilder(args).start()

% # Option B: Unity JNI helper double-fork
% nativeStartDetached(dst, ["-r", <counters>])

% function nativeStartDetached(path, args):
%     pid1 = fork()               # first fork
%     if pid1 > 0: return         # parent returns
%     setsid()                    # child detach
%     pid2 = fork()               # second fork
%     if pid2 > 0: exit(0)        # first child exits
%     execv(path, [path] + args)  # Execute in grandchild 
% \end{lstlisting}
%In both cases, the GPU profiler runs independently over a defined period of time and monitors GPU performance metrics, logging them into a local file.
This file can be later analyzed or transmitted when the malicious app is relaunched by the user.
Our attack does not need any privilege escalation or access to sensitive sensors, as the GPU profiler is directly accessible from the XR devices.

\begin{comment}

\update{[R4] As an in-app attack workflow, the adversary downloads the OVR Metric Tool packages~\cite{meta_ovr_metrics_download}. Then uses the Android NDK to cross-compile a \sysprof by linking \texttt{libOVRMetricTool.so}.
%
When the app launches, it copies the binary into the private data directory and then starts \sysprof by either invoking \texttt{ProcessBuilder}~\cite{java_processbuilder} in Java or calling a Unity JNI helper~\cite{unity_androidjnihelper} that performs a double-fork method~\cite{stackoverflow_detached_fork}. Therefore, the profiler continues running after the app exits. 

In both proof-of-concept cases, \sysprof writes GPU counter metrics to a file even with the sandbox environment.}
%

\begin{lstlisting}[caption={Example GPU Profiling Script},label=lst:gpu_profiler]
#!/bin/bash
# Nohup command ensure persistent execution
nohup ovrgpuprofiler -r gpu_freq,texture_l2_miss > log.txt &
\end{lstlisting}

\end{comment}

\shortsectionBf{Classification.} 
\system tracks the pre-selected 30 GPU metrics from the built-in GPU profiler. %that showed the most significant changes during our feature selection process (Section~\ref{subsec:metric_selection}).
The real-time metric values, $M_i$ are collected for $n$ seconds, where $i$ is the metric index. 
The collected values generate an individual fingerprint, where each metric is represented as $M_i = \{t_1, t_2,\dots,t_n\}$. 
The fingerprint uniquely characterizes the behavior of specific AR/VR apps and user interactions over time.
Then, the fingerprint is preprocessed by normalizing the metrics using standard normalization to remove the baseline and noise. 

We evaluate the collected dataset using four classification models: Convolutional Neural Network (CNN), Long Short-Term Memory (LSTM), Random Forest (RF), and Support Vector Machine (SVM). 
%
%In addition to the model's test accuracy, we compute the F-1 score, Precision, and Recall values for each model. 
%
For CNN and LSTM models, we input the entire time series by padding multiple metrics for each execution to ensure a total size of $n \times i$ across all samples. %This padding standardizes input dimensions for batch processing while preserving the distinct patterns present across different metrics, leading to more distinctive fingerprints that enhance classification performance.
In contrast, RF and SVM models require a fixed-length feature vector. %Simply feeding the time series into these models would create an excessively large and flat feature vector, leading to overfitting.
Therefore, we compute statistical values ($\mu,\ \sigma, \ \max,\ \min$) for each metric across all time steps, then concatenate these values into a single feature vector.

The preprocessed fingerprint is fed into pre-trained ML models to classify the standalone AR/VR or WebXR app currently running in the foreground.
Once the app is identified, the collected GPU metric values are leveraged to infer the VR object render in both AR/VR scenes.
%
%ML models are trained and tailored to the characteristics of GPU profiling data to achieve attack goals.
%
To ensure the generalizability of \system, we conduct experiments across a diverse set of devices within the Meta Quest family, including cross-series models such as Quest 2, 3, and 3S.

%% file: Sections/5_Evaluation.tex
\section{Evaluation Through Case Studies}\label{sec:casestudies}

%\reham{Instead of case studies we can name the section evaluation. We can add here the evaluation questions we discussed last meeting, we can then make case studies a subsection (based on one of the questions) or we can choose to make it a separate section to balanace the size of sections.}
%\seonghun{we decide to have an evaluation for each case study}
%\reham{As a naming convention, we can still name the section Evaluation and have an introduction that briefly explains how we evaluate the system using the 4 case studies}\\
%\reham{the sentence is missing something here, do we need to mention Meta Quest devices?}\seonghun{fixed}
%
Through four case studies, we evaluate the practical implications of \system framework on Meta Quest devices, which prohibit concurrent standalone app execution (as detailed in Section~\ref{sec:DesignChallenges}, \Cone), a challenge largely unexplored in prior AR/VR side-channel research.
%
% We evaluate the practical implications of \system framework on Meta Quest devices through four case studies, which prohibit concurrent standalone app execution (as detailed in Section~\ref{sec:DesignChallenges}, \Cone) and remain largely unexplored in prior AR/VR side-channel research.
%
%\update{with the most challenging (\Cone) and unexplored Meta Quest series}\reham{briefly detail instead of mentioning "most challenging and unexplored" e.g. "with Meta Quest series since they adopt the most restrictive policies, where no concurrent apps are allowed (as detailed in Section~\ref{x}) and they remain largely unexplored in literature."}.\seonghun{updated}
%
Each case study compromises a distinct privacy and security aspect of AR/VR interaction. Collectively, they address and overcome the primary design challenges (\Cone-\Cthree) posed by our threat model. 
%In particular, we demonstrate how an attacker can bypass Meta Quest's restriction on concurrent application execution and execute a malicious code snippet in the background without requiring additional permissions and installation of additional SDKs (\Cone). We further show that even a low-resolution profiler can still extract sensitive information from user activities (\Cthree). 
%
%These design challenges are solved in all of our four case studies. 
%
%As described in Section~\ref{sec:AttackWorkflow},
In the following sections, we detail each case study. \system proceeds in two stages: first, the attacker determines which (i) AR/VR standalone app or (ii) WebXR website has been launched; and second, it targets user privacy within these environments to extract sensitive information, including (iii) rendered VR objects and (iv) meeting room participants. 
%
%\chandrika{Open World User Interaction is also part of evaluation. We can consider adding one sentence at the end here about the user study and make the corresponding section under evaluation.}
%\seonghun{only if we add underneath the evaluation section}

%second, it focuses on compromising user privacy (iii), (iv) within those apps.  
%As described in Section~\ref{sec:AttackWorkflow}, \system proceeds in two stages. First, the attacker determines which (i) AR/VR  application or (ii) WebXR website has been launched. Once the applications or websites are classified, the second stage focuses on compromising user privacy. Specifically, we demonstrate how an adversary can (iii) detect VR objects within a shopping application and (iv) infer the number of participants in a private AR/VR meeting. %\reham{We can merge figure 2 with threat model figure, not necassary to refer to the fire here.}

\subsection{Case Study I: AR/VR Standalone App Fingerprint~}\label{sec:casestudy1}
%\chandrika{Maybe AR/VR Standalone Application (WebXR also build AR/VR app)}
%\seonghun{fixed!}
The first stage of our attack performs AR/VR application fingerprinting as the victim launches a standalone application within the immersive environment of the Meta Quest series, without requiring any user interaction.
%The first stage of our attack implements AR/VR application fingerprinting as the victim launches the standalone application in the immersive environment provided by the Meta Quest series without any interaction. %with a web browser.
%
%\chandrika{interaction with a web browser: I think can be removed.}\seonghun{fixed}

%In this case study, \texttt{OVRSeer} framework runs in the background without requiring additional permissions and successfully performs a standalone application fingerprinting attack (RQ1) without launching any concurrent applications or installation of additional SDKs. Additionally, \texttt{OVRSeer} effectively classify Applications under (RQ2) low-resolution (1 second) sampling rate of the \texttt{ovrgpuprofiler}.

\shortsectionBf{Experiment Design.} We target the top 100 most popular free applications from Meta Quest App Store\footnote{https://www.meta.com/experiences/view/1321443348416166}, which consist of six different categories: Gaming (31\%), Entertainment (21\%), Fitness (10\%), Social (7\%), Productivity (27\%), and \update{Mixed} Reality (4\%). %\reham{We can add percentages beside each category e.g. Gaming (31\%) }.\seonghun{fixed}
%
%Although these categories represent a diverse set of apps, gaming is the most popular use case for AR/VR headset~\cite{popular_meta_apps}, resulting in a large portion of the App Store containing gaming apps.
%
Although these categories encompass a diverse set of applications, gaming remains the most popular use case for AR/VR headsets~\cite{popular_meta_apps}, where it dominates a large portion of the App Store.
%
%Even though these categories represent a diverse set of applications, gaming applications make up a significant portion of the App Store since AR/VR devices are designed to deliver an immersive and interactive experience that aligns with the purpose of gaming \reham{instead we can say "since gaming is the most popular use case for AR/VR headsets" and give a reference}.\seonghu{fixed} 
%
%However, many applications have functionalities that span multiple categories. For instance, some applications may offer both social interaction and gaming elements. Thus, it is difficult to strictly classify applications into seven different categories. 
%
We list those 100 apps with their main purpose categories in  Appendix Table~\ref{table:AppList}.
%We categorize the 100 targeted AR/VR apps based on their main purposes in Appendix Table~\ref{table:AppList}.

\shortsectionBf{Data Collection.} After selecting the target applications, we employ a malicious script file to run the \system framework. Specifically, we automatically launch the target applications to collect the app fingerprinting dataset.
%by leveraging the \texttt{monkey} tool~\cite{androidMonkey}, a command-line tool that operates on Android devices. 
%This \texttt{monkey} tool easily helps us to automatically launch the interest applications by the application package name.
%
%We simultaneously monitor 30 selected GPU metrics provided from the \texttt{ovrgpuprofiler} tool for 30 seconds per app execute, as outlined in Section~\ref{sec:leakageVectors}. %\reham{
For each measurement, we run an app for 30 seconds and simultaneously monitor the 30 selected GPU metrics as outlined in Section~\ref{sec:leakageVectors}.
%
%\reham{Please revise to make sure terms are consistent, e.g. sample vs measurement}\seonghun{fixed}
%
Each app is profiled 20 times, %\reham{
resulting in a dataset consisting of 2000 measurements across the Meta Quest family (Quest 2 and 3).
For each device, we randomly split the dataset into balanced subsets using 80\% for training and 20\% for testing.
%
%We randomly split the datasets into 80\% for training and 20\% for testing \update{in each Meta Quest series} \reham{in each device?}. \reham{do we need the last sentence?} 
%
%It is not strictly necessary to stratify sampling since our dataset is balanced. \reham{We can say "For each device, we randomly split the dataset into balanced subsets, with 80\% for training and 20\% for testing. "}\seonghun{fixed}

%\reham{did we use stratified sampling?}\seonghun{its balanced dataset, which means 20 labels with 100 for each measurement, so it's not strictly necessary to do stratified sampling}. % on our classification tasks. 
% using a selected set of 30 metrics out of the 72 available GPU metrics, as outlined in Section~\ref{sec:leakageVectors}. 
%
%Simultaneously, we monitor the metrics provided by the \texttt{ovrgpuprofiler} tool for 30 seconds. Each AR/VR application is profiled 20 times with selected 30 metrics among 72 total available GPU metrics as described in Section~\ref{sec:leakageVectors}. 

\begin{figure}[t]
  \centering

  \subfloat[Gorilla Tag\label{fig:AppFP_subfig1}]{%
    \includegraphics[width=0.15\textwidth,height=2.3cm]{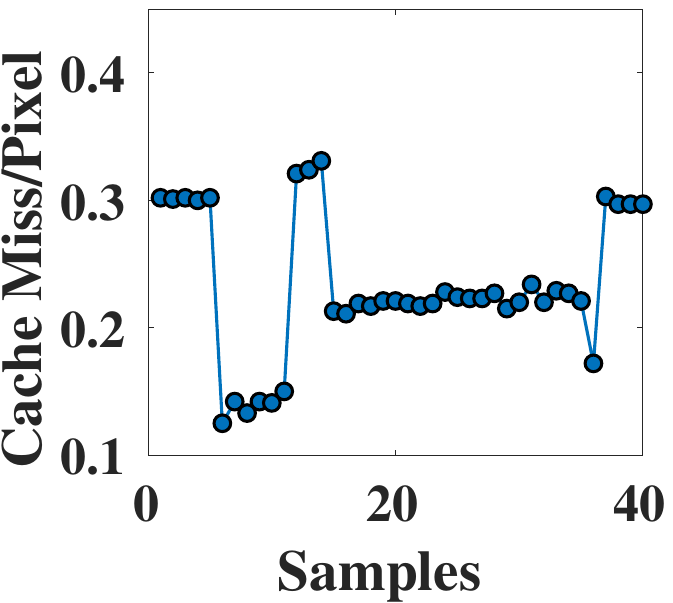}%
  }
  \hfill
  \subfloat[Bigscreen Beta\label{fig:AppFP_subfig2}]{%
    \includegraphics[width=0.15\textwidth,height=2.3cm]{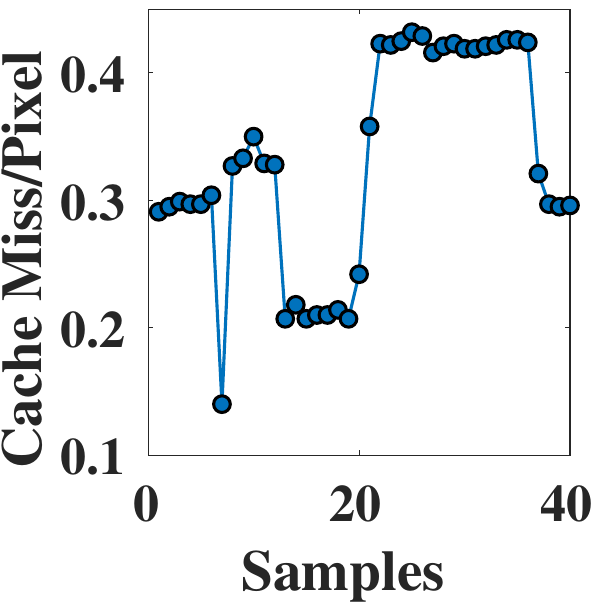}%
  }
  \hfill
  \subfloat[Multiverse\label{fig:AppFP_subfig3}]{%
    \includegraphics[width=0.15\textwidth,height=2.3cm]{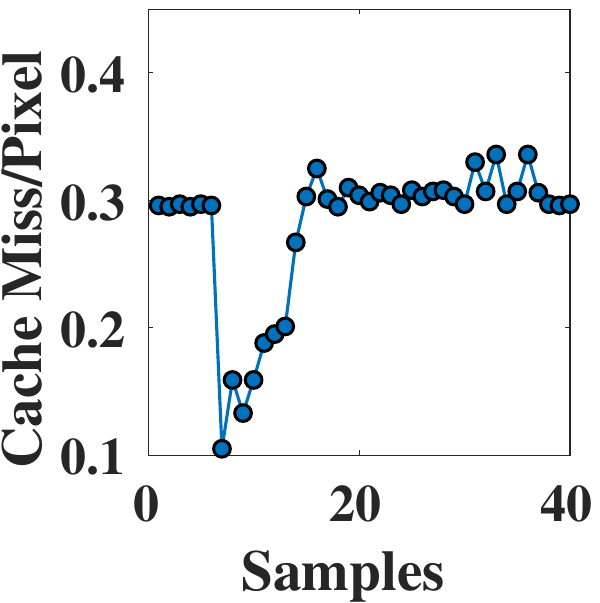}%
  }

  \caption{Example of fingerprints collected from \update{Meta} Quest 2 for AR/VR standalone apps based on the \texttt{L1 Texture Cache Miss Per Pixel} metric: (a) Gorilla Tag, (b) Bigscreen Beta, and (c) Multiverse.}
  \label{fig:AppFPExample1}
\end{figure}

%
\shortsectionBf{Results.} We observe that each AR/VR app exhibits a unique fingerprint due to its immersive graphics and resource-intensive design. 
Such high rendering complexity leads to distinct GPU usage patterns. Examples of these fingerprints are shown in Figure~\ref{fig:AppFPExample1}. In particular, before any application is launched, there is a constant baseline of GPU usage attributed to the system due to the background load, such as AR (passthrough) view or VR (immersive) view. Once an AR/VR app starts, GPU usage rises sharply due to the demands of rendering immersive 3D graphics. During the app's runtime, GPU usage remains elevated to handle real-time interactions and dynamic scene updates.
When the user closes the application, GPU usage returns to its baseline level associated with either the AR view or the VR view. 

\begin{table}[t]
\caption{AR/VR standalone app fingerprinting performance on 100 standalone apps on Quest 2, using a combination of 30 GPU metrics (average scores).
%: average F1, precision, recall, and test accuracy.
%\chandrika{Mention Quest2? if we are mentioning everywhere, otherwise remove everywhere}\seonghun{changed}
}
  \centering
  \setlength{\tabcolsep}{0.3em}   
  \renewcommand{\arraystretch}{1}% adjust row padding
  \begin{tabular}{lcccc}
    \toprule
    \textbf{Model} & \textbf{F1 (\%)} & \textbf{Precision (\%)} & \textbf{Recall (\%)} & \textbf{Accuracy (\%)} \\
    \midrule
    CNN   & 98.3 & 98.5 & 98.4 & 99.3 \\
    LSTM  & 98.7 & 98.9 & 98.9 & 98.6 \\
    RF    & 99.4 & 99.5 & 99.6 & 99.5 \\
    SVM   & 85.1 & 86.9 & 87.6 & 86.8 \\
    \bottomrule
  \end{tabular}
% \caption{AR/VR standalone app fingerprinting performance on 100 standalone apps on Quest 2, using a combination of 30 GPU metrics: average F1, precision, recall, and test accuracy.
% %\chandrika{Mention Quest2? if we are mentioning everywhere, otherwise remove everywhere}\seonghun{changed}
% }
  \label{table:AppFP_acc}
\end{table}

As shown in Table~\ref{table:AppFP_acc}, CNN, LSTM, and RF models achieve accuracies of 99.3\%, 98.6\%, and 99.5\% , respectively, in classifying 100 different AR/VR standalone apps while utilizing a selected combination of 30 metrics.
%
%\reham{edited below, please check}\seonghun{checked and modified}
Specifically, test accuracy shows the overall percentage of correctly classified samples across all class labels. For each class label, we compute the F-1 score, precision, and recall metrics individually, and we report the average values. %The reported F1 score, precision, and recall are averaged metrics for each class individually. 
%
%where test accuracy is the overall proportion of correct prediction across all F-1, precision, and recall. \reham{do you mean "average accuracy across all class labels?" Is the F-1, precision and recall computed per class label? Please check LocIN paper for the definitions } \seonghun{fixed but i don't know why we need this explanations...}
%
%Furthermore, the confusion matrix for the RF model, illustrated in Appendix Figure~\ref{fig:ConfusionMatrix}, shows a high true positive rate and low false positive rates, confirming the classification model's ability to accurately distinguish between different applications based on GPU metrics.
%
%\reham{
Additionally, we 
%selected the top 10 metrics based on the individual classification accuracy on RF model} \reham{we can mention here top 10 based on what?} and 
evaluated the classification performance using each GPU metric individually. Notably, each 10 metrics achieves more than 90\% F-1 score (Appendix Table~\ref{table:AppFP_indiv}) using the RF model, indicating that even a single metric from the \texttt{ovrgpuprofiler} tool is feasible to perform an AR/VR app fingerprinting attack.
%
%\reham{we need to explain what is cross-device setting and individual accuracy}\seonghun{added}

Overall, \system successfully identifies the foreground app with 99.5\% accuracy using 30 metrics on Quest 2 and maintains over 90\% accuracy even with a single metric. 
Furthermore, we achieve similar 95.8\% classification accuracy throughout the cross-device setting (Appendix Table~\ref{tab:cross_series}). This indicates that the pre-trained model is feasible to identify user rendering AR/VR apps across different device setups. % as shown in the \seonghun{Appendix Section~\ref{sec:appendixAppFP}}. %Furthermore, the confusion matrix for the CNN model, illustrated in Appendix Figure~\ref{fig:ConfusionMatrix}, shows a high true positive rate and low false positive rates, confirming the classification model's ability to accurately distinguish between different applications based on GPU metrics.

\subsection{Case Study II: WebXR App Fingerprint}\label{sec:casestudy2}
% \reham{Let's be consistent in the shortsection titles among different case studies}
%
WebXR API enables users to explore and engage with virtual environments directly through the browser of any HMD.
%
% The lack of dependence on separate app installations enables WebXR apps to reach a wider audience.
%
Users can access a WebXR app in a manner similar to accessing a standard website through a browser.
Before entering the 3D mode, the app loads within the 2D browser screen displayed on the HMD.
By default, WebXR enables the rendering of WebGL scenes, allowing websites and browsers to utilize the GPU for rendering. 
%
%\reham{This capability makes our attack feasible for fingerprinting WebXR apps.?}\seonghun{fixed}
%This capability makes our attack feasible for all WebXR apps.
This capability makes our attack feasible for fingerprinting WebXR apps.
%
%
% We adopt an approach similar to fingerprinting standalone AR/VR applications.
%Similar to the fingerprinting of standalone AR/VR apps, we adopt a parallel approach to fingerprinting WebXR apps. \reham{what is parallel approach?}
%
% To launch the attack, \system operates in the background without requiring additional permissions, running any concurrent apps, or installing an additional SDK.
% %
% \system can effectively classify WebXR apps under low resolution of 1 second sampling rate of the \sysprof.

%
\shortsectionBf{Experiment Design.} 
We selected 100 popular and free WebXR apps based on recommendations from social media lists~\cite{awesome_webxr, webxr_metaverse, vrsites, webxr_games} and example showcases from development platforms such as A-Frame~\cite{aframe_examples}, WebXR~\cite{webxr_samples}, as well as free apps hosted on Glitch~\cite{glitch}.
%
% We target the 100 popular free WebXR applications based on social media sources (e.g., Reddit lists) and development platforms (e.g., A-Frame, Three.js, WebXR), as detailed in Appendix Table~\ref{table:webxr_apps}.  
%
We selected applications from five distinct categories: Gaming \& Entertainment~(19\%), Art \& Creativity~(11\%), Tours \& Exploration~(18\%), Health \& Fitness~(8\%), and Demonstration~(44\%) as detailed in Appendix Table~\ref{table:webxr_apps}.
%
% \reham{does this refer to the demonstration apps (44\%) or other apps as well?}
% \seonghun{Not sure about this @Chandrika}
Although most of the applications are showcase examples from standard WebXR sites (Demonstration category (44\%)), they effectively demonstrate the diverse capabilities offered by WebXR.
These capabilities include anchors~\cite{webxr_anchors}, positional audio~\cite{immersive_web_positional_audio}, animation~\cite{aframe_animation_raw}, as well as minimal demonstrations such as shopping~\cite{aframe_shopping_showcase} and reading~\cite{aframe_comicbook_showcase}.
%
% \reham{We can make the links as references to save space}

%

\begin{figure}[t]
  \centering
  \subfloat[VARTISTE\label{fig:WebXRFP_subfig1}]{%
    \includegraphics[width=0.15\textwidth,height=2.3cm]{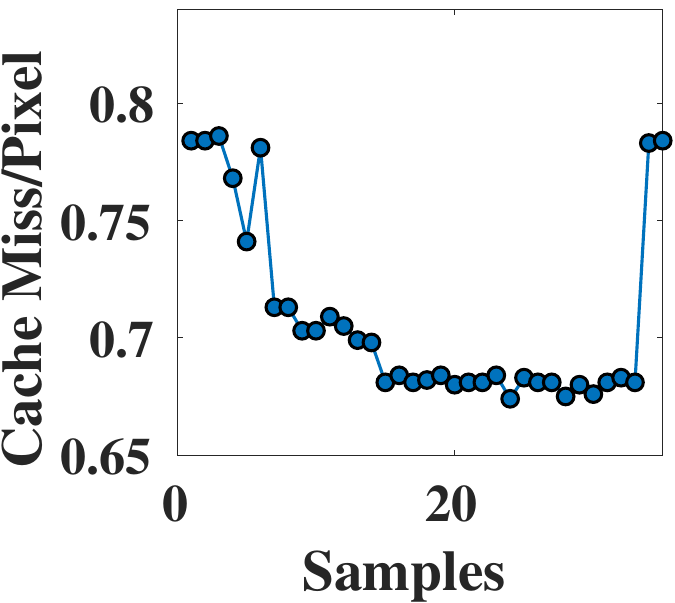}%
  }
  \hfill
  \subfloat[Moon Rider\label{fig:WebXRFP_subfig2}]{%
    \includegraphics[width=0.15\textwidth,height=2.3cm]{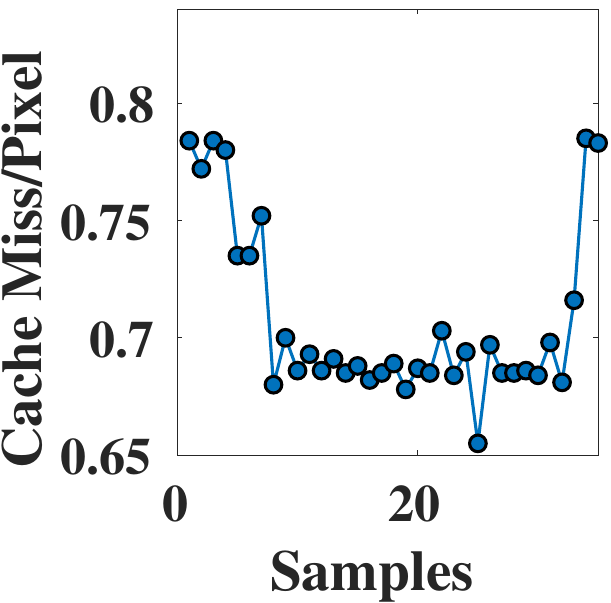}%
  }
  \hfill
  \subfloat[Towermax Fitness\label{fig:WebXRFP_subfig3}]{%
    \includegraphics[width=0.15\textwidth,height=2.3cm]{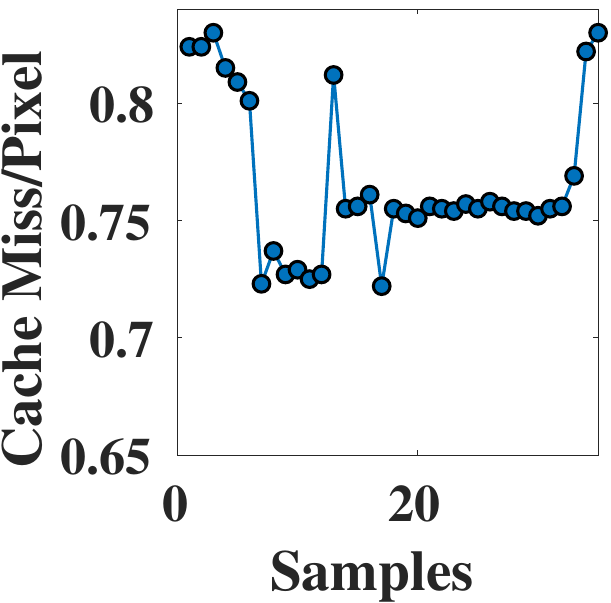}%
  }

  \caption{Example fingerprints for WebXR apps based on the \texttt{L1 Texture Cache Miss Per Pixel} metric. (a) Vartiste, (b) Moon Rider, and (c) Towermax Fitness.}
  \label{fig:webxr_app_fingerprint}
\end{figure}

\shortsectionBf{Data Collection.}
After selecting the target applications, we execute the \system framework.
%malicious script files to run the \system framework. 
% \reham{please make sure we are consistent in using the term capitalized}
% \chandrika{ADB is captitalized in other places.}
%We launch WebXR apps using the ADB shell commands~\cite{meta_horizon_android} and Android intent~\cite{android_intent_reference}.
%
Similar to the standalone AR/VR apps, we monitor the metrics provided by \sysprof for 30 seconds.
We collect 30 metrics out of the 72 available GPU metrics using a similar methodology, as detailed in Section~\ref{sec:leakageVectors}, repeating the process 20 times for each selected WebXR app.
%
% When a WebXR app is launched, the graphics rendering process causes a sharp spike in GPU usage from its baseline level.
% %
% Upon termination of the app, GPU usage returns to the initial baseline level.
% %
% This behavior is observed in both the passthrough view and the fully immersive view.

%\input{Tables/WebXR_Apps/classification_quest3}
%
%\input{Tables/WebXR_Apps/classification_quest2}

\begin{table}[ht]
\caption{WebXR app fingerprinting performance on 100 standalone apps on Quest 2, using a combination of 30 GPU metrics (average scores).}
  \centering
  \setlength{\tabcolsep}{0.3em}   
  \renewcommand{\arraystretch}{1}
  \begin{tabular}{lcccc}
    \toprule
    \textbf{Model} & \textbf{F1 (\%)} & \textbf{Precision (\%)} & \textbf{Recall (\%)} & \textbf{Accuracy (\%)} \\
    \midrule
    CNN   & 97.5 & 98.3 & 97.6 & 97.6 \\
    LSTM  & 96.1 & 96.7 & 96.4 & 96.4 \\
    RF    & 99.2 & 99.4 & 99.0 & 99.0 \\
    SVM   & 72.8 & 74.8 & 74.5 & 74.5 \\
    \bottomrule
  \end{tabular}
  % \caption{WebXR app fingerprinting performance on 100 standalone apps on Quest 2, using a combination of 30 GPU metrics: average F1, precision, recall, and test accuracy.}
  \label{table:webxr_classification_quest2}
\end{table}
%
%\chandrika{write down few sentence about the result from both the devices.}
%\seonghun{We need to say that some of the websites are almost the  same but a  few features are different, for example https://immersive-web.github.io/webxr-samples/input-tracking.html and https://immersive-web.github.io/webxr-samples/room-scale.html are almost identical...}

\shortsectionBf{Results.}
We observe distinct fingerprints for each app based on the selected 30 metrics.
When a WebXR app is launched, GPU usage spikes sharply from its baseline and returns to normal upon termination.
For example, based on \texttt{L1 Texture Cache Miss Per Pixel} metric, Figure~\ref{fig:webxr_app_fingerprint} shows distinct fingerprints for three applications, VARTISTE~\cite{vartiste}, Moon Rider~\cite{moonrider}, and Towermax Fitness~\cite{towermax_fitness},
% \reham{make sure these URLs work in the PDF or cite them}
which represent the categories Gaming \& Entertainment, Art \& Creativity, and Health \& Fitness, respectively.

As shown in Table~\ref{table:webxr_classification_quest2}, the RF and CNN models achieves accuracies of 99.0\% and 97.6\%, respectively, in classifying 100 WebXR applications using 30 metrics, while the LSTM model achieves 96.4\%.
%
% The RF model's confusion matrix (Appendix Figure~\ref{fig:webxr_conf}) shows high true positive rates and low false positives, confirming its effectiveness in distinguishing WebXR apps using GPU metrics.
% the CNN, LSTM and RF models achieved high accuracies of $98\%$, $97.8\%$ and
% $97\%$ respectively in classifying 100 WebXR applications using selected 30 metrics.
%
%\reham{The table in appendix has more than 4, do you mean the top 4 in the table?}\seonghun{fixed}
Moreover, the top four metrics listed in Appendix Table~\ref{table:top_met_webxr}  individually achieve over 72\% accuracy in classifying 100 WebXR apps using the RF model.
%
% Moreover, four metrics from these 30 metrics - \texttt{Texture Fetch Stall}, \texttt{GPU Bus Busy}, \texttt{Texture L2 Miss}, and \texttt{L1 Texture Cache Miss Per Pixel} - individually achieved accuracies exceeding 65\%.
%
Additionally, \system achieves over $94\%$ accuracy in the RF model using only the top two metrics and $98.7\%$ accuracy combining the top four metrics, demonstrating the effectiveness of the attack with a reduced set of metrics.
%
%\reham{We can give an average number of the cross-device accuracy}
Furthermore, across Quest 2, 3 and 3S, the RF model maintains an accuracy of 97.15\% (Appendix 
Table~\ref{tab:cross_series}) demonstrating cross-device robustness.
Unlike the fingerprinting of standalone AR/VR applications rendered in 3D mode, this case study demonstrates that even rendering a WebXR application  within a 2D browser interface is sufficient for \system to achieve high classification accuracy, despite operating at a significantly lower sampling resolution compared to prior GPU profiler-based website fingerprinting attacks~\cite{naghibijouybari2018rendered}. %We assume that we will achieve higher accuracy with a single metric in the common commercial Alex's 100 website list, which will have a distinctive website layout, such as format and contents in each website, across the website rather than the WebXR application. 

\subsection{Case Study III: Virtual Object Detection}\label{sec:casestudy3}
%\chandrika{replace furniture with products.}\seonghun{fixed}

After \system infers the standalone AR/VR or WebXR app, it can further compromise the user's privacy by identifying the rendered objects within the immersive environment.
For example, applications like IKEA Place~\cite{ikeaPlace} enable users to virtually place furniture, allowing them to visualize how different pieces would fit into their existing environment using mobile devices.
Similarly, immersive virtual showrooms developed by Demodern enable users to explore and arrange furniture in simulated environments, providing a realistic and interactive platform for home design~\cite{demodernIkeaVR, presentDigitalIkea}.

Inferring details about a user’s home layout and the types of products they explore within a shopping app can be exploited for targeted advertising, unauthorized surveillance, and more advanced privacy attacks.
%Gaining information about the user's home layout and identifying the types of furniture they try in a shopping app can be exploited for targeted advertising, unauthorized surveillance, and more advanced privacy attacks.
%
%\reham{Why did we add proof-of-concept?}\seonghun{because we don't have user study but removed}
In this case study, we evaluate the ability of the \system framework to accurately detect and identify virtual product placement within both AR and VR scenes.
%
%Specifically, we focus on realistic furniture prefabs designed to resemble IKEA furniture commonly used in immersive environments for customers to visualize and place furniture in their homes or offices virtually.
Specifically, we focus on realistic product prefabs designed to resemble IKEA-style furniture, which are commonly used in XR environments to help customers virtually visualize and place products within their homes or offices.

\shortsectionBf{Experiment Design.}
We evaluate the object detection attack in both AR and VR environments. 
%
%\reham{Instead of saying limited, we can directly give the number of prefabs they provide}\seonghun{removed "limited"}
We selected product prefabs from five free asset packages available on Meta Asset Store~\cite{metaAssetStore}, including Toon Furniture~\cite{toonFurniture}, Free Furniture Set~\cite{freeFurnitureSet}, HDRP Furniture Pack~\cite{hdrpFurniturePack}, Apartment Kit~\cite{apartmentKit}, and Chair and Sofa Set~\cite{chairSofaSet}. 
From these assets, we chose a representative subset of 35 items, spanning large pieces (sofas, beds, desks) to smaller home accessories (lamps, coffee machines, toasters) to ensure coverage of diverse shapes, sizes and usage contexts in a shopping scenario.
\update{
%\textbf{[R7]} 
We also select multiple items from the same furniture category.
%, such as different chair models with similar shapes. 
%
This 3D object selection will demonstrate \system's ability to distinguish between visually analogous 3D objects. }

\begin{figure}[t]
  \centering
  % Row 1
  \subfloat[Default VR Scene\label{fig:furnitureFP_subfig1}]{\includegraphics[width=0.23\textwidth,height=2.2cm]{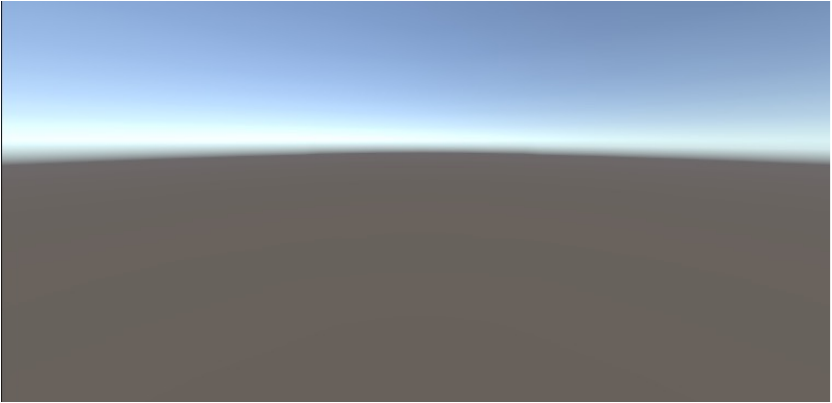}%
  }
  \hfill
  \subfloat[VR Living Room Scene\label{fig:furnitureFP_subfig2}]{\includegraphics[width=0.23\textwidth,height=2.2cm]{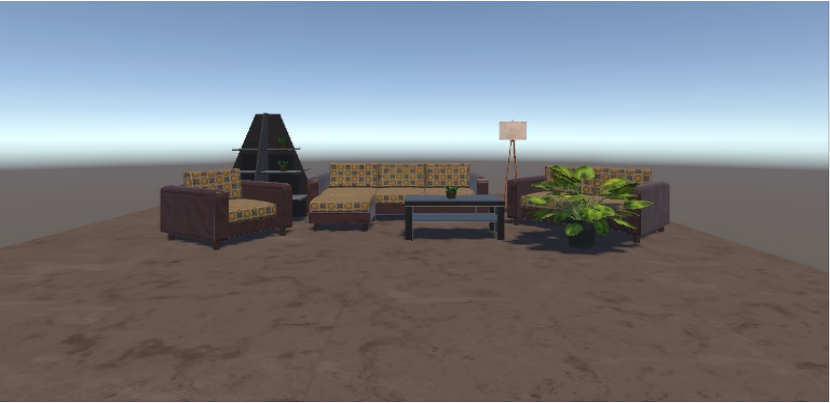}%
  }
  %\vskip\baselineskip

  % Row 2
  \subfloat[AR Office Scene\label{fig:furnitureFP_subfig3}]{%
    \includegraphics[width=0.23\textwidth,height=2.2cm]{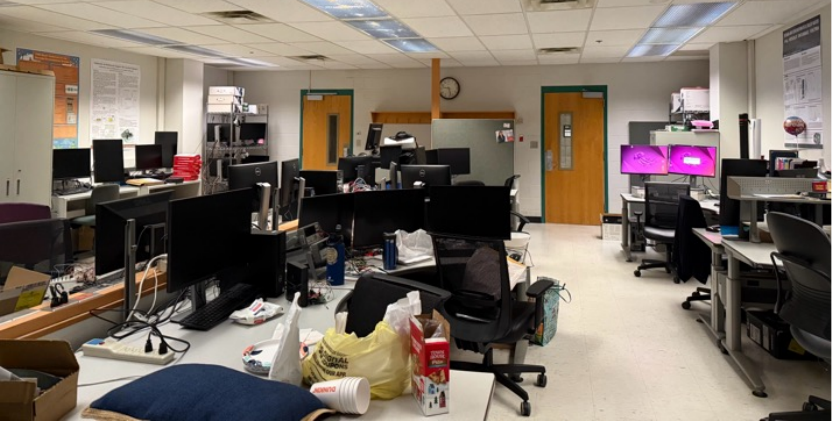}%
  }
  \hfill
  \subfloat[AR Living Room Scene\label{fig:furnitureFP_subfig4}]{%
    \includegraphics[width=0.23\textwidth,height=2.2cm]{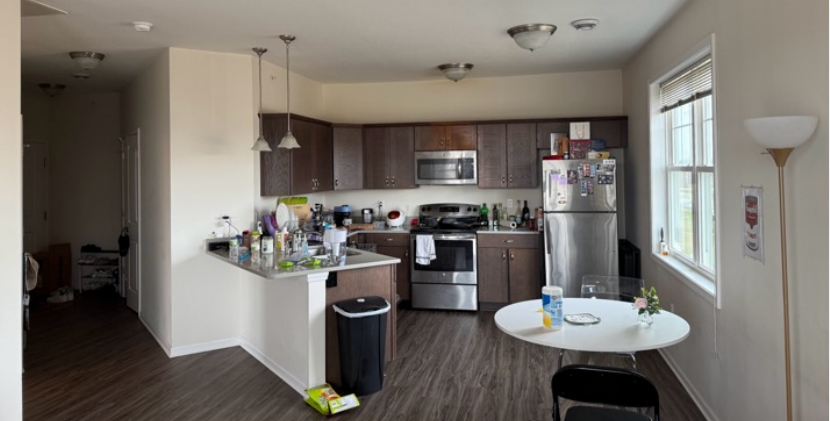}%
  }

  \caption{Experiment scenes for furniture classification. Two VR scene setups (a) without VR objects and (b) with VR objects (living room layout). Two AR scene setups in an (c) office and (d) living room. %\berkay{there is a white space at the top of the images. We should remove those spaces.}
  }
  \label{fig:scenesetup}
\end{figure}

%\reham{can we justify why we selected those 35 if there are others?}\seonghun{We don't have many prefabs from free assets. I changed in a sense of reasoning}
%
%including a wide range of categories\reham{specify the number of categories} \seonghun{since we are not classifying based on the category we remove category parts}
%such as sofas, refrigerators, beds, desks, dressers, chairs, lamps, sinks, and smaller kitchen products like dishes, coffee machines, pots, and toasters, as depicted in Figure~\ref{fig:FurniturePrefabs}. 
%
We design four experimental scenes to simulate real-world usage scenarios: (i) an empty VR environment as a baseline, (ii) a VR living room setup mimicking a typical living room layout with multiple VR objects, (iii) an AR living room where actual objects are placed within the physical living room environment, and (IV) an AR office environment with an actual office space and real furniture such as desks, chairs, and monitors. Figure~\ref{fig:scenesetup} illustrates these four experimental setups.

\shortsectionBf{Data Collection.} %During data collection, we 
We simulate typical user interactions with a virtual product, such as moving items to explore optimal placements or moving the viewpoint (headsets) to better visualize the overall arrangement. In each AR/VR scene, the selected prefab is rendered with speeds of $v = 1,\ 10,\ 50$ $unit/second$ %\reham{please make sure Units/Second is written consistently within the paper}\seonghun{fixed} 
from left to right.  
%\reham{We do not test the static case? }\seonghun{we don't for furnitures..}
Furthermore, we vary the distance values between the user's origin and the virtual product position. 
Specifically, the selected object distance coordinates are set at $z = 0,\ 5,\ 10\ unit$, with the camera positioned at $z = -3\ unit$ relative to the user's default origin.
% Specifically, the selected distance coordinates are $z =0,\ 5,\  10\ unit$, where the camera is located at a coordinate $z=-3\ unit$ relative to the user's default origin. %  \reham{unit? which camera? you can refer to the location of the origin wrt the user } by default. \seonghun{fixed}
%
The variations in distance and movement speed mimic realistic user interaction with virtual objects and exploration of product placement. We collect 20 iterations for each combination of speed and distance values with 35 prefabs.
To automate this process, we developed a Unity code that reads a configuration file, which is saved as a text file in the application directory. This text file specifies the prefab name, position, movement speed, and distance for each iteration.
A bash script then saves this configuration into the app's working folder and invokes \sysprof to start GPU profiling, while the Unity code reads the configuration text file and moves the object.
For each iteration, this unified automation workflow logs GPU metrics to a local file over a total duration of 40 seconds.
Specifically, the first 10 seconds allow the AR/VR scene to fully load and stabilize to ensure consistent tracking and rendering, while the remaining 30 seconds capture the user moving the selected virtual furniture object at one of the predefined speed values to simulate realistic interaction. 
%which contains the scene to stabilize (10 seconds), and the application launches for virtual furniture to move (30 seconds).
%As of result, we come up with total data size with $(Measurement,\ Samples, \ Metrics)$.
%\seonghun{categorizing maybe??}

\begin{table}[ht]
\caption{Fingerprinting performance of virtual products on Quest 2 using a combination of 17 GPU metrics (average scores in \%).
%: average F1 (\%), Precision (\%) and Recall (\%).
}
  \centering
  \scriptsize                              
  \setlength{\tabcolsep}{0.4em}           
  \renewcommand{\arraystretch}{1}        
  \begin{tabular}{l
                  *{3}{c}
                  *{3}{c}
                  *{3}{c}
                  *{3}{c}}
    \toprule
    & \multicolumn{6}{c}{\textbf{VR Scene}} 
    & \multicolumn{6}{c}{\textbf{AR Scene}} \\
    \cmidrule(lr){2-7} \cmidrule(lr){8-13}
    & \multicolumn{3}{c}{Default}
    & \multicolumn{3}{c}{Living room}
    & \multicolumn{3}{c}{Office}
    & \multicolumn{3}{c}{Living room} \\
    \cmidrule(lr){2-4} \cmidrule(lr){5-7}
    \cmidrule(lr){8-10} \cmidrule(lr){11-13}
    \textbf{Model}
    & \textbf{F1} & \textbf{Pre} & \textbf{Rec}
    & \textbf{F1} & \textbf{Pre} & \textbf{Rec}
    & \textbf{F1} & \textbf{Pre} & \textbf{Rec}
    & \textbf{F1} & \textbf{Pre} & \textbf{Rec} \\
    \midrule
    CNN   & 96.5 & 96.6 & 94.3 & 88.3 & 90.6 & 90.2 & 95.1 & 94.6 & 96.7 & 94.4 & 94.6 & 94.5 \\
    LSTM  & 92.6 & 92.3 & 94.3 & 75.8 & 80.1 & 80.9 & 89.8 & 90.9 & 91.3 & 91.4 & 91.1 & 93.0 \\
    RF    & 98.1 & 98.2 & 98.2 & 95.6 & 96.4 & 96.2 & 98.1 & 98.1 & 98.2 & 96.6 & 97.0 & 97.0 \\
    SVM   & 50.6 & 50.7 & 59.6 & 36.7 & 39.6 & 43.3 & 50.6 & 50.7 & 59.6 & 56.0 & 57.6 & 62.6 \\
    \bottomrule
  \end{tabular}
  % \caption{Fingerprinting performance of virtual products on Quest 2 using a combination of 17 GPU metrics: average F1 (\%), Precision (\%) and Recall (\%).}
  \label{table:FurnitureFP_acc}
\end{table}

\shortsectionBf{Results.} We deploy the same ML model types used in Case Study I and II, which are CNN, LSTM, RF with 100 estimators, and SVM, to ensure consistency in our evaluation. % and overall attack workflow shown in Figure~\ref{fig:attackflow}. 
%
%\reham{The accuracies reported below is not accurate, since it is only for the default scene not all of the scenes, we can add a column in the table with the average among all scenes and report this one here.}\seonghun{fixed}
As shown in Table~\ref{table:FurnitureFP_acc}, CNN, LSTM, and RF models achieve the highest accuracies of 96.5\%, 92.6\%, and 98.1\%, respectively, in the default VR scene with 17 selected metrics (Appendix Table~\ref{table:VirtualObject}).
Averaging across all four scenes, these three models maintain robust performance with 93.6\%, 87.4\% and 97.1\%, respectively. 
%across four different scene setups.
%
%\reham{Please refer to the table that summarizes the results before explaining it.}\seonghun{fixed}
%
%From the initial selection, 30 metrics were obtained through the feature selection process (Section~\ref{sec:leakageVectors}).
%
%\reham{Do we have the 17 metric experiment in appendix?}
%\update{We further narrowed the pre-selected 30 metrics down to 17 metrics (\seonghun{Appendix Table~\ref{}}) to demonstrate high classification accuracy on the furniture classification.}
%We further narrowed the pre-selected 30 metrics down to 17 metrics to demonstrate \system's high classification accuracy with a smaller subset of metrics. 
%
Specifically, the selected 17 metrics (out of the 30 collected metrics) individually achieve more than 80\% classification accuracy (CNN) applied in the default VR scene with $v =1\ unit/second$ speed and distance coordinate with $z=0$. 
Notably, \texttt{Non-Base Level Textures} and \texttt{Global Buffer Read L2 Hit} metrics individually achieve 100\% classification accuracy, which demonstrates that a single metric is feasible to classify the VR objects. 
%\seonghun{assuming camera located $x=-3$}

The VR living room scene (Figure~\ref{fig:furnitureFP_subfig2}) %\reham{better name them (a) (b) (c) (d)})\seonghun{fixed}
presents the most challenging environment due to the presence of default VR objects (living room furniture) within the scene. This introduces significant complexity and noise since several overlapping objects
%, varying light conditions \reham{do we test different lighting conditions?}\seonghun{removed}, 
and complex textures make it more challenging to distinguish GPU activities caused by an individual furniture object.
%even with combination of 18 . 
%
%\reham{single metric or 17?}\seonghun{fixed}
In this challenging VR living room scenario, we still maintain 87\% classification accuracy with a single metric, \texttt{Prims Trivially Rejected}. % The results demonstrate the high accuracy of \system even in the challenging scene. 

The GPU metrics show minimal impact due to the scene complexity in AR environments because the objects in the AR are not rendered through GPU components. 
The real-world background is directly taken from the headset's built-in camera feed (passthrough mode) rather than being fully rendered by the GPU. 
Hence, the GPU only handles the virtual objects overlaid on top of the camera feed. 
Consequently, \system achieves nearly 100\% accuracy in both AR scenes (Figure~\ref{fig:furnitureFP_subfig3} and ~\ref{fig:furnitureFP_subfig4}) by leveraging only the \texttt{Non-Base Level Texture} percentage metric that captures the subtle pixel changes as demonstrated in Section~\ref{subsec:CorrelationPixel}. 
%
%Specifically, in AR scenes, rendering focuses primarily on overlaying virtual objects onto the real-world background, which affects GPU metrics. Consequently, specific metrics like \texttt{Non-Base Level Texture} percentage become more indicative of virtual object detection accuracy because they capture the subtle pixel changes as we demonstrated in Section~\ref{subsec:CorrelationPixel}.
%
%\reham{Need a justification why the best metric changes?}\seonghun{fixed}
%Instead, these objects are directly transferred to the AR scene through the built-in camera. Hence, \texttt{OVRSeer} achieves nearly 100\% accuracy in the CNN model with single \texttt{\%Non-Base Level Texture} metric. %This demonstrates that the presence of real-world objects and backgrounds in an AR scene does not significantly change the GPU's metric values relative to virtual object rendering.

When all 17 metrics are combined to create a single fingerprint, the classification models achieve consistently high accuracy across all four experimental scenes. In particular, RF surpasses 95\% accuracy even in the VR living room scene  ($v=1$, $z=0$), %the most challenging %with a combined 17 metrics
while CNN and LSTM each maintain accuracies above 90\% in all four scenes. 
%
% However, SVM lags behind with low accuracy ranging from 37\% to 60\% in four scenes setup, as shown in Table~\ref{table:FurnitureFP_acc}. This performance gap occurs because SVM requires extensive tuning, and it may not fix complex data as effectively as RF or deep neural networks. 
However, SVM shows significantly lower accuracy, ranging from 37\% to 60\% in all scenes, as shown in Table~\ref{table:FurnitureFP_acc}. 
While SVM requires extensive tuning and struggles to handle complex data effectively, the RF model leverages an ensemble of decision trees to handle multidimensional data, and CNN/LSTM models efficiently capture important features in a sequential dataset.

\update{
%\textbf{[R3]} 
Beyond the accuracy-based metric pruning, we apply Pearson correlation-driven selection (Section~\ref{subsec:metric_selection}).
We narrow 5 core metrics (Appendix~\ref{table:PrunedMetrics}) by taking the intersection with our selected 17 metrics in Case Study III (Appendix Table~\ref{table:VirtualObject}) and 11 metrics from correlation-driven metric selection (Appendix Table~\ref{table:pruned_gpu_metrics}). 
We retrain our object detection models on a 5-pruned metric set and achieve 91\%. We gain only a slight drop (4\%) in the RF model in the VR living room scene, while CNN still maintains 90\% accuracy. 
This additional comparison demonstrates that a minimal subset of uncorrelated, high information metrics can obtain robust performance with lower computational overhead. 
%
%\chandrika{similar to the case study I and II, can we add one line for the cross-device setting in this para? - 
%
The cross-device setting across Quest 2, 3, and 3S also demonstrates a high accuracy of 93\%~(Appendix Table~\ref{tab:cross_series}).
%}

%...prune metrics 
%Global Memory Load Instructions
%SP Memory Read (Bytes/Second)
%\% Vertex Fetch Stall
%\% Anisotropic Filtered
%\% Non‑Base Level Textures
}

%Nevertheless, we include SVM alongside RF, CNN, and LSTM for consistency with the overall workflow (Case studies I-III), which those four models are used in earlier case studies  \reham{last sentence not needed}\seonghun{fixed}. 
%Notably, the combined metrics in $1\  Units/Second$ speed with $z=0$  achieve more than 95\% accuracy on the RF model even in the VR living room scene and mostly more than 90\% accuracy with CNN, LSTM and RF model across all four scenes, as shown in Table~\ref{table:FurnitureFP_acc}. 
\begin{comment}

\begin{table}[ht]
  \centering
  \setlength{\tabcolsep}{0.8em}    % adjust column separation
  \renewcommand{\arraystretch}{1.1} % adjust row padding
  \begin{tabular}{lccc}
    \toprule
    \textbf{Speed ($v$)} & \multicolumn{3}{c}{\textbf{Distance ($z$)}} \\
    \cmidrule(lr){2-4}
    & 0 & 5 & 10 \\
    \midrule
    1   & 96.5 & 97.1 & 95.7 \\
    10  & 72.1 & 86.7 & 93.1 \\
    50  & 68.5 & 78.9 & 89.1 \\
    \bottomrule
  \end{tabular}
  \caption{CNN F1 score (\%) across varying speeds ($v$) and distances ($z$) in the default VR scene (Figure~\ref{fig:furnitureFP_subfig1}).}
  \label{table:Furniture_acc}
\end{table}
\end{comment}

Our classification accuracy in the variation of speeds ($v= 1,\ 10,\ 50$) and distance coordinates ($z= 0,\ 5,\ 10$) for the default VR scene reveals that the attacker can detect virtual objects in different scenarios, as shown in the Appendix Table~\ref{table:Furniture_acc}. 
The accuracy drops by roughly 25\% at the extreme speed, $v=50$ with the same distance $z= 0$, but only 4\% drop at the furthest distance $z= 10$. 
%
%\reham{The following sentence is odd here}\seonghun{removed}
%
%Specifically, the speed unit, $units/second$, corresponds to movement relative to the VR scene's coordinate system. 
%
This is because the speed and size of the object are directly related to its distance from the origin. When the object is closer to the user (smaller $z$ values), it occupies a larger portion of the screen, spanning a greater number of units within a specific time frame.
%
%\reham{Please check if this justification is correct and complete it similarly "This is because the speed and size of the object are directly related to its distance from the origin. When the object is closer to the user (smaller $z$ values), it occupies a larger portion of the screen, spanning a greater number of units within a specific time frame."  } \seonghun{make sense and fixed it}
%
%For example, when a VR object moves at $v=50$ and close to the user (smaller $z$ values), it causes rapid changes due to larger on-screen representations and faster traversal across the user's field of view ($v_{screen}$). However, as the distance from the user increases (larger $z$ values), the smaller size of object resulting slower on-screen representations. 
%
This relationship can be expressed as $v_{screen} \propto \frac{s\times v}{z}$, where $s$ is the size of the object. Although high speed changes may reduce distinctive GPU changes due to \system's low resolution, the results demonstrate that the high speed and small size virtual objects are still detectable.

\begin{figure}[t]
  \centering

  \subfloat[
  %\texttt{Non-Base Level Texture}
  \label{fig:AvatarDetection_subfig1}]{%
    \includegraphics[width=0.16\textwidth, height=2.2cm]{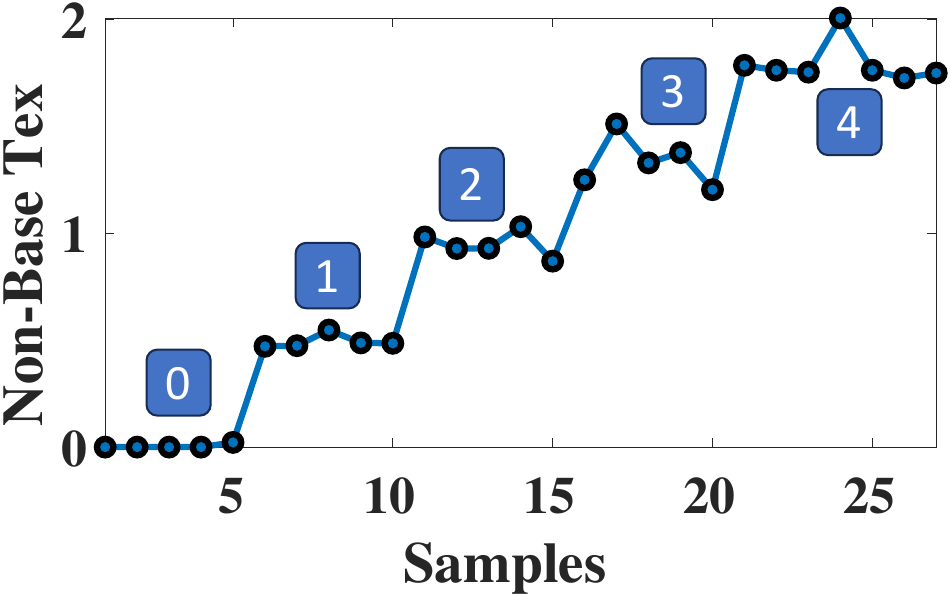}%
  }
  \hfill
  \subfloat[
  %\texttt{Avg Load-Store Instructions/Cycle}
  \label{fig:AvatarDetection_subfig2}]{%
    \includegraphics[width=0.16\textwidth, height=2.4cm]{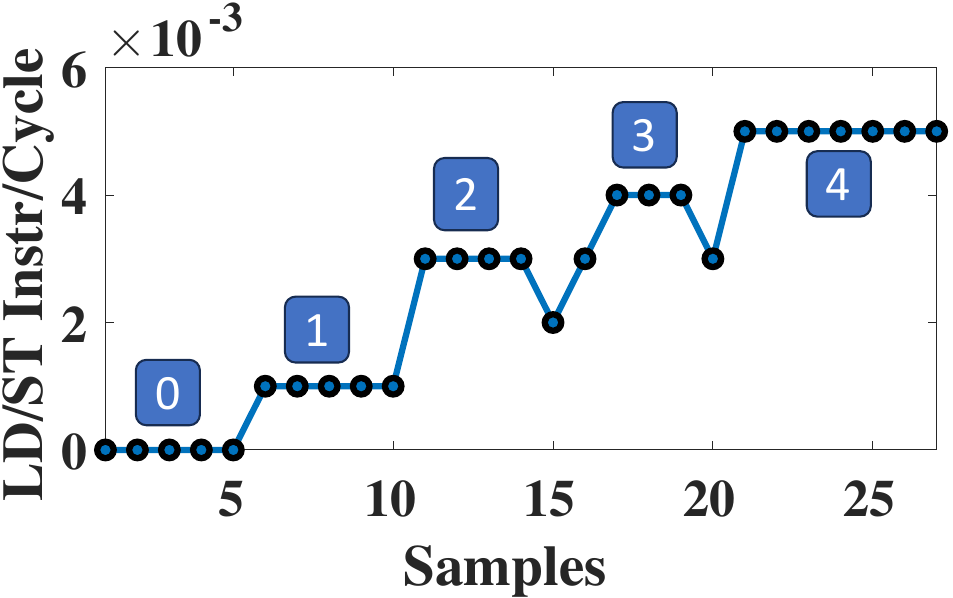}%
  }
  \hfill
  \subfloat[
  %\texttt{Prims Clipped}
  \label{fig:AvatarDetection_subfig3}]{%
    \includegraphics[width=0.16\textwidth, height=2.2cm]{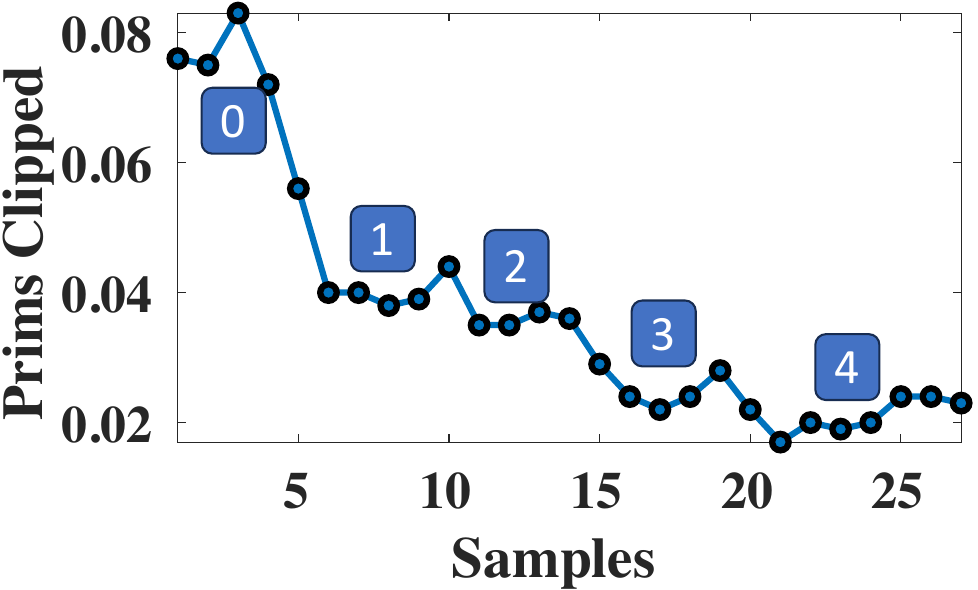}%
  }

  \caption{Example of fingerprint dataset of meeting participant inferences in default VR scene setup (Figure~\ref{fig:furnitureFP_subfig1}). (a) \texttt{Non-Base Level Texture}, (b) \texttt{Avg Load-Store Instructions Per Cycle}, and (c) \texttt{Prims Clipped} metric in AR scene. The number on the graph represents the number of participants joining the target meeting room.}
  \label{fig:AvatarDetection}
\end{figure}

\subsection{Case Study IV: Meeting Room Inference}\label{sec:casestudy4}
%\reham{similarly the introduction needs to be connected with previous case studies}\seonghun{fixed}

%As AR/VR technologies have become an integral component of professional and social interactions, virtual meetings are considered more effective in immersive environments. Unlike traditional video call meetings, applications such as Spatial~\cite{spatial}, AltspaceVR~\cite{altspacevr}, Horizon Workrooms~\cite{horizonworkrooms}, and VRChat~\cite{vrchat} not only offer platforms for remote collaboration, virtual classrooms, and social gatherings but also mimic in-person experiences with a realism. Especially AR meeting applications provide holographic or highly realistic 3D representations of participants along with the actual conference rooms or office environments. These advancements in immersive meeting environments provided by AR/VR devices facilitate remote work, collaborations, and even virtual interviews. However, this realism also brings privacy challenges. For example, during AR/VR meeting environments, sensitive information such as the number of participants or virtual object arrangement can be leaked through GPU usage in AR/VR devices.  

%\berkay{next two paragraphs are too verbose please write it concise.}\seonghun{fixed}
As XR platforms such as Spatial~\cite{spatial}, AltspaceVR~\cite{altspacevr}, Horizon Workrooms~\cite{horizonworkrooms}, and VRChat~\cite{vrchat} become popular venues for hosting social and professional gatherings, concerns over potential leakage of private information related to these meetings become more critical.
The use of realistic AR/VR meeting applications, featuring detailed 3D representations of participants, introduces significant GPU workloads.  These GPU usage patterns can inadvertently expose sensitive information, creating new vectors for privacy leakages.
% As XR platforms such as Spatial~\cite{spatial}, AltspaceVR~\cite{altspacevr}, Horizon Workrooms~\cite{horizonworkrooms}, and VRChat~\cite{vrchat} host increasing numbers of social gatherings. 
%
% However realistic AR meeting application or 3D representation of participants causes high GPU workload and GPU usage patterns creates new privacy risk.}
%
%Building on the insights from the previous case studies on virtual object detection, we now explore another dimension of privacy risks in immersive environments.
%
%As AR/VR technologies have become an integral component of professional and social interactions, virtual meetings on platforms such as Spatial~\cite{spatial}, AltspaceVR~\cite{altspacevr}, Horizon Workrooms~\cite{horizonworkrooms}, and VRChat~\cite{vrchat} increasingly mimic in-person experience with high realism.
%
%In these environments, especially AR meeting applications that provide holographic or realistic 3D representations of participants within actual conference rooms or office settings, sensitive information can be unintentionally leaked through GPU usage patterns.
%

In this case study, we simulate both AR and VR meeting rooms to infer the number of participants by analyzing GPU performance metrics. \system demonstrates the ability to infer the number of meeting participants accurately, revealing meeting attendance without any user interaction.
%\system can accurately infer the number of attendees and revealing meeting attendance and collaboration patterns without any user interaction.}
%
%In this case study, we create both AR and VR meeting scenarios to infer the number of participants by analyzing selected GPU metrics. By counting the number of participants in such scenes, an attacker can obtain meeting attendance information or estimate the level of collaboration within an organization or the number of interviewers participating in private interviews. 

\begin{figure}[t]
  \centering
  \subfloat[Office VR Scene\label{fig:MeetingRoom_subfig2}]{%
    \includegraphics[width=0.23\textwidth,height=2.2cm]{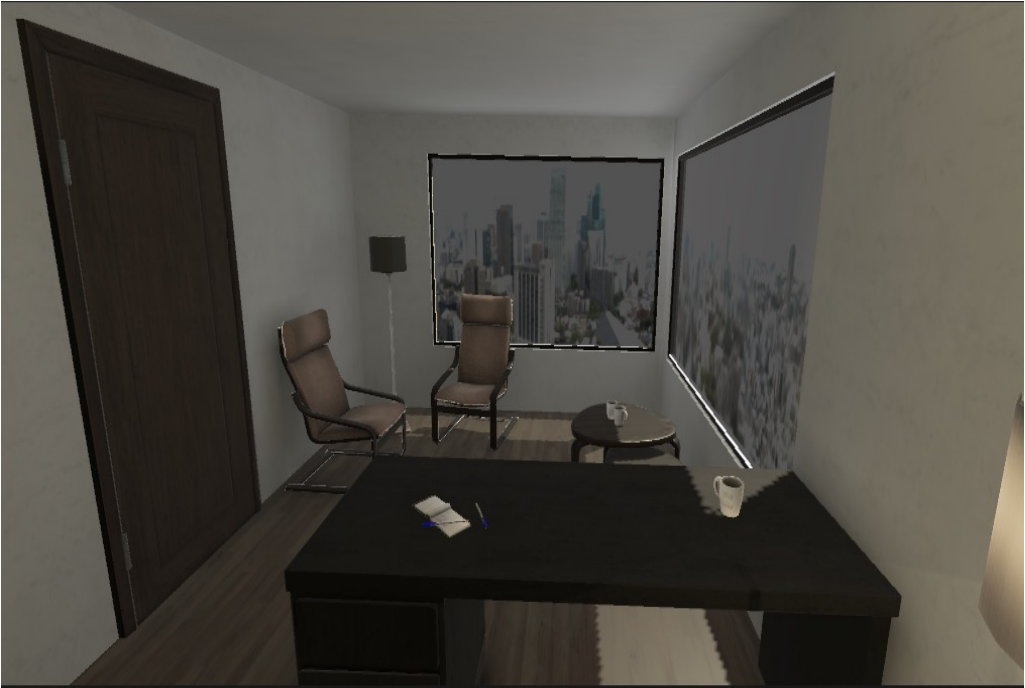}%
  }
  \hfill
  \subfloat[Conference Room AR Scene\label{fig:MeetingRoom_subfig1}]{%
    \includegraphics[width=0.23\textwidth,height=2.2cm]{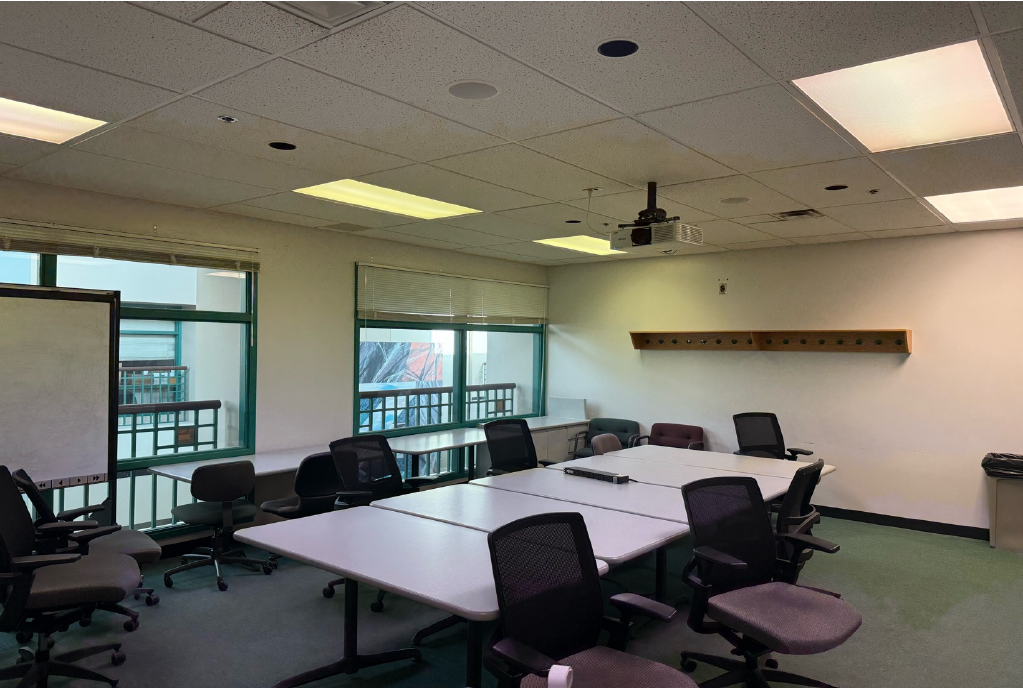}%
  }

  \caption{Experiment scenes for meeting room inference. (a) Office VR scene and (b) conference room scene.}
  \label{fig:meetingroom}
\end{figure}

\shortsectionBf{Experiment Design.} For this case study, we designed three different AR/VR scenes to simulate various meeting environments: (i) the default VR empty environment as shown in Figure~\ref{fig:furnitureFP_subfig1}, 
%a free \reham{what is free?} \seonghun{fixed}
(ii) office room-like VR scene~\cite{officeroom} with limited space size as shown in Figure~\ref{fig:MeetingRoom_subfig2} and (iii) actual conference room as depicted in Figure~\ref{fig:MeetingRoom_subfig1}. 
%\reham{better swap the scenes in the image based on the order you mention them}\seonghun{fixed}
%Note that AR meeting scenarios offer a more realistic and immersive experience because they integrate virtual participants with the actual physical environment, which makes the meeting feel more realistic. 

To simulate different numbers of participants in a virtual meeting, we use standard humanoid avatar~\cite{humanoid} prefabs obtained from the Unity Asset Store. 
These avatars are positioned within each scene to ensure they remain visible from the user's point of view.
In the simulated AR environment (Figure~\ref{fig:MeetingRoom_subfig1}), the relatively larger room accommodates up to four humanoid avatars at a distance coordinate of $z = 3$, all within the user's field of view.
However, the more confined VR scene, resembling a small office room (Figure~\ref{fig:MeetingRoom_subfig2}), can fit only two avatars without overlap, while maintaining visibility from the user’s view. 

To further mimic real-world scenarios, we introduce an additional five humanoid avatars alongside the initial four, resulting in a total of nine avatars within the AR and VR meeting scene. These additional avatars create overlapping situations that commonly occur in practical AR/VR meetings, e.g., when participants move within the environment and some avatars become partially obscured by others.
We selected a total of 9 humanoid avatars to simulate these overlapping situations, ensuring they are positioned to fit entirely within the AR/VR scene without exceeding the spatial constraints.

%To simulate different numbers of participants in a virtual meeting, we use one of the popular humanoid avatar~\cite{humanoid} prefabs obtained from the Unity Asset Store. These avatars are placed within each scene from the user's point of view. In the simulated AR/VR environments, four humanoid avatars occupy the user's point of view with distance coordinate $z=3$. However, in the constrained office room-like VR scene (Figure~\ref{fig:MeetingRoom_subfig2}), only two avatars can be fit into the scene to fill the user's point of view without overlapping other avatars. 
%
%To further mimic a real scenario, we introduced an additional five humanoid avatars alongside the initial four avatars. These extra avatars create overlapping situations that possibly occur in real AR/VR meetings. For example, when participants move in those scenes, some avatars stand behind others, which causes overlapping situations. We selected a total number of 9 humanoid avatars that fit entirely in our AR/VR scenes.  

\shortsectionBf{Data Collection.} We use the same set of 30 metrics selected during the feature selection process (Section~\ref{sec:leakageVectors}).
%
%We maintain this consistent set of metrics to ensure that our experimental workflow is uniform with all case studies, allowing for a reliable \system framework. 
%
During the data collection, we execute \system across different scene setups and render up to 9 humanoid avatars (two avatars in the VR office scene).
%We set the distance coordinate to $z=3$, assuming the user point of view (camera) is at $z=0$.

To simulate participants joining a virtual meeting, avatars are incrementally rendered within the scene. 
Each avatar remains within the user's field of view for 5 seconds before the next avatar appears in the scene. 
In practice, scenes with up to four avatars require a total rendering duration of approximately 30 seconds, accounting for app initialization, avatar transitions, and a final stabilization period. For scenarios involving all nine avatars, the total rendering time extends to 55 seconds to ensure the complete sequence is captured.

We construct datasets for the two scenes, the default VR scene (Figure~\ref{fig:furnitureFP_subfig1}) and the AR meeting scene (Figure~\ref{fig:MeetingRoom_subfig1}). Each dataset is formed of different scenarios, with each scenario representing a specific number of participants in the scene, ranging from 0 to 9. For each scenario, we collect 10 measurements, each with a duration of 70 seconds. 
For the VR office scene (Figure~\ref{fig:MeetingRoom_subfig2}), we construct the scenarios with 0 to 2 participants due to spatial constraints, and similarly, 10 measurements are collected per scenario. 
%We perform a 5-fold cross-validation on our dataset to ensure robust evaluation and avoid the overfitting risk.

%Each avatar stays in the user's view for 5 seconds after the avatar joins the meeting room. Then, another avatar appears in the same scene, increasing the number of participants incrementally. %This ensures distinct changes in GPU metrics are captured as new participants enter the scene. 
%In practice, for scenes with four avatars, the total execution time is approximately 30 seconds, accounting for initialization, avatar transition, and a final stabilization period. For nine avatars, the total execution time is extended to 55 seconds to capture all the avatars. 

\shortsectionBf{Results.} Analysis of the 30 GPU metrics reveals distinct changes based on the number of participants across three distinct scenes. 
In the AR office scene, 22 metrics show noticeable jumps from the baseline values as the number of participants increases, as illustrated in Figure~\ref{fig:AvatarDetection}. 
These jumps indicate a direct correlation between GPU workload and the number of participants. Interestingly, four specific metrics, \texttt{Prims Trivially Rejected}, \texttt{Prims Clipped}, \texttt{Average Vertices / Polygon}, and \texttt{Average Polygon Area}, show decreasing step behavior as the number of participants increases in the scene (Figure~\ref{fig:AvatarDetection_subfig3}). 
 
%\reham{can you explain the term "reject or dismiss a primitive"?}\seonghun{fixed}
Specifically, the \texttt{Prims Trivially Rejected} metric measures the percentage of basic shapes (primitives) the GPU can immediately ignore from the rendering pipeline if they do not contribute to rendering in the immersive scene. 
For example, in the empty scene, many primitives are rejected (ignored), resulting in a high baseline value. As more participants join, more primitives become relevant, so the GPU rejects fewer primitives. This causes the metric value to decrease in noticeable steps. 
%
%\reham{Do we have a reference for this "\texttt{Average Vertices / Polygon} reflects the complexity of shoes" ? }\seonghun{changed}
The \texttt{Prims Clipped} indicates the percentage of primitives cut off by the camera view, \texttt{Average Vertices / Polygon} captures geometric complexity by measuring the average number of vertices per polygon, and the \texttt{Average Polygon Area} measures the screen space each shape covers. Therefore, these metrics exhibit decreasing values if more participants join the scene.
We train the same types of machine learning models to accurately determine the number of participants in VR and AR meeting scenarios and perform a 5-fold cross-validation on our dataset to ensure robust evaluation and avoid the overfitting risk. %We leverage the same four machine learning models from \system.
%to keep the consistency of the attack workflow.
%\reham{We can move this part to the data collection section}
%We construct datasets for the two scenes the default VR scene (Figure~\ref{fig:furnitureFP_subfig1}) and the AR meeting scene (Figure~\ref{fig:MeetingRoom_subfig1}). Each dataset is formed of different scenarios, with each scenario representing a specific number of participants in the scene, ranging between 0 to 9. For each scenario, we collect 10 measurements, each with a duration of 70 seconds. 
%
%For the VR office scene (Figure~\ref{fig:MeetingRoom_subfig2}), we construct the scenarios with 0 to 2 participants due to spatial constraints, and similarly, 10 measurements are collected per scenario. 
%We perform a 5-fold cross-validation on our dataset to ensure robust evaluation and avoid the overfitting risk.
%
%For each metric, we collect 10 measurements corresponding to 10 different participants (0-9) in the default VR scene (Figure~\ref{fig:furnitureFP_subfig1}) and the AR meeting scene (Figure~\ref{fig:MeetingRoom_subfig1}). We also collected 10 measurements for 3 different participant cases (0-2) for the VR office scene (Figure~\ref{fig:MeetingRoom_subfig2}). %Unlike earlier case studies that pad or concatenate multiple metrics, we simply evaluate each metric individually. 
% We perform a 5-fold cross-validation training on our dataset to ensure robust evaluation and avoid the overfitting risk. 
We achieve 100\% accuracy in detecting the number of participants by monitoring 20 metrics individually with the RF model in the default VR scene and 18 metrics in both the VR office scene and AR meeting room setup.
We identified 13 common individual metrics resulting in 100\% accuracy in inferring the number of participants across all three scenes by employing the RF classifier.
%(Appendix Table~\ref{table:Meeting_indiv}). 
%
Furthermore, from cross-device setup, we achieve 100\% accuracy as mentioned in the Appendix Table~\ref{tab:cross_series}.
These overall results demonstrate that \system can perfectly infer the number of participants in both complex VR and the realistic AR meeting setup.

Across all four case studies, \system achieves consistently high accuracy on Meta Quest 2, 3, and 3S by leveraging the GPU metrics.
This demonstrates that our method bypasses Meta Quest's strict concurrent app policy (\Cone), operates effectively with a minimal set (1-3) of metrics with avoiding \update{profiling} overhead and data loss (\Ctwo), and extracts user activity at a low 1Hz sampling rate (\Cthree).
\begin{comment}
This case study demonstrates that \system bypasses the concurrent app restriction (\Cone), individual 13 metrics correlated with participant inference (\Ctwo) across diverse scene setup, and one-second resolution is feasible to infer the number of participants in the dataset (\Cthree).
\end{comment}

%\seonghun{table included Quest2, Quest3, Quest3S}

%\seonghun{----finished addressing comments------}

\begin{comment}

\begin{table}[ht]
\centering
\begin{tabular}{c|c|c|c|c|}
\cline{2-5}
     & \textbf{CS 1} & \textbf{CS 2} & \textbf{CS 3} & \textbf{CS 4} \\ \hline\hline
\multicolumn{1}{|c|}{Quest 2}  & $\checkmark$  &$\checkmark$  & $\checkmark$  & $\checkmark$        \\ \hline
\multicolumn{1}{|c|}{Quest 3} & $\checkmark$  & $\checkmark$ & $\checkmark$  & $\checkmark$        \\ \hline
\multicolumn{1}{|c|}{Quest 3S}   &  $\checkmark$ & $\checkmark$ & $\checkmark$ &$\checkmark$\\ \hline
\end{tabular}
\caption{This is nothing...}
\label{table:-}
\end{table}

\end{comment}

%% file: Sections/5-1_UserStudy.tex
\section{Open World User Interaction}\label{sec:userstudy}

\update{
%\textbf{[R2]} 
AR/VR applications depend heavily on natural user movements and gestures, yet our earlier case studies rely on developer-designed applications and evaluation in a controlled environment.
Furthermore, AR/VR shopping and home-design apps such as IKEA Place~\cite{ikeaPlace} let users preview virtual furniture in their own living spaces, but they are only accessible in selected retail locations~\cite{demodernIkeaVR}.
To demonstrate that \system remains effective when users drive the experience, we evaluate its performance in an open-world AR setting using the pre-built Meta application called Meta Layout~\cite{meta_layout_experience}. 

This app enables users to browse, scale, rotate, and position realistic furniture models directly within their physical environment via AR passthrough, which mirrors the IKEA experience.
We leverage the Layout app with real users as they freely place, hold, and remove virtual objects. We demonstrate that \system can distinguish virtual objects under real-use conditions with the 1Hz, low-resolution GPU profiling tool. 
%The Layout app mirrors the IKEA experience by supporting intuitive drag-and-drop gestures and in-screen panel selection, and it runs on any Quest headset without installation. 

\shortsectionBf{Data Collection.} We obtained Institutional Review Board (IRB) approval and recruited six volunteers from our university. 
Participants are diverse in age, height, weight, and prior AR/VR experience, with more details provided in the Ethical Considerations section.
Prior to the experiment, they received a brief training session on Quest 3S and the Meta Layout app.
%
%
%All data were collected on the most recent Meta Quest 3S headsets. 
%All data were collected on the most recent Meta Quest 3S headsets under the Institutional Review Board (IRB).
%
%\seonghun{need mention participants}
%
We design two separate scenarios for data collection. 

(i) In the static scenario, participants hold a Quest controller button to generate one of the five 3D furniture items (chair, sofa, table, desk, and bed) into the user's view for 5 seconds, then make the 3D object disappear for another 5 seconds by pressing the button on the controller.
This render-remove cycle takes 10 seconds and is repeated 10 times for each furniture item, a total of 100 seconds of GPU profiling traces per participant.
%\reham{does each participant load the 5 items? how do you calculate the 2 minutes?}\seonghun{fixed}
%
%\chandrika{difference between dynamic and static scenario is not clear. You can replace the line with the suggested.}\seonghun{changed!}
%\chandrika{

%
(ii) As users are expected to interact with virtual objects in real-world VR applications, we also designed a dynamic scenario in which the participant opens the in-scene panel in the Layout app, selects a furniture item, and drags it into their field of view for 10 seconds.
%}
%(ii) In the dynamic scenario, the participant opens the in-scene panel in the Layout app to select a 3D object \reham{Are these the same objects as the static scenario? if yes, please use the term "furniture items" to be consistent} and drag-and-drop it into their field of view for 10 seconds. 
%
After 10 seconds, the participant drags the 3D object out of their view. % Each 3D object takes 5 trials and 40 seconds to complete one task. 
%\reham{Please check if the following sentence is correct}\seonghun{correct!}
This scenario takes around 40 seconds and is repeated 5 times for each 3D object. 
Including brief rest breaks and menu navigation, the dynamic scenario requires approximately 25 minutes per participant to complete all 3D object interactions. 

\shortsectionBf{Results.} 
%We parse each iteration into twelve ten-second windows for the static scenario and compute the mean and standard deviation for normalization. 
%
We leverage the 17 selected metrics from Case Study III~\ref{sec:casestudy3} (Appendix Table~\ref{table:VirtualObject}) and employ RF, XGBoost, and SVM models to distinguish the objects. 
During data collection, we noticed that participants sometimes pressed or released the controller button early or late.
Therefore, GPU readings didn't line up exactly with our intended 5-second cycles. 
To handle these misalignments in the dynamic scenario, we first count how many samples each trace actually captures per second, then extract the 10-second window 
%\reham{is this referring to the dynamic scenario only? If yes, please specify this} \seonghun{fixed}
as the object fingerprint. 
For each window, we compute the per-metric mean and standard deviation and concatenate these values into our feature vector.
%
%\seonghun{need fix-}We discard partial or misaligned samples so that every fracture covers a full user interaction period. \seonghun{-}
%

In the static scenario, all the fingerprints from each user are combined to generate the dataset, where 20\% of the dataset is used for the test dataset while the remaining is used for training. The RF model achieves 83\% accuracy while XGBoost reaches 82\%. 
%The 3D Obeject wise F1 score are 0.67 (bed), 0.87 (bookshelf), 0.92 (chair), 0.83 (sofa), and 0.85 (table) 
%
We then train all models and perform hyperparameter tuning via grid search with 5-fold cross-validation. Under this evaluation, the RF model shows the robustness by achieving $86\% \pm 4$ and $88\%\pm6$ accuracy for the XGB model. 

In the dynamic scenario, where users incorporate additional real-time movement, including dragging, dropping, head movement, and continuous cursor movement, the overall classification accuracy slightly declines.
We achieve $81\%$ accuracy with the RF model on the same 80/20 test split.  Applying the same 5-fold cross-validation approach, the RF achieves $77\%\pm5$ as shown in Table~\ref{tab:userstudy_models}. The accuracy drop from the static to dynamic scenario shows the impact of extensive user interaction, which introduces noise in each object's fingerprint.
%
%\reham{We did not mention transformer before, we need to be consistent, either report it with all methods or remove it. Also, we did not report CNN in this section} 
%
%However, due to the lack of participants, we achieve at most $72\%$ accuracy in the leave-one-participant-out (LOPO) cross validation with the Transformer model and $64\%$ with the CNN model.
%
%\reham{Suggestion: 

Additionally, we evaluate the accuracy for unseen users by applying the leave-one-participant-out (LOPO) cross-validation method. We achieve an accuracy of $72\%$ using the transformer model. These results demonstrate the effectiveness of the \system attack in practice, which can be further improved through training on a larger and more diverse user base. 
%}

%Overall, these results demonstrate that with low-resolution (1Hz) profiling GPU metrics and under real-user conditions, \system can distinguish between 3D objects in both controlled and more realistic interaction scenarios. We believe that collecting more participants will improve our model's robustness and yield better accuracy. 
%in the LOPO cross-validation.
%, especially in the LOPO cross-validation.
%RF model with 

}
\begin{table}[t!]
\caption{Accuracy of RF, XGB, and SVM in the static and dynamic user‑study scenarios on Meta Quest 3S.}
  \centering
  \setlength{\tabcolsep}{0.5em}
  \renewcommand{\arraystretch}{1}
  %\def\arraystretch{0.5}
%  \resizebox{\columnwidth}{!}{%
%\resizebox{0.70\columnwidth}{!}{
  \begin{tabular}{llcc}
    \toprule
    \textbf{Scenario} 
      & \textbf{Model} 
      & \textbf{Test (\%)} 
      & \textbf{5‑Fold (\%)} \\
    \midrule
    \makecell[l]{Static Interaction} 
      & RF   & 83 & $86\pm6$ \\
      & XGB  & 82 & $88\pm6$ \\
      & SVM  & 30 & $38\pm4$ \\
    \midrule
    \makecell[l]{Dynamic Interaction} 
      & RF   & 81 & $77\pm5$ \\
      & XGB  & 67 & $66\pm9$ \\
      & SVM  & 78 & $80\pm10$ \\
    \bottomrule
  \end{tabular}%
%}
  % \caption{Accuracy of RF, XGB and SVM in the static and dynamic user‑study scenarios on Meta Quest 3S.}
  \label{tab:userstudy_models}
\end{table}

%% file: Sections/8_Discussion.tex
\section{Countermeasures and Limitations}\label{sec:discussion} 
%In this section, we outline several countermeasures that can be implemented to mitigate the \system attack. Additionally, we discuss the limitations of our proposed attack, providing insights to potential improvement and future research.

\subsection{Countermeasures}
\shortsectionBf{Restrict GPU Profiler Access.}
The \sysprof tool utilizes Performance Interface Library (PIL), a low-level on-device library within the Oculus OS that exposes real-time GPU metrics. 
Because developers depend on this profiler for legitimate performance tuning, we cannot simply remove or disable it without breaking valid workflows.
Instead, PIL calls should be restricted to apps running in developer mode and the app store's vetting process should block unapproved uses of these APIs.
%
%Simply disabling the tool is insufficient to prevent the attacks, whereas the PIL library must be restricted for userspace applications. 
%
This aligns with previous works to prevent access to OS-level sensors such as frequency scaling~\cite{liu2022frequency,dipta2022df}, power consumption~\cite{lipp2021platypus,zhang2021red}, and API accesses to performance counters~\cite{zhang2023s,naghibijouybari2018rendered}.
\shortsectionBf{Dynamic Detection.} 
%Benign applications typically do not continuously profile GPU usage via the \sysprof tool. 
A defense mechanism can monitor system calls using tools such as \textit{perf trace} to detect repetitive access patterns to the GPU profiler. Upon detection, the system can inform the user with a warning on the screen. Since the profiler samples GPU metrics at a low resolution, this detection approach incurs minimal performance overhead.

%Benign applications is not expected to profile the GPU usage constantly through the \texttt{ovrgpuprofiler} tool. 
%
%Hence, a defense mechanism can be designed to profile the system calls of running processes in AR/VR devices. Such a mechanism can be created by incorporating a system call tracer such as \textit{perf trace} tool to monitor the system calls invoked on the device. Once a repetitive access pattern to the GPU profiler tool is detected, the user is informed with a warning on the screen. Since the \texttt{ovrgpuprofiler} tool only allows applications to sample the GPU metrics every second, the defense mechanism would incur minimal performance overhead with a low-resolution detection approach.

%
\shortsectionBf{Noise Injection.} Introducing artificial noise into the GPU components by executing dummy instructions can disrupt fingerprinting attacks.
Such a mechanism can be built by rendering random, non-intrusive objects, similar to noise injection strategies used against website fingerprinting in network traffic~\cite{shan2021real} and browsers~\cite{cook2022there,seonghun2023defweb}. However, noise injection method introduces additional overhead, potentially slowing down object rendering.
Hence, the trade-off between the attack success rate reduction and performance overhead needs to be considered for the noise injection countermeasure.
%
%This can be achieved 
%
%Another potential countermeasure is to introduce artificial noise into the GPU components by executing dummy instructions in the foreground application. 
%
%Such a mechanism can be built around rendering random objects on the screen that have no effect on the user experience. This approach has been implemented against website fingerprinting attacks in the network traffic~\cite{shan2021real} and browsers~\cite{cook2022there,seonghun2023defweb} to degrade attackers' success rate.
%
%On the other hand, noise injection introduces additional overhead to the applications, which could eventually slow down the object rendering process. Hence, the trade-off between the attack success rate reduction and performance overhead needs to be considered for the noise injection countermeasure.
%

\subsection{Limitations}
\shortsectionBf{Applicability to Diverse Devices.}
%Our evaluation has focused on the most popular Meta Quest series, showing how \system can effectively fingerprint both apps and objects within their immersive environments using low-resolution (1Hz) GPU metrics.
%
%In this work, we demonstrate our attack on the popular Meta Quest devices, showing how \system can effectively fingerprint both apps and objects within their immersive environments using low-resolution GPU metrics.
%
%However, 
XR devices from other vendors expose different profiling interfaces. For example, Apple Vision Pro~\cite{apple2025visionpro} supports concurrent app execution and offers extensive GPU counters via Xcode \textit{Instruments} (over 150 metrics)~\cite{apple2025gpuoptimization}, while Microsoft HoloLens~\cite{microsoft2025hololens} provides profiling through \textit{Windows Performance Recorder} (WPR)\cite{microsoft2025wpr} and PIX\cite{microsoft2025pix} for detailed GPU profiling.
%These devices support concurrent app execution and offer metrics through tools like Apple’s Xcode \textit{Instruments}~\cite{apple2025gpuoptimization}, with over 150 GPU counters, and Microsoft’s \textit{Windows Performance Recorder} (WPR)\cite{microsoft2025wpr} and PIX\cite{microsoft2025pix} for detailed GPU profiling.
%
%Although the official documentation does not specify the exact number of available metrics for WPR and PIX, it can provide valuable insight into GPU behavior.

\begin{figure}[t]
  \centering
  \subfloat[Algorithmic Nature\label{fig:HololensAFA_subfig1}]{%
    \includegraphics[width=0.15\textwidth,height=2.2cm]{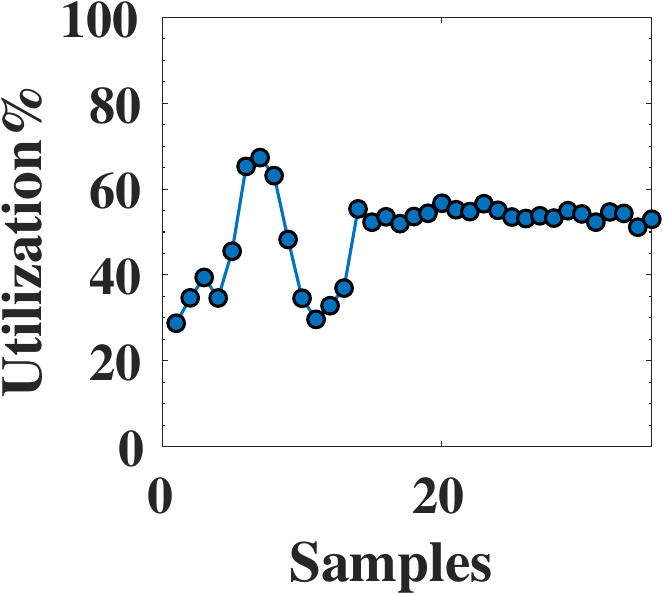}%
  }
  \hfill
  \subfloat[Graffiti 3D\label{fig:HololensAFA_subfig2}]{%
    \includegraphics[width=0.15\textwidth,height=2.2cm]{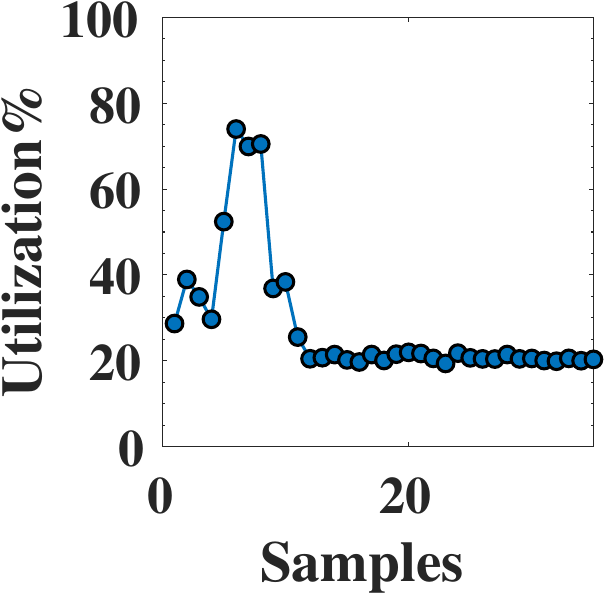}%
  }
  \hfill
  \subfloat[Insight Heart\label{fig:HololensAFA_subfig3}]{%
    \includegraphics[width=0.15\textwidth,height=2.2cm]{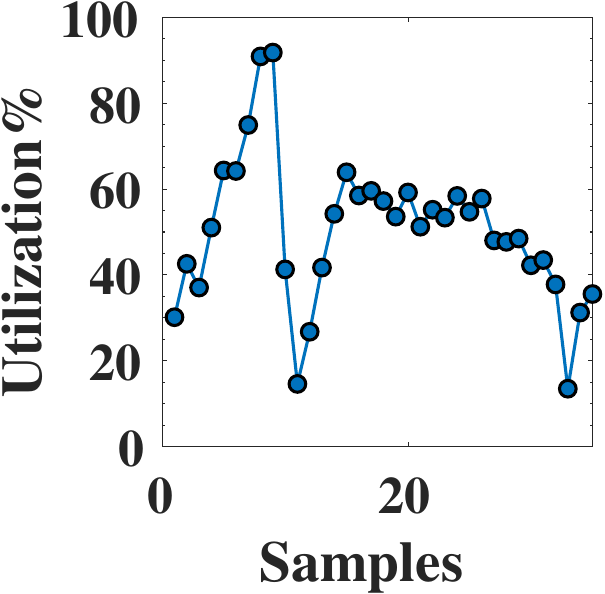}%
  }

  \caption{Example fingerprints for standalone MR apps on  HoloLens 2 based on the \texttt{GPU Utilization} metric. (a) Algorithmic Nature, (b) Graffiti 3D, and (c) Insight Heart.}
  \label{fig:hololens_app_fingerprint}
\end{figure}

To validate \system's broader applicability, we also tested it on Microsoft HoloLens 2 by running publicly available MR apps from the Microsoft Store.
By collecting the device's built-in GPU metrics such as \texttt{Utilization}, \texttt{Dedicated Memory}, \texttt{System Memory}, and \texttt{System Memory Used}, we confirmed that our methodology can capture rendered app behavior as shown in the Figure~\ref{fig:hololens_app_fingerprint}.
%
%The similarity in GPU profiler usage patterns across different platforms (Figure~\ref{}) demonstrate that attckers could exploit built-in GPU profiler as side channels 
%
In future work, we aim to expand our attack framework to encompass various AR/VR devices, including the Apple Vision Pro's Xcode \textit{Instruments} and other vendor-specific tools.
%, further demonstrating its versatility across the evolving AR/VR landscape.

\shortsectionBf{Generalizability.}
Our case studies (Sections~\ref{sec:casestudy1} and \ref{sec:casestudy2}) utilize publicly available standalone AR/VR and WebXR applications, demonstrating their generalizability.
In Sections~\ref{sec:casestudy3} and ~\ref{sec:casestudy4}, we selected a diverse set of objects and varied parameters (speed and distance) to prove the concept of object and avatar identification.
%to address our challenges, providing the proof-of-concept for identifying different objects and avatars.
%
Moreover, our setup focuses on a single-app scenario. However, in real-world scenarios, users may open multiple browsers or switch between apps, which can affect attack accuracy.
Additionally, rapidly moving ($< 1s$) virtual objects within the scenes might influence attack effectiveness.
Future work will extend our evaluation to multi-app and highly dynamic environments.

%% file: Sections/9_Conclusion.tex
\section{Conclusion}\label{sec:conclusion}
\update{
%\textbf{[R1]}
We introduce \system that exploits real-time GPU profiling metrics to extract sensitive user activities from AR/VR users. 
This finding invalidates the common defense of simply lowering profiler resolution.
%
%\berkay{please fix the sentence below. There is something wrong. Throughout does not make sense; you need a comma before \system.}\seonghun{fixed}
Across app detection, website classification, object detection, meeting participant counting, and open-world user interaction with 3D objects, \system achieves high accuracy with a 1Hz sampling rate.  
Our findings underscore the need for more restrictive permission models in AR/VR operating systems to safeguard user privacy against sophisticated side-channel attacks.
%
% Our findings highlight the need for more restrictive permission models in AR/VR OSes to protect user privacy against sophisticated side-channel attacks.
}

%% file: Sections/10_Acknowledgment.tex
\section*{Acknowledgment}\label{sec:acknowlegdment}
This work was partially supported by the National Science Foundation (NSF) under grants CNS-2144645 and IIS-2229876. Grant 2229876 was also supported in part by funds from the Department of Homeland Security and IBM. Any opinions, findings, and conclusions or recommendations expressed in this material are those of the author(s) and do not necessarily reflect the views of the NSF or its federal and industry partners. 

%We thank our shepherd and the anonymous reviewers for their valuable feedback and suggestions.
%
%This work was partially supported by the National Science Foundation (NSF) under grants CNS-2144645 and IIS-2229876. 
%
%The views expressed are those of the authors and do not necessarily represent those of the NSF.
% We appreciate the valuable feedback and suggestions from
% our shepherd and anonymous reviewers.
% %
% This research was
% partially funded by the National Science Foundation (NSF)
% through grants CNS-2144645 and IIS-2229876. The findings,
% conclusions, and recommendations presented in this paper are
% solely those of the authors and do not necessarily represent
% the views of the NSF.
%
% This research was partially funded by the National Science Foundation (NSF) through grants CNS-2144645 and IIS-2229876.

%% file: Sections/EthicalConsiderations.tex
\section*{Ethical Considerations}\label{sec:ethical}
\update{
%[R5]
%
We introduce \system, a GPU-based side-channel attack on AR/VR headsets that abuses vendor-provided GPU metrics and exposes sensitive user activities, undermining both user privacy and service-provider trust.
To address these ethical concerns, we responsibly disclosed our findings to Meta via their bug bounty program~\cite{meta_bug_bounty} on January 15, 2025.

In March 2025, Meta confirmed that user-space applications can indeed access basic GPU performance counters and awarded us a Bug Bounty for identifying this side-channel vulnerability.
We are waiting for their response on the upcoming firmware update.

As of June 4, 2025. Meta has tightened developer access requirements for native Android AR/VR development~\cite{meta_android_setup}. Developer mode now requires users to join or create a verified developer organization via the Meta Horizon Developer Dashboard. 
While these changes may not directly prevent misuse of performance metrics, they raise the barrier for arbitrary access by requiring stronger identity validation for developer mode access.

We conducted a user study to validate the attack's effectiveness in a real-world scenario, recruiting six participants (age 18+) from our university through an IRB-approved process.
To ensure participants' well-being during Mixed Reality interaction, interested individuals were asked to complete an online screening survey assessing medical conditions such as vision or hearing impairments, seizures, motion sickness, and neurological disorders.
Only individuals without any of the listed conditions were invited to the in-lab study.
Throughout the study, participants were monitored for safety and informed of their right to withdraw at any time.
To protect participant privacy, no personally identifiable information (PII) was collected during the in-lab sessions.
}

%% file: Sections/Appendix.tex
%\appendix
%\section{Appendix}
%\label{sec:app_guide}
\begin{comment}

\section{Background Thresholding}\label{sec:appendixBGThresholding} 
\begin{figure}[h]
    \centering
    {
    
    \includegraphics[width=\linewidth]{Source/Sections/Figures/LeakageVectors/BGThresholding_v2.png}
    }
    \caption{Background thresholding technique used to isolate the target 3D object from the background. (a) A basic cube object rendered in the original scene, and (b) a designated layer used to isolate the cube object from the background.}
    \label{fig:BlackBG}
\end{figure}

%\newpage
\end{comment}

\section{GPU Metrics Selection by Pearson Correlation}\label{sec:appendixMetricSelection}
%\chandrika{instead of writing the method in table caption, 1-2 sentences can be added under the heading.}

\begin{table}[H]
    \caption{Pruned 11 GPU metrics and reasons after pair-wise Pearson correlation analysis. If the threshold is redundant, drop the second metric in each pair. Metrics that never appear in a high‑correlation pair are kept (Uncorrelated).}
  \scriptsize
  \centering
  \setlength{\tabcolsep}{6pt}
  \renewcommand{\arraystretch}{0.5}
  \begin{tabular}{@{}l p{0.50\linewidth}@{}}
    \toprule
    \textbf{Metric Name} & \textbf{Justification} \\
    \midrule
    GPU Bus Busy
      & Prims Clipped (r = -0.995) \\ Vertex Fetch Stall
      & \makecell[l]{Texture Fetch Stall (r = 0.911)\\
                    Texture L2 Miss (r = 0.907)\\
                    Stalled on System Memory (r = 0.913)\\
                    Prims Trivially Rejected (r = 0.908)\\
                    Nearest Filtered (r = 0.906)\\
                    Avg Bytes / Fragment (r = 0.907)\\
                    Global Image Uncompressed Data Read\\  (r = 0.918)} \\

    Anisotropic Filtered
      & Uncorrelated \\

    Non-Base Level Textures
      & Uncorrelated \\

    SP Memory Read (Bytes/Second)
      & \makecell[l]{Global Buffer Read L2 Hit (r = -0.932)\\
                    Bytes Data Actually Written (r = -0.909)\\
                    Global Buffer Data Read BW (r = 1.000)} \\

    Preemptions / second
      & Global Buffer Data Read Request BW (r = 0.919) \\

    Avg Preemption Delay
      & Uncorrelated \\

    Global Memory Load Instructions
      & Uncorrelated \\

    Local Memory Store Instructions
      & Uncorrelated \\

    Avg Load-Store Instructions Per Cycle
      & Uncorrelated \\

    Bytes Data Write Requested
      & Uncorrelated \\

    \bottomrule
  \end{tabular}
  % \caption{Pruned 11 GPU metrics and reasons after pair-wise Pearson correlation analysis. If the threshold is redundant, drop the second metric in each pair. Metrics that never appear in a high‑correlation pair are kept (Uncorrelated).}
  \label{table:pruned_gpu_metrics}
\end{table}

\section{GPU Metrics in Relation to Speed and Depth}\label{sec:appendixCorrelation}
%We further evaluate other GPU metrics in relation to speed and depth. As depicted in Figure~\ref{fig:SpeedDistacne_l2miss}, these observations are the foundation for practical case studies.
\begin{figure}[H]
  \centering

  \subfloat[$v=1, z=2$\label{fig:SpeedDistacnesubfig1_l2miss}]{
    \includegraphics[width=0.145\textwidth,height=2.1cm]{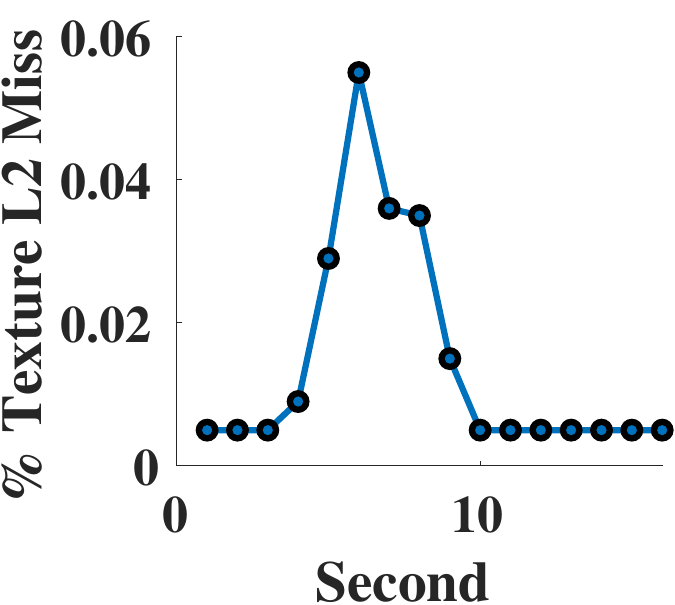}
  }
  \hfill
  \subfloat[$v=2, z=2$\label{fig:SpeedDistacnesubfig2_l2miss}]{
    \includegraphics[width=0.145\textwidth,height=2.1cm]{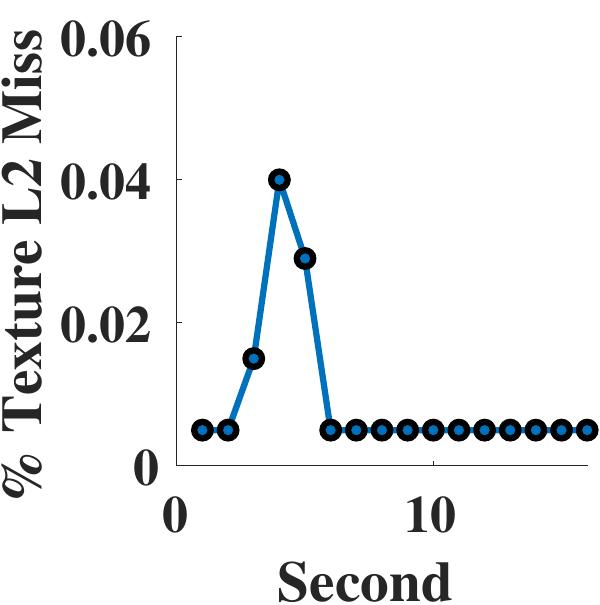}
  }
  \hfill
  \subfloat[$v=1, z=3$\label{fig:SpeedDistacnesubfig3_l2miss}]{
    \includegraphics[width=0.145\textwidth,height=2.1cm]{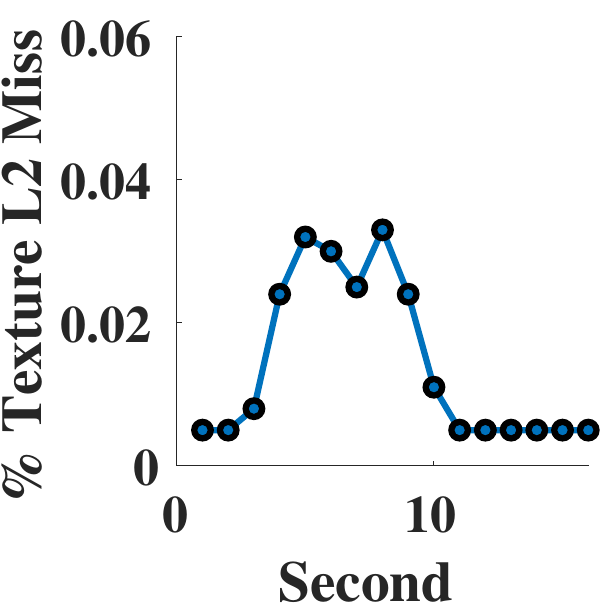}
  }

  \caption{ Correlation between \texttt{Texture L2 Miss} metric and VR Cube object rendering with (a) the speed, $v$ = 1 $unit/second$ and distance coordinate $z=2\ unit$, (b) increasing speed to $v$ = 2 $unit/second$, and (c) move object further from the point of view, $z=3\ unit$.}
  \label{fig:SpeedDistacne_l2miss}
\end{figure}

%\newpage

\section{AR/VR Standalone App Fingerprint}\label{sec:appendixAppFP}

\begin{table}[h]
\caption{Accuracy and F1 scores for the top-performing metrics employing a Random Forest classifier on 100 standalone AR/VR apps.}
\scriptsize
\centering
\begin{tabular}{|l|c|c|}
\hline
\textbf{Metric Name} & \textbf{Accuracy} & \textbf{F-1 Score} \\ \hline\hline
Average Vertices / Polygon  & 0.945 & 0.935              \\ \hline
Vertex Memory Read (Bytes/Second) & 0.940  & 0.931              \\ \hline
Prims Trivially Rejected  & 0.935  & 0.921              \\ \hline
Average Polygon Area & 0.920  & 0.905              \\ \hline
Vertex Instructions / Second & 0.920 & 0.912              \\ \hline
L1 Texture Cache Miss Per Pixel  & 0.915    & 0.900              \\ \hline
Prims Clipped  & 0.910  & 0.885              \\ \hline
Global Image Uncompressed Data Read BW & 0.902             & 0.868              \\ \hline
Texture L2 Miss                                 & 0.900             & 0.881              \\ \hline
Pre-clipped Polygons/Second                        & 0.897             & 0.890              \\ \hline
Avg Bytes / Fragment                               & 0.895             & 0.860              \\ \hline
GPU Bus Busy: Accuracy                          & 0.890             & 0.860              \\ \hline
\end{tabular}
% \caption{Accuracy and F1 scores for the top-performing metrics employing a Random Forest classifier on 100 standalone AR/VR apps.}
\label{table:AppFP_indiv}
\end{table}

\newpage
\section{WebXR App Fingerprint}\label{sec:appendixWebXRFP}
%To determine how well a single GPU metric can distinguish among 100 standalone WebXR applications on Meta Quest 2, we evaluate accuracy with a single metric and report on Table~\ref{table:top_met_webxr}.
\begin{table}[h]
\caption{Accuracy and F1 scores for the top-performing metrics employing a Random Forest on 100 WebXR apps.}
\scriptsize
\centering
\begin{tabular}{|l|l|l|}
\hline
\textbf{Metric Name}            & \textbf{Accuracy} & \textbf{F-1 Score} \\ \hline\hline
Average Polygon Area            & 75.7              & 75.2               \\ \hline
Average Vertices / Polygon      & 73.25             & 72.9               \\ \hline
\% Texture L2 Miss               & 72.75             & 72.21              \\ \hline
\% Texture Fetch Stall           & 72.0              & 71.1               \\ \hline
\% Prims Clipped                 & 68.5              & 67.8               \\ \hline
L1 Texture Cache Miss Per Pixel & 68                & 68.1               \\ \hline
GPU \% Bus Busy                  & 65.7              & 65.6               \\ \hline
\end{tabular}

% \caption{Accuracy and F1 scores for the top-performing metrics employing a Random Forest on 100 WebXR apps.}
\label{table:top_met_webxr}
\end{table}
%

%\newpage
\section{Virtual Object Detection}\label{sec:appendixFurnitureDetection}

\begin{table}[h]
\caption{CNN F1 score (\%) across varying speeds ($v$) and distances ($z$) in the default VR scene (Figure~\ref{fig:furnitureFP_subfig1}).}
  \centering
  \setlength{\tabcolsep}{0.8em}    % adjust column separation
  \renewcommand{\arraystretch}{0.5} % adjust row padding
  \begin{tabular}{lccc}
    \toprule
    \textbf{Speed ($v$)} & \multicolumn{3}{c}{\textbf{Distance ($z$)}} \\
    \cmidrule(lr){2-4}
    & 0 & 5 & 10 \\
    \midrule
    1   & 96.5 & 97.1 & 95.7 \\
    10  & 72.1 & 86.7 & 93.1 \\
    50  & 68.5 & 78.9 & 89.1 \\
    \bottomrule
  \end{tabular}
  % \caption{CNN F1 score (\%) across varying speeds ($v$) and distances ($z$) in the default VR scene (Figure~\ref{fig:furnitureFP_subfig1}).}
  \label{table:Furniture_acc}
\end{table}

\begin{table}[h]
\caption{Accuracy and F1 scores for the top‐performing 17 metrics employing a Random Forest classifier on 35 virtual products.}
\scriptsize
\centering
\begin{tabular}{|l|c|c|}
\hline
\textbf{Metric Name} & \textbf{Accuracy} & \textbf{F-1 Score} \\ \hline\hline
Avg Bytes / Fragment                              & 0.993 & 0.992 \\ \hline
Global Image Uncompressed Data Read BW & 0.979 & 0.978 \\ \hline
Global Memory Load Instructions                    & 0.964 & 0.961 \\ \hline
\% Non-Base Level Textures                         & 0.964 & 0.968 \\ \hline
\% Texture L2 Miss                                 & 0.957 & 0.960 \\ \hline
Global Buffer Data Read Request BW     & 0.943 & 0.941 \\ \hline
SP Memory Read                       & 0.893 & 0.866 \\ \hline
Global Buffer Data Read BW             & 0.893 & 0.866 \\ \hline
\% Prims Clipped                                   & 0.864 & 0.806 \\ \hline
Reused Vertices / Second                           & 0.864 & 0.855 \\ \hline
\% Prims Trivially Rejected                        & 0.836 & 0.779 \\ \hline
Pre-clipped Polygons/Second                        & 0.779 & 0.744 \\ \hline
Vertex Memory Read                   & 0.750 & 0.743 \\ \hline
\% Vertex Fetch Stall                              & 0.736 & 0.686 \\ \hline
Vertices Shaded / Second                           & 0.736 & 0.695 \\ \hline
Vertex Instructions / Second                       & 0.714 & 0.651 \\ \hline
\% Anisotropic Filtered                            & 0.621 & 0.649 \\ \hline
\end{tabular}
% \caption{Accuracy and F1 scores for the top‐performing 17 metrics employing a Random Forest classifier on 35 virtual products.}
\label{table:VirtualObject}
\end{table}

\begin{table}[h]
\caption{Core GPU metrics retained after correlation‑driven pruning in Case Study III~\ref{sec:casestudy3}}
\scriptsize
\centering
\begin{tabular}{|l|}
\hline
\textbf{Metric Name} \\ \hline\hline
\% Vertex Fetch Stall \\ \hline
SP Memory Read (Bytes/Second) \\ \hline
Global Memory Load Instructions \\ \hline
\% Anisotropic Filtered \\ \hline
\% Non‑Base Level Textures \\ \hline
\end{tabular}
% \caption{Core GPU metrics retained after correlation‑driven pruning in Case Study III~\ref{sec:casestudy3}}
\label{table:PrunedMetrics}
\end{table}

\section{Per-Device and Cross-Device Evaluation }\label{sec:appendixCrossDevice}
%\berkay{why do you not have these experiment results in the evaluation section, but in the Appendix? We have space. Let's discuss this on Slack.}\seonghun{it would be too complicated if we put it in the evaluation section }

To verify that \system generalizes across the entire Meta Quest family, we repeated all four case studies (AR/VR standalone (I) and WebXR (II) app fingerprint, virtual object detection (III), and meeting room participant inference (IV)) on Quest 2, Quest 3, and the most recent Quest 3S headsets.
For each device, we collected the same GPU metrics under identical scene setups, focusing on the top 3-4 metrics identified based on our prior analysis.

First, we performed per-device evaluation by splitting each device's dataset into 80\% for training and 20\% for testing. Then we trained four classifiers (CNN, LSTM, RF, and SVM) independently on each device (Table~\ref{tab:per_device_accuracy}). Because Quest 3 and Quest 3S share the same GPU specification (Section~\ref{sec:HW_sec}), we report standalone (I) app fingerprinting only on Quest 2 and Quest 3.
Next, to account for hardware variability even within the same model, we conducted a two-device test on Quest 2 with Case Study 1 to demonstrate effectiveness on app fingerprinting attacks. We trained our models on data from one Quest 2 headset and saved the pretrained weights. Then we evaluated 15 standalone apps without retraining the ML models (Table~\ref{tab:quest2_app_fp}).
Finally, we merged 80\% of the training and 20\% of the testing data from all three headsets to train a single cross-series model (Table~\ref{tab:cross_series}).

Our results show that \system remains highly effective across hardware generations and headset variations, using as few as 3–4 GPU metrics.

\begin{table}[h]
    \caption{Per‑device accuracies for each model across Quest 2, Quest 3, and Quest 3S.
  % \chandrika{Should we write a reason why we don't do experiements in Quest3S - case study1?}
  }
  \centering
    \setlength{\tabcolsep}{0.3em}
    \def\arraystretch{1}
     % \resizebox{\columnwidth}{!}{%
  \begin{tabular}{llccc}
    \toprule
    \textbf{\makecell{Case Study \\(\# of GPU Metrics)}} & \textbf{Model} 
      & \multicolumn{3}{c}{\textbf{Accuracy (\%)}} \\
    \cmidrule(lr){3-5}
      & & Quest 2 & Quest 3 & Quest 3S\\
    \midrule
    \makecell[l]{I. AR/VR Standalone App (4)}
      & LSTM & 95.70    & 89.25    & —    \\
      & CNN  & 97.31    & 87.90    & —    \\
      & \textbf{RF}   & \textbf{98.39}    & \textbf{95.97}    & —    \\
      & SVM  & 94.89    & 80.65    & —    \\
    \midrule
    II. WebXR App  (3)
      & LSTM & 83.74 & 83.25 & 83.50 \\
      & CNN  & 87.62 & 87.75 & 91.00 \\
      & \textbf{RF}   & \textbf{92.72} & \textbf{96.25} & \textbf{93.50} \\
      & SVM  & 40.29 & 38.00 & 36.00 \\
    \midrule
    III. Virtual Object Detection (3)
      & LSTM & 90.71 & 60.00 & 72.14 \\
      & CNN  & 94.29 &  68.57 & 95.71 \\
      & \textbf{RF}   & \textbf{99.29} & \textbf{92.86} & \textbf{100.00} \\
      & SVM  & 75.71 & 19.29 & 28.57 \\
    \midrule
    IV. Meeting Room Inference (3)
      & LSTM & 80.00 & 80.00 & 70.00 \\
      & \textbf{CNN}  &\textbf{100.00} &\textbf{100.00} &\textbf{100.00} \\
      & \textbf{RF}   &\textbf{100.00} &\textbf{100.00} &\textbf{100.00} \\
      & SVM  &100.00 & 95.00 & 90.00 \\
    \bottomrule
  \end{tabular}%
  %}
  % \caption{Per‑device accuracies for each model across Quest 2, Quest 3, and Quest 3S.
  % % \chandrika{Should we write a reason why we don't do experiements in Quest3S - case study1?}
  % }
  \label{tab:per_device_accuracy}
\end{table}

\begin{table}[H]
    \caption{Two‑device cross‑evaluation on Quest 2 for AR/VR standalone app fingerprinting (Case Study I).  
    %\berkay{why do you use def arraystretch{0.5} Please use {1} in all tables.? It makes the spacing between rows too compact.}\seonghun{fixed}
    }
  \centering
    \setlength{\tabcolsep}{0.4em}
    \def\arraystretch{1}
      \resizebox{\columnwidth}{!}{%
  \begin{tabular}{llcccc}
    \toprule
    \textbf{\makecell{Case Study\\ (\# of GPU Metrics)}} & \textbf{Model} 
      & \multicolumn{4}{c}{\textbf{Quest 2}} \\
    \cmidrule(lr){3-6}
      & & \textbf{Accuracy (\%)} & \textbf{Precision} & \textbf{Recall} & \textbf{F1} \\
    \midrule
     \makecell[l]{I. AR/VR Standalone (4)}               
      & LSTM & 70.67 & 0.596 & 0.707 & 0.621 \\
      & \textbf{CNN}  & \textbf{80.00} & \textbf{0.693} & \textbf{0.800} & \textbf{0.730} \\
      & RF   & 37.33 & 0.306 & 0.373 & 0.310 \\
      & SVM  & 53.33 & 0.435 & 0.533 & 0.461 \\
    \bottomrule
  \end{tabular}
  }
  % \caption{Two‑device cross‑evaluation on Quest 2 for AR/VR standalone app fingerprinting (Case Study I). }
  \label{tab:quest2_app_fp}
\end{table}

%\FloatBarrier
\begin{table}[H]
\caption{Cross‑device evaluation of \system: final test results (Accuracy, Precision, Recall, F1) for four classifiers (CNN, LSTM, RF, SVM) across the four case studies on the entire Meta Quest series.}
  \centering
    \setlength{\tabcolsep}{0.2em}
    \def\arraystretch{0.9}
  \begin{tabular}{llcccc}
    \toprule
    \textbf{\makecell{Case Study\\ (\# of GPU Metrics)}} & \textbf{Model} & \textbf{Accuracy} & \textbf{Precision} & \textbf{Recall} & \textbf{F1} \\
    \midrule
    I. \makecell[l]{AR/VR Standalone App (4)}                & LSTM & 89.38\% & 0.913 & 0.894 & 0.889 \\
                                         & CNN  & 89.92\% & 0.924 & 0.899 & 0.899 \\
                                         & \textbf{RF}   & \textbf{95.83\%} & \textbf{0.963} & \textbf{0.958} & \textbf{0.958} \\
                                         & SVM  & 72.72\% & 0.755 & 0.727 & 0.716 \\
    \midrule
    \makecell[l]{II. WebXR App (3)}              & LSTM & 68.09\%     & 0.668       & 0.667        & 0.639       \\
                                         & CNN  & 70.66\%     & 0.734       & 0.703        & 0.672       \\
                                         & \textbf{RF}   & \textbf{97.15\%}     &\textbf{ 0.968}       &\textbf{ 0.969}        & \textbf{0.966}       \\
                                         & SVM  & 13.39\%     & 0.124       & 0.153        & 0.127       \\
    \midrule
    III. Virtual Object Detection (3)       & LSTM & 67.62\%     & 0.731       & 0.657        & 0.649       \\
                                         & CNN  & 37.38\%     & 0.398       & 0.407        & 0.331       \\
                                         & \textbf{RF}   & \textbf{93.57\%}     & \textbf{0.933}       & \textbf{0.935}        & \textbf{0.930}       \\
                                         & SVM  & 24.76\%     & 0.209       & 0.263        & 0.209       \\
    \midrule
    IV. Meeting Room Inference (3) & LSTM & 90.00\% & 0.850       & 0.900        & 0.867       \\
                                         & \textbf{CNN}  &\textbf{100.00\%}     &\textbf{ 1.000}       &\textbf{ 1.000}        &\textbf{ 1.000}       \\
                                         & \textbf{RF}   & \textbf{100.00\%}     &\textbf{ 1.000}       &\textbf{ 1.000}        &\textbf{ 1.000}       \\
                                         & SVM  & 88.33\%     & 0.878       & 0.883        & 0.876       \\
    \bottomrule
  \end{tabular}
    % \caption{Cross‑device evaluation of \system: final test results (Accuracy, Precision, Recall, F1) for four classifiers (CNN, LSTM, RF, SVM) across the four case studies on the entire Meta Quest series.}
  \label{tab:cross_series}
\end{table}

%\onecolumn

%\newpage

\section{AR/VR Standalone Application List}\label{sec:appendixAppFPList} 

\begin{table}[ht]
    \caption{Categorization of top 100 AR/VR standalone applications from the Meta Quest app store.}
  \centering
  \scriptsize
  \setlength{\tabcolsep}{3pt}
  \renewcommand{\arraystretch}{1.1}
  \begin{tabularx}{\columnwidth}{@{}l X@{}}
    \toprule
    \textbf{Category} & \textbf{Application Name} \\
    \midrule
    \textbf{Gaming (31)} &
      Gorilla Tag, PokerStars VR, GunRaiders, CardsTankards, Supernatural Shootout, ForeVR Cornhole, Roblox, Yeeps, Innerworld, SHARKS, Big Ballers VR, Noclip VR, Monkey Doo, HyperDash, Primal Apes, Beastcraft, Hell Horde:Mixed Reality Survival, Dissection Simulator:Frog Edition, Truck Parking Simulator VR Demo, Gun Shooting Range with Pistol, Running Monkeys, A2RL VR, Animal Company, Scary Baboon, Untangled, Gorilla Warzone, Modified Bed Bath Simulator, WIN Reality Baseball, Machine Shop Simulator, NeVR Fear The Dentist, Preflight Simulator \\
    \addlinespace
    \textbf{Entertainment (21)} &
      Bigscreen Beta, Anne Frank House VR, Epic Roller Coasters, Amazon Prime Video, Oktoberfest, 4XVRVideoPlayer, Pianogram, Campfire, VENTA X, Notre‑Dame de Paris, StartVR Streaming Video Player, FloatVR Relaxation and Focus, Mobile VR Station, PlayAniMaker Beta, Wolvic, ecosphere, Fluid, Spaceframe, Prism, Maloka, Viso \\
    \addlinespace
    \textbf{Fitness (10)} &
      Xtadium, Alcove, TRIPP, FitXR, Shado Running, VRFS, Litesport, XRWorkout, Baseball Softball Training, Hoame \\
    \addlinespace
    \textbf{Social (7)} &
      Horizon Worlds, Multiverse, VRChat, WhatsApp, Twitch:Live Streaming, MeetinVR, Flipside \\
    \addlinespace
    \textbf{Productivity (27)} &
      Immersed, Horizon Workrooms, Meta Quest Browser, Engage, Spatial, Human Anatomy VR, Lab Monkey, GRAB, vSpatial, ALVR, Hard Drive, Nanome, VR Anatomy Lab, SketchUp Viewer, IMMERSE–Language Learning, Arkio, ShapesXR, Doodle Board, Gravity Sketch, Resolve, Mistika VR Connect, DelTrain Adult ICU Delirium, Arthur, Noda, Naer, Masterpiece, Open Brush–3D Painting \\
    \addlinespace
    \textbf{Mixed Reality (4)} &
      A2RL VR, Hell Horde:Mixed Reality Survival, Hello Dot, Hololight Space \\
    \bottomrule
  \end{tabularx}
  % \caption{Categorization of top 100 AR/VR standalone applications from the Meta Quest app store.}
  \label{table:AppList}
\end{table}

\newpage
%\onecolumn
\section{WebXR Application List}\label{sec:appendixWebXRFPList} 
\begin{table}[ht]
    \caption{Categorization of 100 WebXR applications.}
  \centering
  \scriptsize
  \setlength{\tabcolsep}{3pt}
  \renewcommand{\arraystretch}{1.1}
  \begin{tabularx}{\columnwidth}{@{}l X@{}}
    \toprule
    \textbf{Category} & \textbf{Application Name} \\
    \midrule\textbf{Gaming \& Entertainment(19)} & moonrider.xyz/, webvr.soundboxing.co/, plockle.com/play, spiderman.webvr.link/, jorgefuentes.net/projects/halloVReen/, aboveparadowski.com/, anumberfromtheghost.com/, jorgefuentes.net/projects/vuppets/vuppets\_6DOF, spacerocks.moar.io, blocksarcade.xyz/, xrpet.me/, crossthestreet.fun/game/, aframe.io/a-blast/, micosmo.com/trajectilecommand, beatknightxr.web.app, play.js13kgames.com/human-not-found/, worldsdemolisher.totalviz.com, slime-freighter.glitch.me/?autoplay=true, play.js13kgames.com/teleport/ \\ \addlinespace
    \textbf{Art \& Creativity (11)} & aframe.io/a-painter/, brushworkvr.com/paint, vartiste.xyz/, esc.art/, flowerbed.metademolab.com/, artsalad.net/, castle.needle.tools/, cecropia.github.io/thehallaframe, demos.littleworkshop.fr/track, hatsumi.netlify.app, framevr.io/frame-tutorial \\
    \addlinespace
    \textbf{Tours \& Exploration (18)} &
      immersive-web.github.io/webxr-samples/360-photos.html, immersive-web.github.io/webxr-samples/stereo-video.html, aframe.city/, jorgefuentes.net/projects/puppetrilla, a-frame--360-vr-tour.glitch.me, forestwave.glitch.me, 3xr.space, travisbarrydick.github.io/vr-planets/dist, xrdinosaurs.com, msub2.github.io/hello-webxr, live.arrival.space/welcome, codercat.xyz/ma, anvropomotron.com, codercat.xyz/dying-to-find, a.flow.gl/flow/i2nxns/display, codercat.xyz/three-body, new-cesium-a-frame-test.glitch.me, modelo-3d-a-frame.glitch.me \\
    \addlinespace

    \textbf{Health \& Fitness (8)} & towermax.fitness/tower/, towermax.fitness/peakfreak/, towermax.fitness/reaction/, towermax.fitness/punchingballgames/, towermax.fitness/slingsurf/, towermax.fitness/towerhockey/, towermax.fitness/stepuprun/, towermax.fitness/drilltrack/ \\ \addlinespace
    \textbf{Demonstration (44)} & From \url{aframe.io/aframe/examples/}—anime-UI, shopping, comicbook, spheres-and-fog, dynamic-lights, hand-tracking-grab-controls, boilerplate/ar-hello-world, test/fog, test/visibility, performance/animation-raw, animation/arms, animation/pivots, animation/unfold, performance/multiview-extension/?multiview=on, test/shadows, primitives/torus, docs/aincraft; From \url{immersive-web.github.io/webxr-samples}—Reduced-bind-rendering.html, room-scale.html, input-tracking.html, input-selection.html, controller-state.html, positional-audio.html, anchors.html; From \url{https://glitch.com/}—a-frame-particules-rain.glitch.me, tree-by-a-frame-by-jenjira-santaw.glitch.me, a-frame-goofs.glitch.me, ma3120-a-frame-space-image-02.glitch.me, ma3120-a-frame-image-layered-repetition.glitch.me, ma3120-a-frame-texture-and-grab.glitch.me, military-heathered-creator.glitch.me, fluttering-robust-kiwi.glitch.me, a-frame-spinosaurus-for-vr.glitch.me, a-frame-physics-entities-balls-fall-from-the-sky.glitch.me, a-frame-wolf.glitch.me, a-frame-stationary-caustics.glitch.me, celda-de-trabajo-automatizada-en-a-frame.glitch.me, a-frame-snowman-rv.glitch.me, simple-solarsystem-a-frame.glitch.me, city-whit-a-frame-5a.glitch.me, lmc-2700-a-frame-project.glitch.me, a-custom-a-frame-scene.glitch.me, dune-a-frame-navmesh.glitch.me, a-frame-tree-model-loading.glitch.me \\
    \bottomrule
  \end{tabularx}
  % \caption{Categorization of 100 WebXR applications.}
  \label{table:webxr_apps}
\end{table}